\documentclass{aa}  

\usepackage{graphicx}
\usepackage{txfonts}
\usepackage{mhchem, color}
\newcommand{\yr}{\,\mathrm{yr}}
\newcommand{\au}{\,\mathrm{au}}
\newcommand{\Kelvin}{\,\mathrm{K}}
\newcommand{\unit}[1]{\,\mathrm{#1}}
\newcommand{\pc}{\unit{pc}}
\newcommand{\um}{\unit{\mu m}}
\newcommand{\mum}{\um}
\newcommand{\cm}{\unit{cm}}
\newcommand{\kms}{\unit{km}\unit{s}^{-1}}
\newcommand{\erg}{\unit{erg}}
\newcommand{\ergs}{\unit{erg}\unit{s}^{-1}}
\newcommand{\Myr}{\unit{Myr}}
\newcommand{\rad}{\unit{rad}}
\newcommand{\eV}{\unit{eV}}
\newcommand{\Msun}{\,M_\odot}

\newcommand{\Rsun}{\,R_\odot}
\newcommand{\Lsun}{\,L_\odot}
\newcommand{\LFUV}{\,L_{\rm  FUV}}

\newcommand{\fref}[1]{Figure~\ref{#1}}
\newcommand{\tref}[1]{Table~\ref{#1}}
\newcommand{\secref}[1]{Section~\ref{#1}}
\newcommand{\eqnref}[1]{Eq.\eqref{#1}}

\newcommand{\deltae}{E_{\rm up}}
\newcommand{\Trot}{T_{\rm rot}}
\newcommand{\nh}{n_{\rm H}}
\newcommand{\barNHmol}{\bar{N}_{\rm H2}}
\newcommand{\dotMacc}{\dot{M}_{\rm acc}}

\newcommand{\cs}{c_{\rm s}}

\newcommand{\dgratio}{\mathcal{DG}}
\newcommand{\openingangle}{\theta_{\rm open}}
\newcommand{\aveopeningangle}{\bar{\theta}_{\rm open}}
\newcommand{\Rgeo}{R_{\rm geo}}
\newcommand{\rext}{r_{\rm ext}}
\newcommand{\abn}[1]{y_{\rm #1}}

\newcommand{\e}[1]{\times10^{#1}}
\newcommand{\dd}{\mathrm{d}}
\newcommand{\braket}[1]{\left(#1\right)}

\newcommand{\revision}[1]{{#1}}

\newcommand{\NameOfTauxxxx}{Tau~042021}

\begin{document}

   \title{Photoevaporation Can Reproduce Extended \ce{H2} Emission from Protoplanetary Disks Imaged by JWST \revision{MIRI}}


   \author{R. Nakatani
          \inst{1}
          \and
          G. Rosotti
          \inst{1}
          \and
          B. Tabone
          \inst{2}
          \and
          A. Sellek
          \inst{3}
          }

   \institute{Dipartimento di Fisica, Universit\`a degli Studi di Milano, Via Celoria, 16, I-20133 Milano, Italy\\
              \email{ryohei.nakatani@unimi.it}
            \and 
            Universit\'e Paris-Saclay, CNRS, Institut d'Astrophysique Spatiale, 91405, Orsay, France
         \and
             Leiden Observatory, Leiden University, PO Box 9513, 2300 RA Leiden, the Netherlands
             }


   \date{Received xxx; accepted xxx}

 
  \abstract
   {Understanding dispersal of protoplanetary disks remains a central challenge in planet formation theory. Disk winds, driven by magnetohydrodynamics (MHD) and/or photoevaporation, are now recognized as primary agents of dispersal. With the advent of James Webb Space Telescope (JWST), spatially resolved imaging of these winds, particularly in \ce{H2} pure rotational lines, have become possible, revealing X-shaped morphologies and integrated fluxes of $\sim 10^{-16}$–$10^{-15}\ergs\cm^{-2}$.}
   {However, the lack of theoretical models suitable for direct comparison has limited interpretation of these features. To address this, we present the first model of photoevaporative \ce{H2} winds tailored for direct comparison with JWST observations.}
   {Using radiation hydrodynamics simulations coupled with chemistry, we derive steady-state wind structures and post-process them to compute \ce{H2} level populations and line radiative transfer, including collisional excitation and spontaneous decay.}
   {Our synthetic images reproduce the observed X-shaped morphology with radial extents of $\gtrsim 50$–$300\au$ and semi-opening angles of $\sim 37^\circ$–$50^\circ$, matching observations of \NameOfTauxxxx{} and SY~Cha. While the predicted line fluxes are somewhat lower than the observed values, 
   they remain broadly consistent for lower-$J$ transitions, despite the model not being specifically tailored to this source. 
    }
   {These results suggest that photoevaporation is a viable mechanism for reproducing key features of observed \ce{H2} winds, including morphology and fluxes, though conclusive identification of the wind origin requires source-specific modeling.
   This challenges the reliance on geometrical structures 
   alone to distinguish between MHD winds and photoevaporation. 
    Based on our findings, we also discuss alternative diagnostics of photoevaporative winds.
    This work provides a critical first step toward interpreting spatially resolved \ce{H2} winds and motivates future modeling efforts. 
    }
  
   \maketitle
%

\section{Introduction} \label{sec:intro}

Understanding how and when protoplanetary disks (PPDs) disperse is a fundamental challenge in planet formation. Resolving this issue is crucial since disk dispersal limits the time and material available for building planets and alters disk-planet interactions, ultimately influencing planetary system architectures.

PPDs lose mass not only through accretion onto the central star but also via winds launched from their surfaces, driven by either magnetic forces \citep[e.g.,][]{1982_BlandfordPayne, 1992_PelletierPudritz, 2009_SuzukiInutsuka} or photoevaporation \citep[e.g.,][]{1993_Shu_b, 1994_Hollenbach}. 
Understanding the mechanisms and relative roles of these winds across evolutionary stages is key to addressing the disk dispersal problem.   



Observational evidence for disk winds has accumulated over decades, starting with the detection of blueshifted forbidden lines in optical spectra \citep[e.g.,][]{1983_Jankovics, 1984_Appenzeller, 1987_Edwards}, with further advances in infrared spectroscopy, particularly with Spitzer and ground-based facilities (e.g., Very Large Telescope and Keck). 
In Class~II sources, most evidence comes from spatially unresolved, blueshifted forbidden lines (e.g., [\ion{O}{I}], [\ion{S}{II}], [\ion{N}{II}], [\ion{Ne}{II}]), commonly seen in T~Tauri stars.
These lines typically show a high-velocity component (HVC; $> 50\kms$) associated with jets, and a low-velocity component (LVC; $\lesssim30\kms$) attributed to disk winds \citep[e.g.,][]{1995_Hartigan, 1995_KwanTademasu, 1999_Cabrit}.

The LVC is often decomposed into a broad component (BC; $\mathrm{FWHM} \sim 100\kms$, $v > 10\kms$) and a narrow component (NC; $\mathrm{FWHM} \sim 30\kms$, $v \sim 3\kms$ ) 
\citep{2013_Rigliaco, 2016_Simon, 2018_Mcginnis, 2018_Fang, 2019_Banzatti}. 
Assuming Keplerian broadening, the BC likely traces inner MHD winds ($0.05$--$0.5\au$), while the NC likely arises from farther out ($0.5$--$5\au$), consistent with either MHD or photoevaporative winds. 
However, the distinction remains debated \citep{2019_Banzatti, 2020_Weber, 2021_Whelan}. 
%


In addition to ionic and atomic emission, blueshifted molecular lines, such as \ce{H2} ($v=1$--$0$~S(1); $2.12\um$) and CO ($\Delta v = 1$; $\sim 4.7\um$), have revealed slow, warm molecular winds launched at radii typical of the inner disk. This \ce{H2} line traces slow winds ($< 20\kms$) launched from $0.05$--$20\au$ \citep{2020_Gangi}, with their origin debated between MHD winds \citep{2012_Panoglou} and photoevaporation \citep{2022_Rab}. 
In contrast, CO emission line profiles generally favor an MHD origin from the inner disk \citep{2011_Pontoppidan, 2013_Brown, 2022_Banzatti}. Narrower blue-shifted absorption features also point to the presence of an outer disk wind \citep{2013_Brown, 2022_Banzatti}, with the origins remaining unclear.

Prior to JWST, spatially resolved observations have also captured low-velocity winds for a handful of sources via near-infrared and UV emissions of \ce{H2} \citep{2008_Beck, 2013_Schneider, 2014_Agra-Amboage, 2019_BeckBary, 2023_Melnikov}, as well as submillimeter CO emission detected by ALMA \citep{2018_Gudel, 2018_Louvet, 2020_deValon, 2023_Launhardt}. Spectro-astrometry has further revealed their spatial extents via CO infrared emission and optical forbidden lines from \ion{O}{I} and \ion{S}{II} \citep{2011_Pontoppidan, 2021_Whelan}. 
More recently, VLT/MUSE observations have also provided spatially resolved maps of the optical forbidden lines \citep{2023_FangNature, 2023_FloresRivera}. 
These winds typically show a wide, less-collimated geometry surrounding faster, narrower atomic and ionic jets/winds, forming nested onion-like kinematic structures.

Overall, despite significant progress, distinguishing MHD winds and photoevaporation
remains a central challenge. 
This distinction is critical for understanding where and at what rates disk material is removed, identifying accretion-driving mechanisms---since MHD winds can remove angular momentum  \citep{2015_Gressel, 2016_Bai, 2023_Lesur},
and testing disk evolution scenarios \citep{2001_Clarke, 2016_Suzuki, 2020_Pascucci, 2020_Kunitomo, 2022_Tabone, 2023_Manara}. 
Notably, direct confirmation of photoevaporative winds has remained elusive, though the LVC of [\ion{Ne}{II}] has been highlighted as a promising tracer of photoevaporative winds \citep{2008_Alexander, 2009_PascucciSterzik, 2011_Pascucci, 2014_Alexander, 2020_Ballabio, 2020_Pascucci}. 

The advent of JWST offers a major opportunity to address these issues, thanks to its high spatial resolution and high sensitivity mid-infrared spectroscopy mode. 
Various wind tracers are accessible with JWST/MIRI-MRS and NIRSpec, including [\ion{Ne}{II}], [\ion{Ne}{III}], [\ion{Ar}{II}], [\ion{Ar}{III}], \ce{H2}, and CO \citep{2024_Bajaj, 2024_SellekBajaj, 2024_Arulanantham,2024_Anderson, 2025_Pascucci, 2025_Schwarz}. 
Among them, pure rotational lines of \ce{H2} are especially compelling: not only as the primary constituent of disk gas, but these lines can trace spatially extended ($\gtrsim 10^2\au$) warm winds ($\lesssim 10^3\Kelvin$), unveiling the nature of molecular winds. 
Indeed, recent JWST observations have revealed such extended \ce{H2} emission \citep{2024_Arulanantham, 2024_Anderson, 2025_Pascucci, 2025_Schwarz}.

On the theoretical side, radiation (magneto)hydrodynamical models coupled with chemistry are now available \citep{2017_Wang, 2018_Nakatani, 2018_Nakatanib, 2019_Wang, 2020_Gressel, 2021_Komaki, 2024_Sellek, 2025_Hu}, enabling detailed comparisons with JWST data. 
Notably, purely hydrodynamical simulations have predicted photoevaporative \ce{H2} winds with structures 
broadly consistent with observations \citep{2018_Nakatani, 2018_Nakatanib, 2021_Komaki, 2024_Sellek}, but it remains unclear whether they can quantitatively reproduce the observed morphology and fluxes.

In this paper, we explore how photoevaporative \ce{H2} winds manifest in JWST observations of pure rotational lines, aiming to establish the first theoretical framework for interpreting current and future JWST observations. 
We pose two core questions: (1) what morphologies do these \ce{H2} lines from photoevaporative winds exhibit?, and (2) what fluxes do they produce? 
To this end, we generate synthetic \ce{H2} intensity maps and line fluxes, leveraging radiation-hydrodynamical simulations with post-processing non-LTE level population calculation and line radiative transfer. 

The structure of this paper is as follows: \secref{sec:methods} describes the computational methods; 
\secref{sec:results} presents the results, answering the core questions; 
\secref{sec:discussions} discusses model limitations and generalization, as well as comparisons with recent JWST observations; and finally, \secref{sec:summary} summarizes the main findings.

\section{Methods} \label{sec:methods}
To generate synthetic \ce{H2} intensity maps and fluxes,
we adopt a two-step approach. 
First, we perform a radiation hydrodynamics simulation of a photoevaporating PPD with non-equilibrium thermochemistry, to obtain a self-consistent physical structure featuring \ce{H2} winds. 
We then post-process the result with RADMC-3D \citep{2012_Dullemond}, computing non-LTE \ce{H2} level populations and performing line radiative transfer for the pure rotational transitions $v = 0$--$0$ S(1)--S(9), including the dust opacity. 

The hydrodynamics simulation employs a modified version of the PLUTO code \citep{2007_Mignone}, 
incorporating thermochemistry and radiation transfer modules \citep[][see also Appendix~\ref{sec:detailed_description_chemistry}]{2010_Kuiper, 2013_Kuiper, 2018_Nakatani, 2018_Nakatanib, 2020_Kuiper, 2021_Nakatani}.
The chemical network includes key PDR reactions as well as extreme-ultraviolet (EUV) and X-ray photoionization and secondary ionization processes. 
The thermochemistry module has been benchmarked against standard PDR codes and the DALI code (Appendices~\ref{sec:benchmark:roellig} and \ref{sec:benchmark:dali}). 

To keep this section concise, we summarize the main features of the setup in \secref{sec:methods:hydro}, with details in Appendix~\ref{sec:detailed_sim_setup}. 
Post-processing methods are described in \secref{sec:methods:radmc}.
We assume axisymmetry and midplane symmetry throughout.

\subsection{Hydrodynamics Model: Overview}    \label{sec:methods:hydro}

\begin{table}[htp]
    \caption{Simulation setup of the fiducial model}
    \centering
  \begin{tabular}{lc}   
    \hline \hline
    [{\it Stellar properties}]                                                       \\
    Mass $(M_*)$     				& $1 \Msun$ \\
    Accretion rate $(\dotMacc)$     & $10^{-10}\Msun\yr^{-1}$ \\
    Radius $(R_*)$                 	& $2.61\Rsun$ \\
    Effective Temperature           & $4278\Kelvin$\\
    FUV luminosity $(\LFUV)$        & $9.20\e{29}\ergs$\\
    EUV luminosity $(L_{\rm EUV})$  & $2.04\e{29}\ergs$\\
    X-ray luminosity $(L_{\rm X})$	& $2.82\e{30}\ergs$   \\	
    \hline
    [{\it Disk properties}]                                                       \\
    Disk mass $(M_{\rm disk})$  	&    $10^{-2}\Msun$\\
    Carbon abundance $(\abn{C})$                &  $1.35\e{-4}$ \\
    Oxygen abundance $(\abn{O})$                &  $2.88\e{-4}$\\
    Small-dust-to-gas mass ratio $(\dgratio)$	&  0.0015 \\
    PAH abundance w.r.t. ISM $(f_{\rm PAH})$     & 0.1 \\ 
    \hline 
    [{\it Numerical configuration}]               \\
    Computational domain &  $r \approx [1.8, 350] \au $\\
                         &  $\theta = [0, \pi/ 2] \rad $ \\
    \hline
  \end{tabular}
    \label{tab:fiducialmodel}
\end{table}

Our fiducial model is a PPD with mass $M_{\rm disk} = 10^{-2}\Msun$ orbiting a solar mass star ($M_* = 1\Msun$).
We assume a relatively evolved system with a low accretion rate $\dotMacc = 10^{-10}\Msun \yr^{-1}$, motivated by the facts that (1) at the early stages, stellar ultraviolet (UV) and X-ray can be heavily screened by disk winds from the inner regions; (2) photoevaporation is expected to dominate disk dispersal at later stages \citep{2020_Pascucci, 2022_Pascucci, 2020_Kunitomo, 2023_Weder}; 
and (3) low $\dotMacc$ is reported for systems with \ce{H2} winds: $6.6\e{-10}\Msun\yr^{-1}$ for SY~Cha, $2\times 10^{-11}\Msun\yr^{-1}$ for \NameOfTauxxxx{}, and $7.1\e{-10}\Msun\yr^{-1}$ for CX~Tau, though 
the first two estimates have been recently revised and therefore remain uncertain (see \secref{sec:jwst_obs}).
Higher $\dotMacc$ cases are explored in \secref{sec:discussions:luminosity_dependence}.
Note that in our setup, $\dotMacc$ only affects the accretion-generated UV emission as defined below, although in reality it likely also impacts EUV and X-ray emission \citep{2025_Shoda}.

The disk is initially in hydrostatic equilibrium and fully molecular (Appendix~\ref{sec:hydro:initial_condition} for more details). We evolve the system until it reaches a quasi-steady state by solving the equations for gas density $\rho$, velocities $\vec{v} = (v_r,~ v_\theta, ~ v_\phi)$, energy, and chemical abundances $\{y_i | i = \ce{H, H^+, H2, ...}\}$. 
Our approach naturally includes advection of chemical species, which is crucial for accurately capturing photoevaporative \ce{H2} winds, where photochemical timescales can be comparable to wind crossing time (\secref{sec:results:wind_properties}).

Our chemical network includes $\sim 140$ reactions (\tref{tab:chem_reac_list}) among 27 gas-phase chemical species:
H, \ce{H+}, \ce{H2}, \ce{H2+}, \ce{H-}, O, C, \ce{C+}, CO, \ce{e-},
\ce{He}, \ce{He+}, \ce{H3+}, CH, \ce{CH+}, \ce{CH2}, \ce{CH2+}, \ce{CH3+}, \ce{CO+}, 
\ce{HCO+}, OH, \ce{OH+}, \ce{H2O}, \ce{H2O+}, \ce{H3O+}, \ce{O+}, 
and \ce{O2}.
We assume the elemental abundances of $\abn{C} = 1.35\e{-4}$ and $\abn{O} = 2.88\e{-4}$ for the gas-phase carbon and oxygen, respectively \citep{2006_Jonkheid, 2009_Woitke}.

Heating processes include: EUV/X-ray photoionization heating \citep{1996_Maloney, 2000_Wilms, 2004_Gorti, 2018_Nakatani, 2018_Nakatanib}, 
far-ultraviolet (FUV) grain photoelectric heating \citep{1994_BakesTielens}, 
FUV \ce{H2} photodissociation heating and pumping \citep{1979_HollenbachMcKee, 1996_DraineBertoldi}, 
and FUV \ion{C}{I} ionization heating \citep{1987_Black, 2004_Jonkheid, 2012_UMIST}. 
Cooling processes include: 
\ion{H}{II} radiative recombination \citep{1978_Spitzer}; 
\ion{H}{I} Ly${\rm \alpha}$ cooling \citep{1997_Anninos}; 
fine-structure line cooling by \ion{O}{I}, \ion{C}{I}, and \ion{C}{II} \citep{1989_HollenbachMcKee,1989_Osterbrockbook,2006_SantoroShull};
and \ce{H2}/CO rovibrational cooling \citep{1998_GalliPalla,2010_Omukai}. 
Additional heating/cooling includes dust-gas collisional heat exchange \citep{2002_CazauxTielens}
and chemical heating/cooling \citep{1979_HollenbachMcKee, 2000_Omukai}.

FUV, EUV, and X-ray transfer is solved by 1D ray tracing along the radial direction \citep[][]{2018_Nakatani, 2018_Nakatanib}.
For optical and infrared radiation, a 2D hybrid scheme is used: stellar irradiation via ray tracing, and dust (re-)emission via flux-limited diffusion \citep{2010_Kuiper, 2013_Kuiper, 2020_Kuiper}.

We construct the spectrum of our $1\Msun$ central star by summing three components: photospheric emission, accretion-generated UV emission, and high-energy radiation (EUV and X-ray) from the hotter atmospheric components. 
The photospheric emission is modeled as a blackbody with a bolometric luminosity of $L_* = 2.34\Lsun$ and the effective temperature of $T_{\rm eff} = 4278\Kelvin$ \citep{2000_Siess, 2009_Gorti}. These are typical for a young star ($\sim 1\Myr$), but our results are not very sensitive to the exact values.

Accretion-generated UV is also represented by a $T_{\rm eff} = 10^4\Kelvin$ blackbody. 
With $\dotMacc = 10^{-10}\Msun\yr^{-1}$ and a stellar radius of $R_* = 2.61 R_\odot$ \citep{2000_Siess, 2009_Gorti}, the resulting accretion luminosity is $L_{\rm acc} = GM_* \dotMacc /R_* \approx 1.2\e{-3}\Lsun  \approx 4.6\e{30}\ergs $. 

For the X-ray component, we adopt the modeled X-ray spectrum of TW~Hya from the DIANA project \citep{2019_Woitke}. 
Note that TW~Hya has a relatively soft X-ray spectrum, which leads to efficient heating at lower column densities.
The EUV spectrum is approximated by a linear interpolation between the X-ray flux at $100\eV$ and the combined photospheric and accretion-originated flux at the Lyman limit ($\approx 13.6\eV$). 

\begin{figure}
    \centering
    \includegraphics[width=\linewidth, clip]{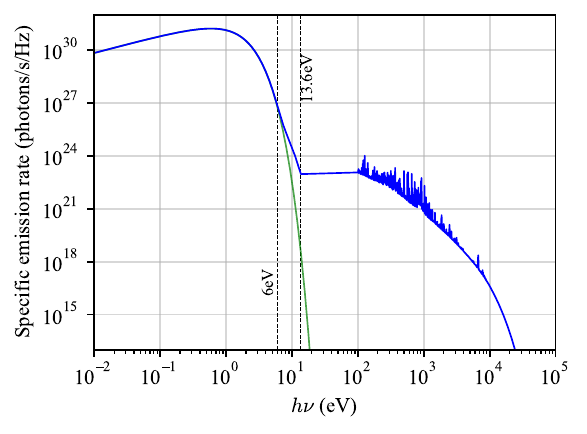}
    \caption{
    Stellar spectrum adopted in our fiducial model. The vertical dashed lines mark the lower energy limits of the FUV and EUV bands. The lower limit of the X-ray band is $100\eV$. (See also \tref{tab:fiducialmodel} for the integrated luminosities.)
    The thin green line shows the photospheric emission, which overlaps with the blue line below $6\eV$, for reference. 
    }
    \label{fig:stellar_spectrum}
\end{figure}

The synthesized stellar spectrum is shown in \fref{fig:stellar_spectrum},
with the total band-integrated luminosities of $L_{\rm FUV}=9.2\e{29}\ergs$, $L_{\rm EUV} = 2.0\e{29}\ergs$, and $L_{\rm X} = 2.8\e{30}\ergs$. 

We assume a dust-to-gas mass ratio of 0.01, with 85\% of the dust mass in larger grains that have settled toward the midplane. This reduces the abundance of small grains in the upper disk surface, where photoevaporative winds arise. 
Accordingly, we scale the rates of dust-gas collisional heat transfer, grain-catalyzed \ce{H2} formation by a factor of 0.15. 
The dust opacity is likewise scaled (see Appendix~\ref{sec:detailed_description_chemistry} for the effects on FUV-driven photochemistry). 

We also consider the depletion of polycyclic aromatic hydrocarbons (PAHs), which are detected in only 10\% of PPDs around low-mass stars and are typically underabundant by factors greater than 10 compared to the ISM \citep{2007_Geers,2010_Oliveira,2013_Vicente}. 
Since PAHs contribute to grain photoelectric heating \citep{1994_BakesTielens}, we scale this heating rate by a depletion factor $f_{\rm PAH}$. In this study, we adopt $f_{\rm PAH} = 0.1$. Even lower values do not significantly impact our results at the assumed $L_{\rm FUV}$.

The simulations are carried out in 2D spherical polar coordinates $(r, \theta)$.
The computational domain spans $r = [0.2, 40]\times R_{\rm g} \approx [1.8, 350]\au$ and $\theta = [0, ~\pi/2] {\rm \, rad}$, where
\begin{equation}
    R_{\rm g} \equiv \frac{GM_*}{\braket{10\kms}^2} \approx 8.9\au \braket{\frac{M_*}{\Msun}}.
\end{equation}
is the gravitational radius for gas at $T\approx 10^4\Kelvin$. 
We use $N_r \times N_\theta = 144 \times 160$ grid cells, with logarithmic spacing in the radial direction.
In the polar direction, uniform spacing is applied, but with different resolutions below and above $\theta = 1\rad$:
each region is divided into 80 cells to provide the finer resolution necessary to resolve the sharp pressure gradient within the disk.

Integration is performed using operator splitting: the hydrodynamics update is applied first---where advection and adiabatic cooling are added---followed by radiative transfer and thermochemistry. In the thermochemistry step, temperature and chemical abundances are updated simultaneously using an implicit Newton-Raphson scheme.

\subsection{Post-Process}    \label{sec:methods:radmc}
We post-process the steady-state structures obtained from our simulations to generate synthetic intensity maps and fluxes for \ce{H2} pure rotational lines, specifically from S(1) to S(9) transitions. This involves calculating non-LTE \ce{H2} level populations and performing subsequent line radiative transfer using RADMC-3D \citep{2012_Dullemond}.

The level populations are computed using the ``Optically Thin non-LTE'' method, appropriate for the typically optically thin nature of the \ce{H2} lines. 
Only collisional excitation, de-excitation, and spontaneous decay are considered. We neglect \ce{H2} pumping by UV, X-ray, and chemical reactions, though the hydrodynamics simulation includes FUV \ce{H2} pumping and chemical heating as heating sources (\secref{sec:methods:hydro}); the impacts of this simplification are discussed in \secref{sec:discussions:pumping}.
It is worth noting that while the population calculations do not incorporate line excitation or de-excitation, self-absorption is accounted for in the subsequent line transfer.

Molecular data for \ce{H2}, including Einstein A coefficients and collisional rate coefficients for collisions with H, para-\ce{H2}, ortho-\ce{H2}, \ce{e-}, and \ce{H+} \citep{2019_Roueff, 2015_Lique, 2021_Gonzalez-lezana, 2000_Flower, 2021_Flower}, are taken from the EMAA database (\url{https://emaa.osug.fr} and \url{https://dx.doi.org/10.17178/EMAA}). 
In the post-processing, we input the density, velocities, temperature, \ce{H2} abundance, and the abundances of the collision partners obtained by the simulations. 
The para-\ce{H2} and ortho-\ce{H2} abundances are specified by assuming an ortho-to-para ratio of 3. 

Unless otherwise noted, we assume a distance to the source of $d = 140\pc$.
We explore various disk inclinations and both dust-free and dust-obscured line emission cases.

For the fiducial case, we adopt an edge-on view ($i = 90^\circ$) and neglect dust obscuration to isolate the intrinsic morphology and fluxes. This serves as a baseline for comparisons. 

Dust-obscured models use RADMC-3D's default silicate opacity and assume a dust-to-gas mass ratio of 
0.01, which is about ten times higher than in the hydrodynamics simulation, to probe the upper-end effects of disk obscuration.
Comparing these results with the dust-free case provides insight into the impact of intermediate dusty levels.
Increasing dust opacity does not alter the thermochemical structures of the \ce{H2} winds in our simulation, so this treatment does not affect the internal consistency.

\section{Results}   \label{sec:results}
We present the hydrodynamics simulation results and discuss the steady-state structures in \secref{sec:results:wind_properties}, followed by an analysis of the system energetics in \secref{sec:results:energetics}. 
Synthetic intensity maps and fluxes at $i = 90^\circ$ without disk obscuration are shown in Sections~\ref{sec:results:images} and \ref{sec:results:fluxes}, respectively. 
\secref{sec:results:prop_traced_by_line} examines the density and temperature regimes traced by each line, followed by a rotation diagram analysis in \secref{sec:results:rot_diagram}.
The effects of disk inclination and disk obscuration are addressed in \secref{sec:results:inclinations}.
\revision{Finally, we discuss spectrally resolved line profiles from the models with disk obscuration in \secref{sec:results:velocity_profiles}. }

\subsection{Properties of Photoevaporative \ce{H2} Winds}
\label{sec:results:wind_properties}
\begin{figure}[h!tbp]
    \centering
    \includegraphics[width=\linewidth, clip]{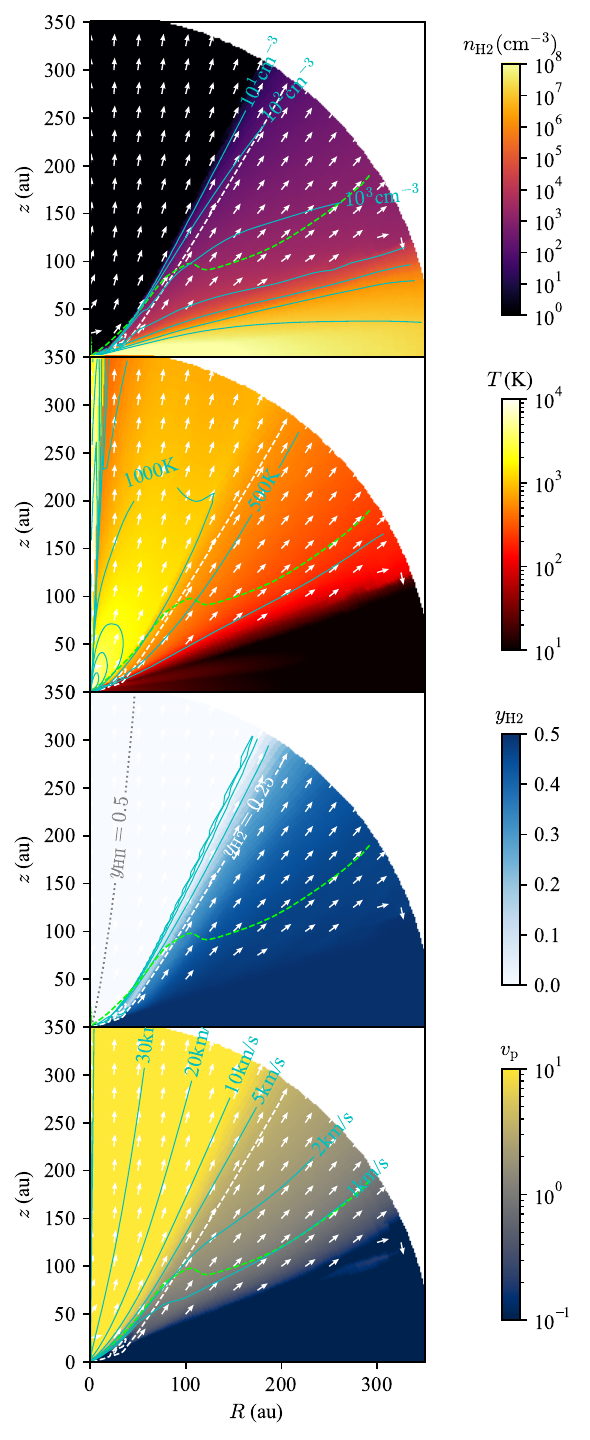}
    \caption{
    Snapshot of \ce{H2} density (top), gas temperature (second), \ce{H2} abundance (third), and poloidal velocity (bottom).  
    Cyan contours indicate isosurfaces for each quantity: $n_{\rm H2} = 10, 10^2, 10^3, 10^4, 10^5, 10^6, 10^7 \unit{cm}^{-3}$; $T = 200, 500, 1000, 2000, 3000, 5000 \unit{K}$; $y_{\rm H2} = 10^{-4}, 10^{-3}, 10^{-2}, 10^{-1}$; and $v_{\rm p} = 1, 2, 5, 10, 20, 30 \unit{km}\unit{s}^{-1}$.
    In all panels, the white and green dashed contours denote the \ce{H2} dissociation front ($y_{\rm H2} = 0.25$) and the isothermal sonic surface, respectively. 
    White arrows indicate the velocity field (direction only, not its magnitude), and are omitted where $T < 100 \unit{K}$. 
    In the \ce{H2} abundance map, the gray dotted contour marks the \ion{H}{I} ionization front ($y_{\rm HII} = 0.5$).
    }
    \label{fig:hydro}
\end{figure}
\fref{fig:hydro} shows a snapshot of the hydrodynamics simulation after the system has reached a quasi-steady state at $t \approx 10^4\yr$. 
Photoevaporative \ce{H2} winds are evident in the third panel, where the blueish region marks the \ce{H2}-dominated region, and white arrows indicate the velocity field. 
Advection plays a crucial role in shaping this \ce{H2} region, highlighting the importance to couple hydrodynamics with chemistry. 
The typical densities and temperatures of the \ce{H2} winds are $10^2\cm^{-3}\lesssim n_{\rm H2}\lesssim10^5\cm^{-3}$ and $100\Kelvin \lesssim T \lesssim 1000 \Kelvin$, respectively (the first and second panels).
X-ray heating and \ce{H2} cooling dominate in the molecular region, while (hard) EUV heating and adiabatic cooling are more important in the atomic/ionic region.

When compared at a similar distance from the star, wind temperature generally anti-correlates with density: hotter winds are less dense. This is because softer X-rays and EUV, which have larger absorption cross-sections, are absorbed at lower column densities and have shorter heating timescales. Besides, atomic and molecular coolants tend to be more abundant at higher column densities and more efficient due to the higher densities.
\revision{Since photoevaporative winds are pressure driven, higher temperatures generally lead to stronger acceleration and thus higher velocities.}
As a result, lower-density winds exhibit higher poloidal velocities (compare the top and bottom panels). 
The low-density wind component ($n_{\rm H2} \lesssim 10^3 \cm^{-3}$) is mostly supersonic, with poloidal velocities of $v_{\rm p} \gtrsim 1\kms$. 

\begin{figure*}[htbp]
    \centering
    \includegraphics[width=\linewidth, clip]{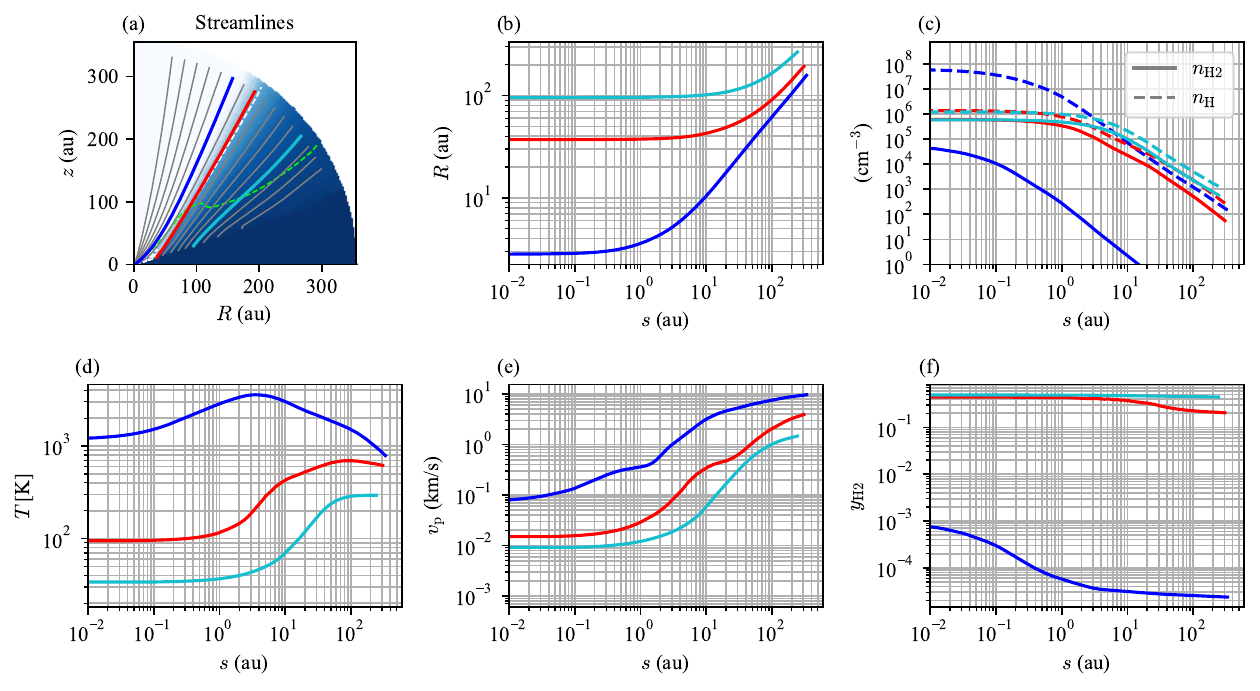}
    \caption{
    Physical properties along streamlines. (a) \ce{H2} abundance map (same as \fref{fig:hydro}) with selected streamlines shown in gray. Three representative streamlines are marked in blue, red, and cyan. (b)--(f) Profiles of cylindrical radius, densities, temperature, poloidal velocity, \ce{H2} abundance along the three streamlines, using consistent color coding. The horizontal axis indicates the distance along each streamline, with $s=0$ defined at the base. In panel (c), solid and dashed lines show \ce{H2} and hydrogen nucleus densities, respectively.
    }
    \label{fig:streamlines}
\end{figure*}
\begin{figure*}[htbp]
    \centering
    \includegraphics[width=\linewidth, clip]{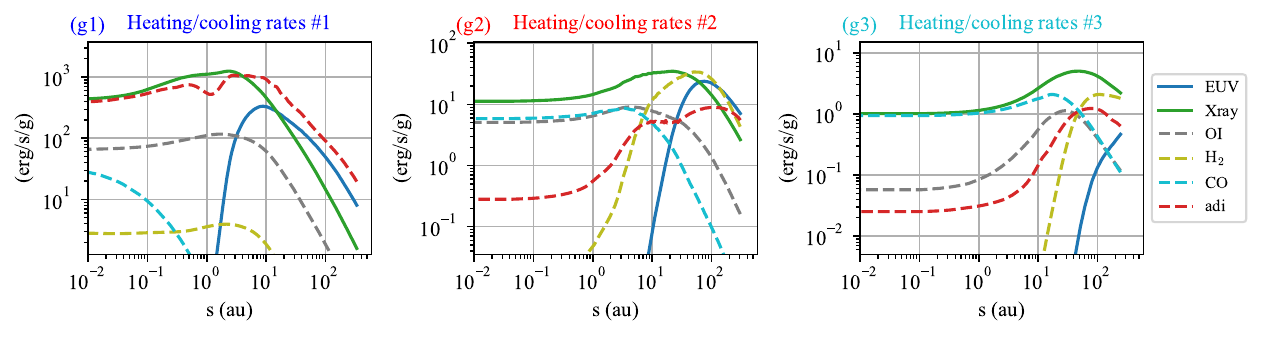}
    \caption{
    Specific rates of major heating and cooling processes along the blue, red, cyan streamlines in \fref{fig:streamlines}, shown from left to right. Labels indicate: EUV photoionization heating (``EUV''); X-ray photoionization heating (``X-ray''); line cooling from \ion{O}{I}, \ce{H2}, and CO; and adiabatic cooling (``adi'').
    (See \fref{fig:streamlines_heatcool_detailed} in Appendix for the heating and cooling rates of other processes.)
    }
    \label{fig:streamlines_heatcool}
\end{figure*}
To examine how physical quantities evolve along the flow, \fref{fig:streamlines} presents their profiles along representative streamlines in the \ion{H}{I} region (blue), near the \ion{H}{I}/\ce{H2} boundary (red), and within the \ce{H2} region (cyan), launched at $\approx 3$, 40, and $100\au$, respectively. 
The profiles are plotted as functions of the streamline coordinate $s$, with $s = 0$ marking the base, defined as the point where 
the Mach number reaches $\mathcal{M} = 0.025$ (see Appendix~\ref{sec:streamlines} for details).
\fref{fig:streamlines_heatcool} similarly shows the specific heating and cooling rates along the representative streamlines, including only the major contributing processes.

The flow near the \ce{H2} dissociation front (e.g., red streamline in \fref{fig:streamlines}) originates from the disk surface at $R \approx 40\au$, consistent with the critical radius at $T \approx 300\Kelvin$.   
This gas travels roughly along the front, accelerating from $< 0.1\kms$ to $4\kms$ (Panel~(e)). It takes $\approx 1000\yr$ for this gas to exit the computational domain after leaving the disk surface.
During the travel, gas density decreases monotonically from $n_{\rm H2} \approx 10^5\cm^{-3}$ to $\approx10^2\cm^{-3}$ (Panel~(c)), while the temperature steadily increases from $\approx 10^2\Kelvin$ to $700\Kelvin$ over a distance of $\approx100\au$, and then levels off, nearly constant from that point to the outer computational boundary (Panel~(d)). 
X-ray and EUV heating are the dominant heating processes (middle panel in \fref{fig:streamlines_heatcool}). 


This gas is launched as fully molecular (Panel~(d)), with \ce{H2} molecules gradually destroyed via X-ray photoionization as well as two X-ray-driven dissociation pathways. 
The first pathway (hereafter Path~1) involves proton transfer and dissociative recombination of hydrogen-bearing species: \ce{H2+ + H2 -> H3+ + H}, followed by \ce{H3+ + e -> 3H}. 
In this sequence, \ce{H2+} is produced via X-ray photoionization, and two \ce{H2} molecules are destroyed per photoionization event. 
The second pathway (Path~2) proceeds through a series of ion-neutral and dissociative recombination reactions of oxygen-bearing ions: (i) \ce{H2 + OH+ -> H2O+ + H}, (ii)\ce{H2 + H2O+ -> H3O+ + H}, and (iii) \ce{H3O+ + e -> OH + 2H}. 
Here, \ce{OH+} in step~(i) is generated through \ce{H+ + OH -> OH+ + H} and \ce{H2 + O+ -> OH+ + H}, where \ce{O+} is created via charge exchange, \ce{H+ + O -> O+ + H}. The hydrogen ion is provided by X-ray photoionization \ce{H + $h\nu$ -> H+ + e} (and to a lesser extent by ion-neutral reaction \ce{H + H2+ -> H2 + H+}). 
In total, \revision{three hydrogen molecules are destroyed per photoionization event, although this number effectively reduces to two when \ce{H2+} ends up with the ion-neutral reaction \ce{H + H2+ -> H2 + H+}.}
FUV photodissociation is negligible under the adopted stellar parameters.

The overall timescale of these dissociation processes is limited by the X-ray photoionization timescale ($\sim 3000\yr$). 
During the outflow, the \ce{H2} abundance drops from $y_{\rm H2}\approx 0.5$ (fully molecular) to $y_{\rm H2}\approx 0.2$ by the time the gas reaches the outer computational boundary, on a timescale of $\approx 1000\yr$. 

Gas launched beyond $R \approx 40\au$ remains largely fully molecular (e.g., cyan streamline in \fref{fig:streamlines}). In this region, X-ray photoionization timescale increases rapidly with radius, due to both geometric dilution and absorption, outpacing the modest rise in the winds' crossing timescales (cf. the fourth panel in \fref{fig:hydro}). The ionization timescale is $ \gtrsim 10^3\yr$, yet X-ray photoionization remains a key \ce{H2} destruction process, along with the ion-neutral reaction \ce{H2+ + H2 -> H3+ + H} (Path~1). 
Along streamlines, the \ce{H2} abundance gradually decreases by $\sim 10\%$ until leaving the computational domain on a timescale of $\approx 4000\yr$. 
The gas density, temperature, and velocity follow similar trends to those \revision{found} near the \ce{H2} dissociation front \revision{described above}, albeit with less steep gradients (cyan lines in \fref{fig:streamlines}).
At these large distances, only X-ray heating remains effective due to the high column density, providing nearly all of the energy input (right panels in \fref{fig:streamlines_heatcool}). 

In contrast, gas launched from within $R \approx 40\au$ undergoes substantial \ce{H2} depletion before being strongly accelerated (e.g., blue streamline in \fref{fig:streamlines}).  
Here, the X-ray photoionization proceeds faster ($\lesssim 10^3\yr$), and the dominant destruction pathway shifts to the oxygen-bearing ion reactions (Path~2). 
In the innermost regions (a few $\au$), the neutral-neutral reaction \ce{H2 + O -> OH + H} also contributes, being efficient at high temperatures (activation energy $\approx 3150\Kelvin$).
This inner atomic gas reaches $T > 10^3\Kelvin$ (cf. the second panel of \fref{fig:hydro}), and shows steep gradients in density, temperature, and velocity along streamlines (blue lines in \fref{fig:streamlines}). 
It attains the highest wind speed of $\approx 10\kms$ and exits the computational domain within $\sim 300\yr$.
In this highly irradiated regime, energy losses via radiative cooling are negligible, and nearly all input energy is converted into the mechanical energy (left panels in \fref{fig:streamlines_heatcool}, and see also \fref{fig:streamlines_heatcool_detailed}).

For reference, the cumulative mass-loss rates are estimated to be $\approx 1\e{-9}\Msun\yr^{-1}$ in total (including helium), and $\approx 0.4\e{-9}\Msun\yr^{-1}$ for \ce{H2}. 
\ce{H2} accounts for nearly all hydrogen-bearing mass loss beyond $\gtrsim 40\au$, contributing about $\approx 70\%$ of the total mass loss there. (See Appendix~\ref{sec:mass-loss_rates} and \fref{fig:mass-loss_rates} for details).

\subsection{Energy Budget} \label{sec:results:energetics}

We examine the energetics of photoevaporative \ce{H2} winds, which can offer a useful diagnostic for assessing a photoevaporative origin (Sections~\ref{sec:discussions:diagnostics} and \ref{sec:jwst_obs}).
The underlying principle is that in a photoevaporating system, a fraction of stellar UV and X-ray photons is absorbed and converted into heat, part of which is subsequently radiated away through \ce{H2} line emission. 
This leads to the energy constraint: 
\begin{equation}
    L_{\rm H2} \leq \epsilon_{\rm heat} L_{\rm UVX} , 
    \label{eq:energy_efficiency}
\end{equation}
where $L_{\rm UVX}$ is the total UV and X-ray luminosity; $\epsilon_{\rm heat} (< 1)$ is the heating efficiency, defined as the fraction of $L_{\rm UVX}$ converted into heat; and $L_{\rm H2}$ is the total \ce{H2} rovibrational line luminosity, or equivalently the total \ce{H2} cooling rate, from the UV- and X-ray-heated region.


In our model, $L_{\rm UVX} = 3.9\e{30}\erg\unit{s}^{-1}$, \revision{and the total luminosity is} dominated by X-rays (\tref{tab:fiducialmodel}). 
Integrating the rates of all photoheating processes throughout the computational domain yields a total heating rate of $1.6\e{29}\erg\unit{s}^{-1}$, with X-rays, EUV, and FUV responsible for $1.1\e{29}\ergs$, $4.7\e{28}\ergs$, and $3.3\e{27}\ergs$, respectively.
This gives $\epsilon_{\rm heat} = 4\%$, consistent with previous radiation hydrodynamics simulations reporting efficiencies of $< 10\%$ \citep{2017_Wang, 2019_Wang}.

The low efficiency reflects two main factors.
First, only $\mathcal{O}(10\%)$ of the emitted photons are actually absorbed. 
For instance, X-rays emitted toward low-polar-angle regions, where the column density is low (the dark areas in the top panel of \fref{fig:hydro}), escape the system with minimal absorption.
Even in the \ce{H2} winds, where the column density is higher, not all photons are absorbed. 
Full attenuation occurs primarily along the boundary between the disk and wind regions, which subtends only a small solid angle from the perspective of the star. 

Second, not all photon energy is converted into heat; a significant portion is lost to ionization and excitation. 
In the case of ionization, energy losses are substantial if photons have energies near the ionization potentials of the absorbers. 
For excitation, energy losses  
typically increase with the photoelectron's energy \citep{1985_ShullSteenberg, 1996_Maloney}. 
For X-ray-generated photoelectrons, only $\approx 10\%$ and $\approx 40\%$ of the primary electron energy is converted into heat in the \ion{H}{I} and \ce{H2} layers, respectively \citep[][see also discussions in \secref{sec:discussions:caveat}]{1991_XuMcCray, 1996_Maloney}.
In the case of FUV photoelectric heating, typically $< 10\%$ of the absorbed energy is converted into heat, and the efficiency can decrease further as the PAH and very small grain abundances reduce.

In our simulation, $L_{\rm H2} \approx 1.3\e{28}\ergs$, accounting for $\approx 9\%$ of the total heating rate ($\epsilon_{\rm heat} L_{\rm UVX} \approx 1.6\e{29}\erg\unit{s}^{-1}$, from X-ray, EUV, and FUV heating at $1.1\e{29}\ergs$, $4.7\e{28}\ergs$, and $3.3\e{27}\ergs$, respectively), and thus $L_{\rm H2} \approx 3.3\e{-3} L_{\rm UVX}$. 
Strictly, $L_{\rm H2}$ includes contribution from the innermost region ($\lesssim 3\au$), where the gas temperature is maintained at $\gtrsim 100\Kelvin$ through dust-gas collisional heat exchange, indicating that the energy source for \ce{H2} emission is not limited to UV and X-ray irradiation, but also includes stellar optical radiation, which heats the dust grains.
However, in our simulation, the contribution from this inner region is negligible compared to that from the UV- and X-ray-heated layers. It is therefore reasonable to consider UV and X-rays as the primary energy sources for $L_{\rm H2}$. 

Note that our analysis does not account for energy absorbed within the inner boundary ($\lesssim 1.8\au$). 
As such, the efficiencies derived here are applicable beyond $\approx 2\au$. 
For consistent comparison with observational estimates, the observed efficiency should likewise exclude contributions from within $\lesssim 2\au$. 
Additionally, while we find $L_{\rm H2} \approx 3\e{-3}L_{\rm UVX}$ in the fiducial model, energy efficiency is generally luminosity-dependent, with typically more deposited energy going into total radiative cooling as luminosity increases \citep{2024_Nakatani}. 
Caution is therefore advised when applying the derived efficiencies to sources with significantly different luminosities.
We revisit this point in \secref{sec:discussions:luminosity_dependence}.






\subsection{Intrinsic \ce{H2} Line Intensity Maps} \label{sec:results:images}
In this section, we present the integrated intensity maps, post-processed with RADMC-3D, to examine the morphology of photoevaporative \ce{H2} winds, one of the central focuses of this study. 
We first analyze the case with $i = 90^\circ$ and no disk obscuration, which reveals the intrinsic emission structure and fluxes. This serves as a reference for more observationally realistic scenarios with different inclinations and disk obscuration, discussed later in \secref{sec:results:inclinations}.

\begin{figure*}[h!tbp]
    \centering
    \includegraphics[width=\linewidth, clip]{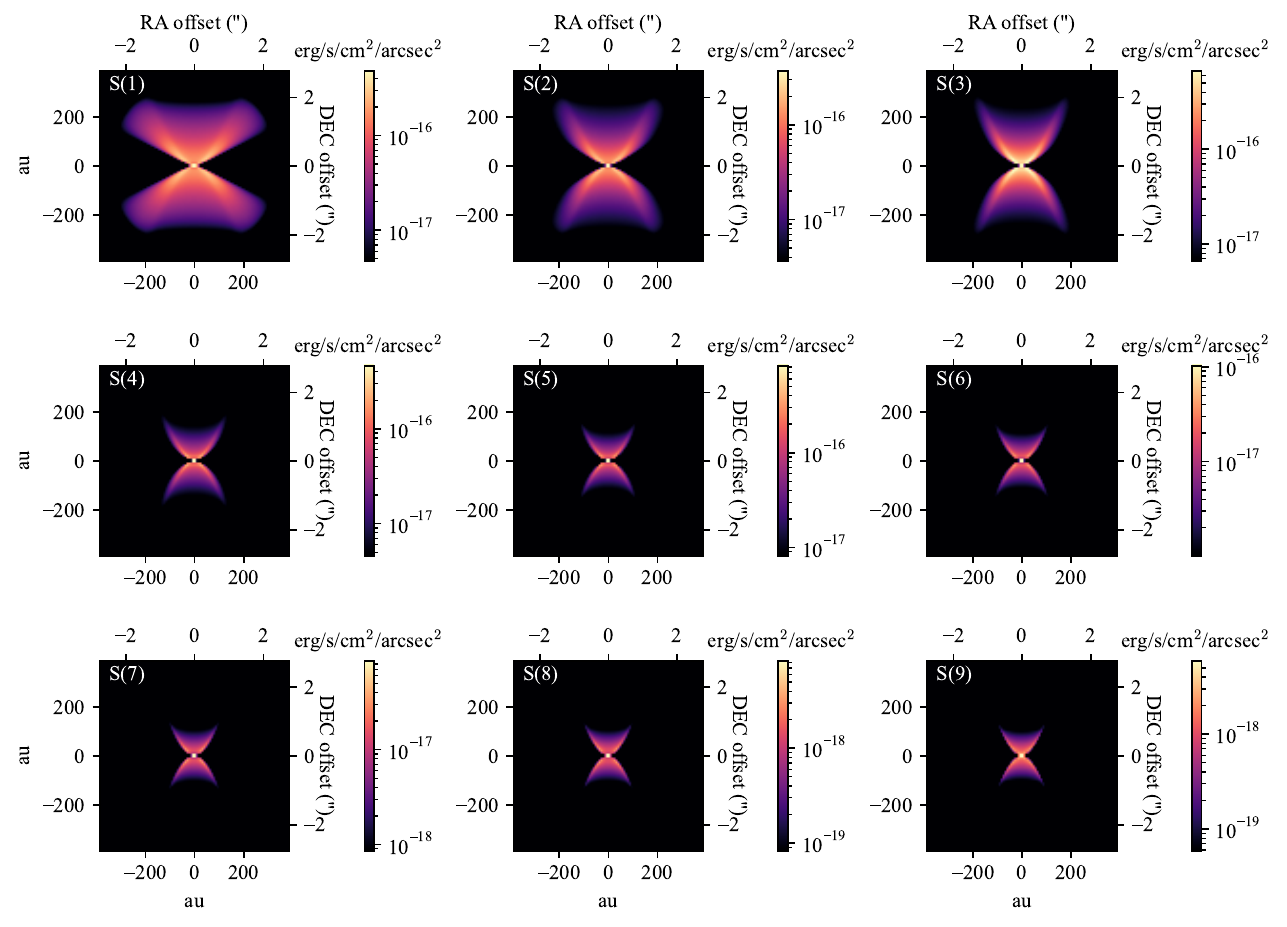}
    \caption{Intrinsic morphologies of the \ce{H2} pure rotational lines (S(1)--S(9)) for a disk viewed edge-on ($i=90^\circ$) without disk obscuration. The colors indicate the frequency-integrated intensities, with individual color bars scaled independently in each panel, 
    Right ascension and declination offsets are shown assuming a source distance of $d = 140\pc$.
    }
    \label{fig:gallery}
\end{figure*}
\begin{table*}[h!tbp]
    \centering
    \begin{tabular}{c c c c|c  c c }
    \hline\hline
    Transition & Line center  & $E_{\rm up}$ & $g_u$ & \multicolumn{2}{c}{Integrated flux ($10^{-16} \unit{erg\,cm^{-2}\,s^{-1}}$)} \\ 
     & ($\unit{\mu m}$) & (K) &  & Model ($i = 90\deg$) & Tau 042021 (total) 
     & \revision{SY Cha (0.5--2.0$\arcsec$) }
     \\ \hline 
    S(1) & 17.0348 & 1015.084 & 21  & $7.2$   & $16$   & $4.8$  \\
    S(2) & 12.2786 & 1681.639 &  9  & $2.8$   & $7.8$  & $5.5 $  \\
    S(3) & 9.66491 & 2503.744 & 33  & $6.1$   & $17$   & $25$  \\
    S(4) & 8.02504 & 3474.498 & 13  & $1.3$   & $5.6$  & $9.1$  \\
    S(5) & 6.90951 & 4586.059 & 45  & $1.9$   & $11$   & $25$ \\
    S(6) & 6.10856 & 5829.842 & 17  & $0.30$  & $1.9$  & $8.3$  \\
    S(7) & 5.51118 & 7196.708 & 57  & $0.37$  & $3.8$  & $17$  \\
    S(8) & 5.05312 & 8677.149 & 21  & $0.053$  & $1.1$  & $5.5$  \\ 
    S(9) & 4.69461 & 10261.45 & 69  & $0.065$  & $4.5$  & -      \\
    \hline
    \end{tabular}
    \caption{
    Molecular data for the pure rotational transitions, along with the integrated fluxes of our model (\fref{fig:gallery}) and observations. 
    The \NameOfTauxxxx{} fluxes represent the total flux from both the western and eastern lobes of the disk. The data are primarily taken from Table~1 of \citet{2024_Arulanantham}, with the S(9) flux adopted from \citet{2025_Pascucci}. 
    Note that the spatial integration areas are larger in our model than in the \NameOfTauxxxx{} observations, so direct quantitative comparisons should be made with caution (see \secref{sec:jwst:Tau042021} for details).
    \revision{The SY~Cha fluxes are adopted from \citet{2025_SchwarzErratum}, measured with a radius aperture of 0.5--2.0$\arcsec$ (note that $d = 180.7\pc$).} 
    }
    \label{tab:fluxes}
\end{table*}
\fref{fig:gallery} displays the frequency-integrated intensity maps for the \ce{H2} S(1)--S(9) lines, all exhibiting an X-shaped morphology with a central emission peak.
The X-shaped structure originates from warm \ce{H2} gas ($\gtrsim 100\Kelvin$; see \fref{fig:hydro}), where thermal excitation is efficient. 
The X-arms trace regions of peak column density in an edge-on view, extending outward with slight collimation and a convex curvature, i.e., the semi-opening angle narrows with increasing distance. 
This morphology reflects the underlying temperature and \ce{H2} distributions (second and third panels in \fref{fig:hydro}). 

The emission becomes increasingly compact and slightly more collimated from S(1) to S(9), reflecting the underlying density distribution and the radial and angular temperature gradients. Higher excitation lines require higher densities and trace hotter gas, which resides at smaller radii and lower polar angles (second panel of \fref{fig:hydro}). 

The central emission peak originates from the hot, high-density atmosphere of the inner disk ($r \lesssim 3\au$). 
Note that our computational domain excludes $r \lesssim 1.8\au$ where hot, even denser atmosphere likely exists, so the peak emission may be underestimated.

\begin{table*}[h!tbp]
    \centering
    \begin{tabular}{c c c c c c c | c c c c | c c}
    \hline\hline
     & \multicolumn{2}{c}{$r_{\rm ext}$}  
     & \multicolumn{2}{c}{$\Rgeo$} 
     & \multicolumn{2}{c}{$\aveopeningangle$} 
     & \multicolumn{2}{c}{$\theta_0$}
     & \multicolumn{2}{c}{$\theta_1$}
     & $\theta_{\rm open}$ in \NameOfTauxxxx{} 
     & \revision{SY~Cha}
     \\ 
     & \multicolumn{2}{c}{($\au$)} 
     & \multicolumn{2}{c}{($\au$)} 
     & \multicolumn{2}{c}{(deg)} 
     & \multicolumn{2}{c}{(deg)}
     & \multicolumn{2}{c}{(deg)} 
     & (deg)
     & (deg)
     \\ 
     & int. & cnv. & int. & cnv. & int. & cnv. & int. & cnv. & int. & cnv. &  & \\
     \hline 
    S(1) &  288  & 300 &   24.4 &  -31.6 &  49.7 & 40.2&  53.1 & 36.2& 12.1 & -26.0 & - & -\\
    S(2) &  270  & 278 &   32.8 &  3.97  &  42.8 & 37.4&  48.1 & 36.4& 22.2 & -9.9  & $35 \pm 5$ & -\\
    S(3) &  239  & 244 &   31.8 &  11.6  &  40.5 & 36.0&  45.8 & 36.5& 23.3 & -0.7  &- & $58\pm 3$\\
    S(4) &  201  & 206 &   29.3 &  13.0  &  39.0 & 35.2&  44.1 & 36.0& 23.0 &  1.7  &- & $52\pm 3$\\
    S(5) &  154  & 159 &   26.9 &  14.5  &  39.1 & 34.6&  43.2 & 35.9& 22.3 &  4.1  &- & $48\pm 2$\\
    S(6) &  127  & 131 &   27.0 &  15.0  &  38.0 & 33.8&  42.4 & 35.2& 21.6 &  3.9  &- & $58\pm 3$\\
    S(7) &  94.7 & 98.7&   26.4 &  14.2  &  37.2 & 33.0&  40.5 & 34.5& 16.7 &  2.6  &- & $64\pm 3$\\
    S(8) &  76.5 & 82.7&   26.0 &  14.6  &  37.3 & 32.2&  40.4 & 33.7& 17.7 &  1.6  &- & -\\ 
    S(9) &  51.1 & 55.2&   25.7 &  18.5  &  37.1 & 32.6&  40.3 & 35.1& 17.9 &  4.7  & $38.5\pm4.5$; $37.6\pm2.6$ & -\\
    \hline
    \end{tabular}
    \caption{
    Geometric properties of the emission for each pure rotational line  (cf. \fref{fig:gallery}): the emission extent $\rext$, the geometric radius $\Rgeo$, and the average semi-opening angle $\aveopeningangle$, along with logarithmic fits, $\openingangle = \theta_0 - \theta_1\log(r/10^2\au)$. 
    ``int.'' and ``cnv.'' refer to measurements based on the intrinsic intensity maps (\fref{fig:gallery}) and the maps convolved with 2D Gaussians representing PSFs, respectively.
    The second rightmost column shows the observed semi-opening angles for the S(2) \citep{2024_Arulanantham} and S(9) \citep{2025_Pascucci} emissions from \NameOfTauxxxx{}. 
    The S(9) values are based on a PSF-deconvolved image, with the two measurements corresponding to the blue- and red-shifted lobes.
    \revision{The rightmost column shows the semi-opening angles of SY~Cha, taken from \citet{2025_Schwarz}.}
    }
    \label{tab:geometry}
\end{table*}
To quantify the morphology, \tref{tab:geometry} lists three key metrics of the images in \fref{fig:gallery}: 
(i) the emission extent $\rext$, defined as the radius enclosing 95\% of the total flux; 
(ii) the geometric radius $\Rgeo$, where a linear fit to the wind emission intersects the disk midplane, following \citet{2025_Pascucci};
(iii) the semi-opening angle $\openingangle$, defined as the polar angle where the emission peaks at each radius.
Since $\openingangle$ varies with the radius, \tref{tab:geometry} reports both the mean value $\aveopeningangle$ and a log-linear fit: $\openingangle = \theta_0 \revision{-} \theta_1  \log (r/100\au)$. The fit uses data from $r > 20\au$ to exclude the inner region. 
For consistency with \citet{2025_Pascucci}, $\Rgeo$ is measured using points $\sim 0.2\arcsec$ above and below the midplane. 

Overall, $\rext$ and $\aveopeningangle$ decrease with decreasing wavelength. $\Rgeo$ also decreases for S(2)--S(9), although S(1) shows a further drop due to its distinct emission geometry (cf. \fref{fig:gallery}).

To evaluate PSF effects on the geometric indicators, we also measure $\rext$, $\Rgeo$, $\openingangle$ from images convolved with 2D Gaussians. 
The FWHMs are set to $0.\arcsec033(\lambda/1\um) + 0.\arcsec106$ for S(1)--S(8) \citep{2023_Law} and $0.\arcsec21$ for S(9), estimated using a STPSF simulation tool \citep[][]{2014_Perrin}. 
For $\Rgeo$ and $\openingangle$, we exclude emission from the inner region ($r < 20\au$) before convolution to isolate the wind component.

\begin{figure*}
    \centering
    \includegraphics[width=0.8\linewidth, clip]{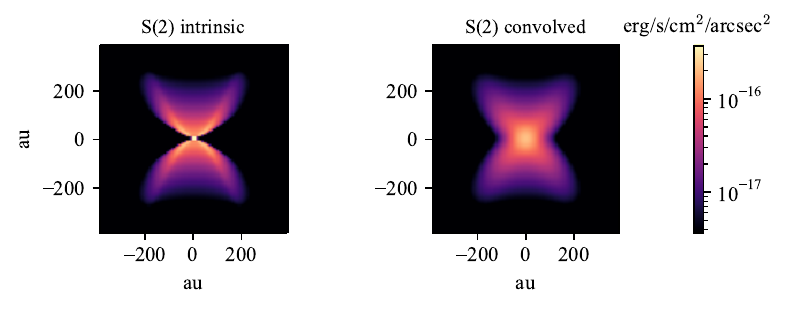}
    \caption{Comparison of emission morphologies between the intrinsic and Gaussian-convolved S(2) images. The left shows the original image from \fref{fig:gallery}, while the right displays the same image convolved with a Gaussian kernel of FWHM $\sim 72\au$. }
    \label{fig:gallery:geometric_comparison}
\end{figure*}

Generally, PSF convolution blurs the convex shapes of the emission, making the morphology appear more like a straight X-shape (\fref{fig:gallery:geometric_comparison}). 
This is also reflected in the smaller absolute values of ``cnv.'' $\theta_1$ than the ``int.'' ones in \tref{tab:geometry}. 

For $\aveopeningangle$, the convolution results in a systematic reduction, due to contributions from emission between the arms of the X-shape.
The reduction $\openingangle$ is more significant at smaller radii ($r \lesssim 50\au$), while at larger radii, $\openingangle$ is mostly unchanged from the intrinsic values.

For $\Rgeo$, the convolution decreases the value across all lines in our model. 
However, whether $\Rgeo$ systematically decreases or increases depends on the intrinsic emission morphology and is not universally predictable. 
The negative $\Rgeo$ for S(1) \revision{originates from the} convolution causing the emission peak at each radius to shift \revision{systematically} upward above the midplane and inward toward the rotational axis, especially within $\lesssim 100\au$. \revision{Near the central region, the convolution mixes} contributions from \revision{both the arms of the} X-shaped structure. 
\revision{The relatively large opening angle further contributes to producing a negative $\Rgeo$.
Similar effects are also responsible for the small $\Rgeo $ measured in the convoluted S(2) image.}

To summarize, PSFs can significantly alter the geometric indicators, and accounting for these PSF-induced biases is essential when interpreting wind-launching points or inferring wind-driving mechanisms from morphology.

\subsection{\ce{H2} Line Fluxes} \label{sec:results:fluxes}
Now we turn to the integrated fluxes derived from the intensity maps (\fref{fig:gallery}), which---alongside morphology---represent a key focus of this paper.

\begin{figure}
    \centering
    \includegraphics[width=\linewidth, clip]{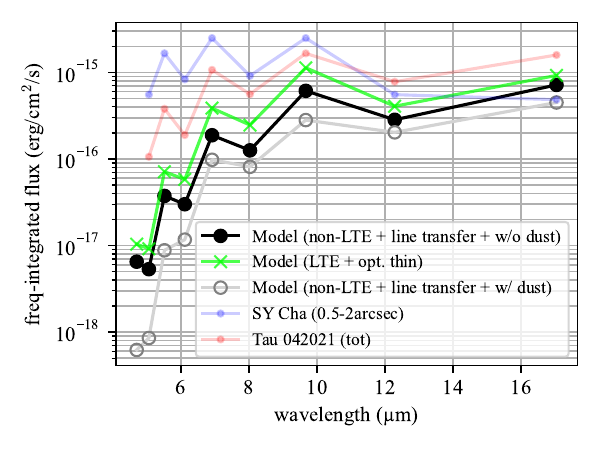}
    \caption{
    Integrated fluxes of the S(1)--S(9) lines for the model with $i = 90^\circ$, without disk obscuration, and assuming $d = 140\pc$ (black dots). 
    For reference, green crosses indicate the fluxes calculated under the assumptions of LTE and optically thin emission. 
    Observed fluxes for \revision{SY~Cha (blue) and} \NameOfTauxxxx{} (red) are also shown for comparison.
    \revision{For SY~Cha, we adopt the fluxes measured with a radius aperture of $0.5$--$2\arcsec$.}
    Since the spatial integration regions differ between our model and the observations of \NameOfTauxxxx{}, direct comparisons should be made with caution (see also \secref{sec:jwst:Tau042021} and \fref{fig:comparison_with_tau}). 
    Open circles correspond to the model fluxes with $i = 90^\circ$ but including disk obscuration (\secref{sec:results:inclinations}; \tref{tab:fluxes_comparison}).
    }
    \label{fig:sed}
\end{figure}
\fref{fig:sed} shows the integrated fluxes (black filled dots) as a function of wavelength, with corresponding values listed in \tref{tab:fluxes} with $d = 140\pc$. 
The S(1)--S(5) lines, which exhibit extended emission, have fluxes on the order of $10^{-16}$--$10^{-15}\ergs\cm^{-2}$. Higher-$J$ lines---emitted primarily from the hot, high-density atmosphere of the inner disk ($r\lesssim 3\au$)---are more compact in emission and show fluxes about an order of magnitude lower.

All lines are generally optically thin; only S(1) becomes marginally optically thick near the center, but this has a minimal impact on the image and flux. 
Most emission arises from the regions close to local thermodynamic equilibrium (LTE), making the optically thin + LTE approximation reasonably accurate. 

This is supported by the green crosses in \fref{fig:sed}, which mark the fluxes computed under the LTE and optically thin assumptions.
These fairly agree with the full non-LTE radiative transfer results (black filled dots) within a factor of two across all lines. 

For comparison, observed fluxes of \NameOfTauxxxx{} \citep{2024_Arulanantham} and \revision{SY~Cha \citep{2025_SchwarzErratum}} are also shown (see also \tref{tab:fluxes}). The overall agreement, especially for lower-$J$ lines, is striking. 
However, we note that the spatial integration domains used to derive the \NameOfTauxxxx{} fluxes are smaller than those in our model, and \revision{the distance to SY~Cha is somewhat larger ($d = 180.7\pc$).}
A more detailed comparison follows in Sections~\ref{sec:jwst:Tau042021} \revision{and \ref{sec:jwst:SYcha}}.


\subsection{Gas Properties Traced by Each Line} \label{sec:results:prop_traced_by_line}
To probe the physical conditions traced by each line, we define \revision{a convenient diagnostic quantity, referred to as} the local integrated luminosity,
\[
    \mathcal{L}_{J} = \mathcal{L}_J (r, \theta) \equiv   h\nu_{ul}A_{ul} n_u r^3 \sin\theta, 
\]
where $\nu_{ul}$ and $A_{ul}$ are the frequency and Einstein A-coefficient for the $J+2 \rightarrow J$ transition, and $n_u$ is the upper-level population. 
This quantity represents the line luminosity contribution from a volume element spanning uniform intervals in $\ln r$, $\theta$, and $\phi$. 

The physical meaning of $L_J$ may become clearer in its volume-integrated form: 
assuming optically thin emission, the integrated line flux is expressed by
\[
    \revision{\mathcal{F}}_J
    =  \frac{1}{4\pi d^2}\int  \mathcal{L}_J \, \dd (\ln r) \dd\theta \dd \phi,
\]
so $\mathcal{L}_J$ approximately indicates the relative contribution of gas at $(r,\theta)$ to the \revision{optically thin} total flux $\revision{\mathcal{F}}_J$ (or equivalently, the total line luminosity). 
Mapping $\mathcal{L}_J$ onto a density-temperature phase diagram provides the physical conditions traced by each line.
\revision{We note that $\mathcal{L}_J$ and $\mathcal{F}_J$ are diagnostic quantities introduced to identify the regions that contribute most to the optically thin total line luminosity and flux, and thus generally differ from those obtained through radiative transfer calculations. However, they become physically equivalent when the line is optically thin and dust extinction is negligible.}

\begin{figure*}
    \centering
    \includegraphics[width=\linewidth]{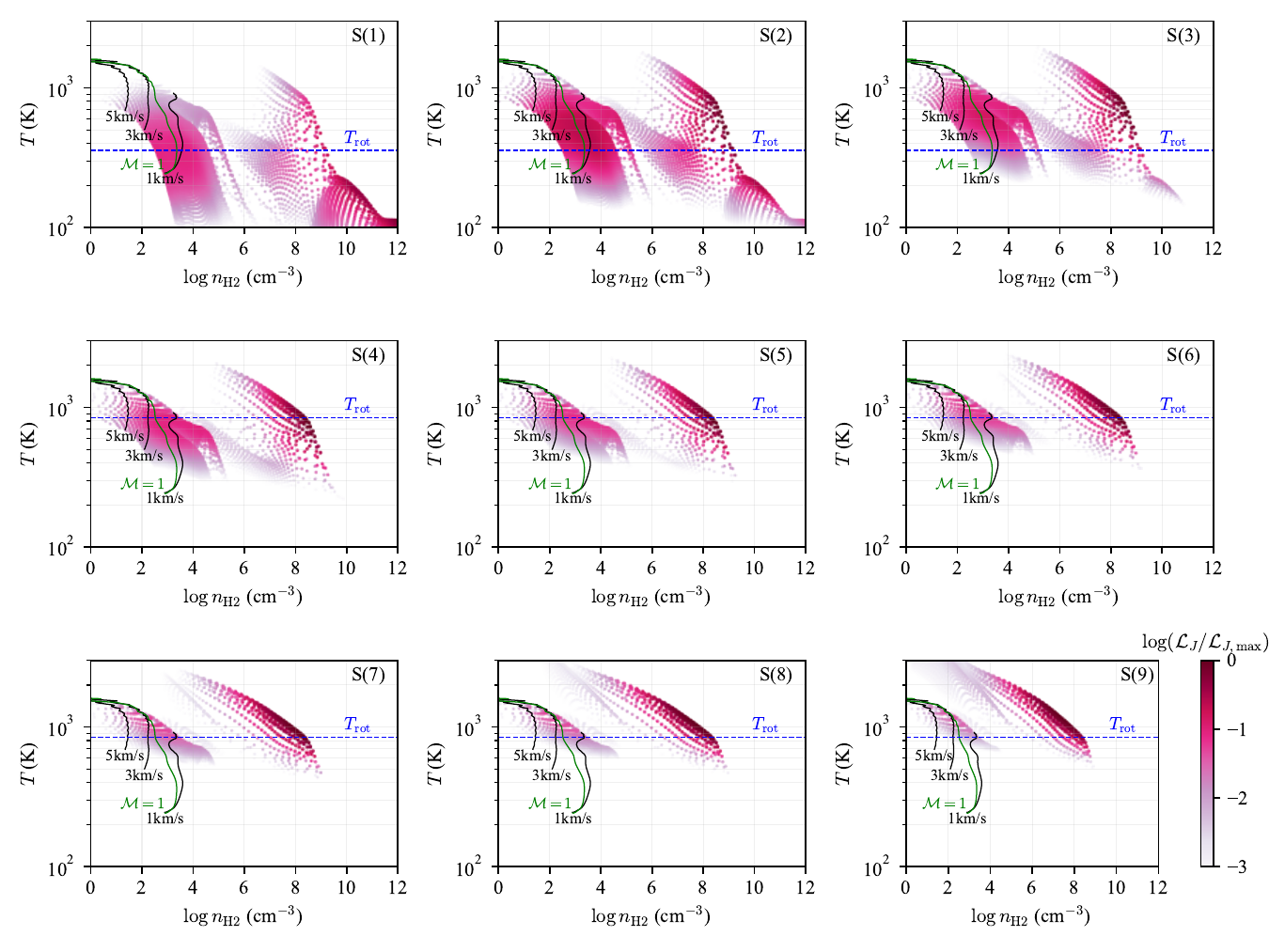}
    \caption{
    Phase-space scatter plots of the normalized local integrated luminosity $\mathcal{L}_J/\mathcal{L}_{J,\rm max}$ for each \ce{H2} line. 
    Each point corresponds to a computational cell in the simulation, with point size scaled by $\mathcal{L}_J/\mathcal{L}_{J,\rm max}$ to reduce overlap and emphasize more emissive regions. 
    The reddish areas highlight the densities and temperatures ranges that most strongly contribute to the total emission. Black solid lines mark approximate contours of poloidal wind velocity ($v_{\rm p}  = 1$, 3, and $5\kms$), and the green contour indicates the isothermal sonic surface. Blue dashed lines represent the best-fit rotational temperatures (\tref{tab:best_fit})
    }.
    \label{fig:phase}
\end{figure*}

\fref{fig:phase} shows scatter plots of normalized $\mathcal{L}_J$ for each computational cell, using the non-LTE level populations computed with RADMC-3D.
The dark reddish points highlight the combinations of $n_{\rm H2}$ and $T$ contributing predominantly to each line. 
In general, emission originates from two components: the hot, high-density atmosphere of the inner disk and the more extended, lower-density photoevaporative winds.
This dual contribution is consistent with the intensity maps \revision{in \fref{fig:gallery}}, which show the brightest emission at the center with the extended diffuse emission. 

These two regions are roughly separated at $r \approx 3$--$7\au$. For S(5)--S(9), the disk component arises primarily from $r \lesssim 3\au$. Note again that our computational domain excludes the innermost region ($r\lesssim 1.8\au$). 

Focusing on the wind ($n_{\rm H2} \lesssim 10^5\cm^{-3}$ on the scatter plots), we find that higher excitation lines trace hotter and more rarefied gas. The contributing conditions range from $(n_{\rm H2}, T) \sim (10^\revision{5}\cm^{-3}, \revision{300} \Kelvin)$ for lower-$J$ lines to $\sim (10^2 \cm^{-3}, \revision{900} \Kelvin)$ for higher-$J$ lines. 
Broadly, the S(1)--S(3) lines primarily trace cooler gas $T \sim 200$--$10^3\Kelvin$, while the S(4)--S(9) lines are sensitive to warmer regions $ 700\Kelvin\lesssim T \lesssim 2000\Kelvin$.
Higher temperatures imply higher sound speeds and faster photoevaporative winds.
This is visualized by the black contours in \fref{fig:phase}, which mark poloidal velocities: S(1)--S(3) emission arises in part from slow wind components ($\sim 1\kms$; cf. \fref{fig:hydro}), while S(4)--S(9) are predominantly associated with faster-moving gas.

\subsection{Rotation Diagram Analysis}   \label{sec:results:rot_diagram}
\begin{figure}

    \centering
    \includegraphics[width=\linewidth, clip]{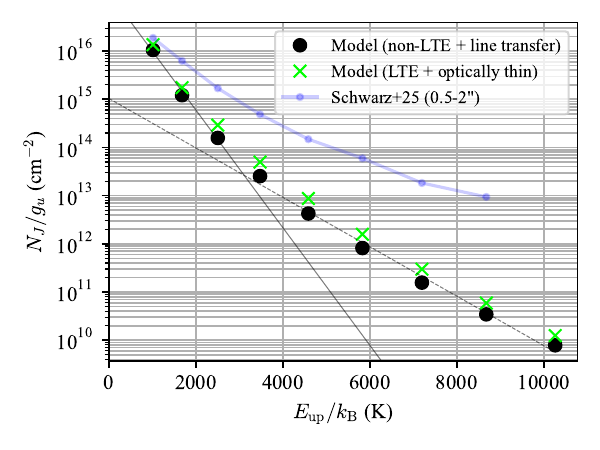}
    \caption{
    Rotation diagram of the \ce{H2} lines. Black dots show the model results from non-LTE calculations with line transfer using RADMC-3D. 
    The solid and dashed black lines are a two-component fit to the lower-$J$ lines S(1)--S(3) and the higher-$J$ lines S(4)--S(9), respectively. 
    Green crosses indicate the values assuming LTE and optically thin emission---i.e., the maximum emission possible within our model setup. 
    \revision{For reference, observational data for SY~Cha are shown with a blue line \citep{2025_SchwarzErratum}. 
    Since the emitting area is not specified in \citet{2024_Arulanantham}, the data for \NameOfTauxxxx{} are omitted to avoid introducing additional uncertainties. }
    }
    \label{fig:rot-diagram_wobs}
\end{figure}
\begin{table*}[htbp]
    \centering
    \begin{tabular}{l c c c c r}
    \hline\hline
    Description & Transitions & $T_{\rm rot} \,({\rm K})$ & $\barNHmol \,({\rm cm^{-2}})$& $M_{\rm H2} \, (M_\odot)$& Note\\ \hline\hline
    Single-component fit    & S(1)--S(9) & 679 & $1.6\e{17}$ & $3.7\e{-8}$ & \\ \hline
    Two-component fit (warm)& S(1)--S(3) & 356 & $1.5 \e{18}$ & $3.5\e{-7}$ & Black solid line in Figure~\ref{fig:rot-diagram_wobs} \\
    Two-component fit (hot) & S(4)--S(9) & 847 & $2.2 \e{16}$ & $5.0\e{-9}$ & Black dashed line in Figure~\ref{fig:rot-diagram_wobs}  \\ \hline
    \end{tabular}
    \caption{Best-fit parameters from the rotation diagram analysis of our model (see also \fref{fig:rot-diagram_wobs}).}
    \label{tab:best_fit}
\end{table*}

We perform a rotation diagram analysis to assess how well the physical properties of photoevaporative \ce{H2} winds can be inferred from observations. We then compare the derived temperatures and \ce{H2} column densities with those in the simulation. 



The analysis fits the integrated fluxes $F_J$, \revision{which are determined by solving radiative transfer,} as a function of $\deltae$ using
\[
    \ln \frac{N_J}{g_u} = -\frac{\deltae}{kT_{\rm rot}} + \ln \frac{\barNHmol}{Q(T_{\rm rot})}. 
\]
where $N_J \equiv {4 \pi F_J }/{h \nu_{ul} A_{ul} \Omega} $, $g_u$ is the statistical weight, $Q$ is the partition function, $\Trot$ is rotational temperature, $\barNHmol$ is the average \ce{H2} column density over the emitting region with solid angle $\Omega$. 

\fref{fig:rot-diagram_wobs} shows the resulting rotation diagram  (black points). 
We fit the data in two ways: using all transitions S(1)--S(9), and using a two-component approach that divides lower-$J$ S(1)--S(3) and the higher-$J$ S(4)--S(9) lines. 
We find that the two-component fit provides a better fit. 
This reflects the presence of multiple temperature components in the wind, as discussed in Sections~\ref{sec:results:wind_properties} and \ref{sec:results:prop_traced_by_line}. 
\fref{fig:rot-diagram_wobs} only shows the two-component fit for clarity. 

The best-fit parameters are listed in \tref{tab:best_fit}. 
While the level populations are not strictly in LTE, the derived $\Trot$ provides reasonable estimates for the temperatures traced by the lower- and higher-$J$ lines (compare the horizontal blue dashed lines and reddish regions in \fref{fig:phase}). 
Similarly, the best-fit $\barNHmol$ agree well with 
simulation values:
$\barNHmol = 3.3\e{18}\cm^{-2}$ for $200\Kelvin \lesssim T \lesssim 10^3\Kelvin$, and $3.7\e{16}\cm^{-2}$ for $T \geq 700\Kelvin$.
These agreements are non-trivial, given the non-LTE nature of the emission, but are expected: most emission arises from regions where LTE is a good approximation.

We note that our two-component fit differs slightly from the method used in \citet{2025_Schwarz}, where S(1)--S(8) fluxes are fitted simultaneously.  
This approach does not require pre-defining the boundary between warm and hot components; instead, it naturally emerges from the fit.
Applying this technique to our fluxes yields a slightly higher hot-component $\Trot$ ($\approx 900\Kelvin$) because S(4) contributes to both components, lowering the flux attributed to the hot component relative to our segmented two-component fit. 
This highlights the need for caution when selecting line ranges in a segmented two-component fit to ensure consistency with simultaneous fitting methods. 

\subsection{Effects of Inclination and disk obscuration}    \label{sec:results:inclinations}
From \secref{sec:results:images} onward, we have focused on an edge-on disk ($i = 90^\circ$) and without disk obscuration. 
Here, we explore how inclination and dust extinction within the disk affect the results.

\begin{table}[htbp]
    \centering
    \begin{tabular}{c | c c c c c c }
    \hline\hline
    Transition &  \multicolumn{6}{c}{Integrated flux ($\times 10^{-16}\unit{erg\,cm^{-2}\,s^{-1}}$)} \\ 
     & \multicolumn{2}{c}{$i = 90^\circ$} & \multicolumn{2}{c}{$i = 60^\circ$} & \multicolumn{2}{c}{$i = 0^\circ$} \\
      &  w/o & w/ & w/o  &  w/  & w/o  &  w/  \\ \hline 
    S(1) &  $7.2$  & 4.5    &$7.4$   & 3.6   &$7.5$  & 3.3       \\
    S(2) &  $2.8$  & 2.0    &$2.8$   & 1.7   &$2.8$  & 1.7       \\
    S(3) &  $6.1$  & 2.8    &$6.1$   & 3.2   &$6.1$  & 3.0      \\
    S(4) &  $1.3$  & 0.81   &$1.3$   & 0.90  &$1.3$  & 0.92      \\
    S(5) &  $1.9$  & 0.98   &$1.9$   & 1.4   &$1.9$  & 1.4      \\
    S(6) &  $0.30$ & 0.12   &$0.30$  & 0.21  &$0.30$ & 0.20         \\
    S(7) &  $0.37$ & 0.088  &$0.38$  & 0.25  &$0.37$ & 0.23         \\
    S(8) &  $0.053$& 0.0085  &$0.053$ & 0.035 &$0.053$& 0.031        \\ 
    S(9) &  $0.065$& 0.0062 &$0.065$ & 0.042 &$0.065$& 0.036        \\
    \hline
    \end{tabular}
    \caption{
    Comparison of model fluxes for different inclination with and without disk obscuration. Models labeled ``w/'' include the effects of disk obscuration, while those labeled ``w/o'' do not. 
    }
    \label{tab:fluxes_comparison}

\end{table}
Inclination alone has little effect on the integrated line fluxes in the absence of disk obscuration, as the emission is optically thin (\secref{sec:results:fluxes}). 
Indeed, the fluxes remain nearly constant across inclinations (\tref{tab:fluxes_comparison}), except for a slight increase in S(1) at lower inclinations due to marginal optical thickness at the center.
We note that this holds for emission beyond $\gtrsim 2\au$, the region covered by our simulation; stronger inclination dependence may appear for inner regions where intense \ce{H2} emission is expected from denser disk atmosphere.

\begin{figure*}
    \centering
    \includegraphics[width=\linewidth, clip]{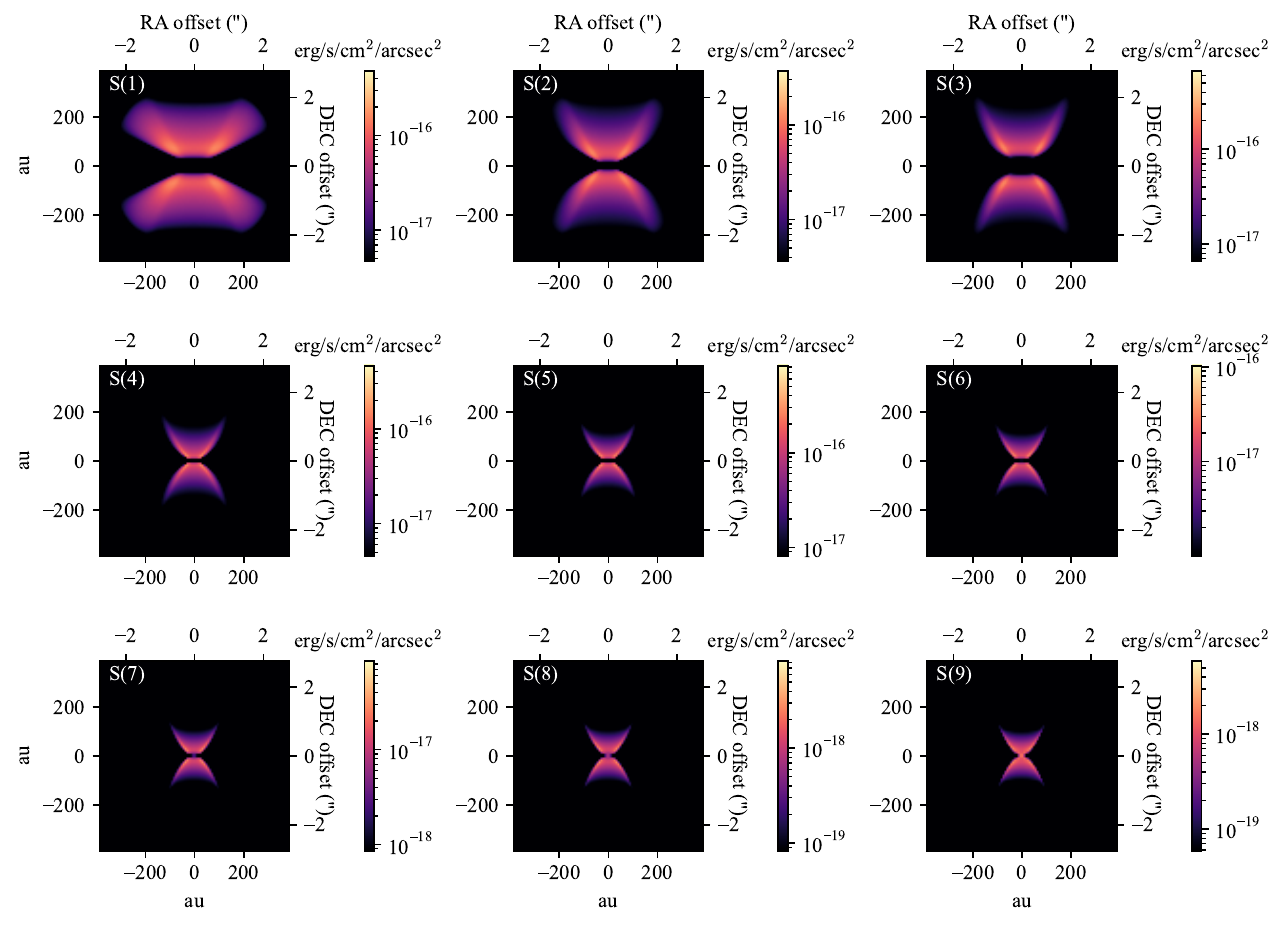}
    \caption{Same as \fref{fig:gallery}, but including dust extinction within the disk. The panels show continuum-subtracted images. 
    }
    \label{fig:gallery_i90_wd}
\end{figure*}

Disk obscuration significantly reduces the flux by obscuring the central and/or far-side regions, depending on viewing angle. 
For nearly edge-on disks, the midplane is obscured, forming a dark lane and leaving only the vertically extended emission visible, in particular for S(1)--S(3) (\fref{fig:gallery_i90_wd}).
This results in a flux reduction by a factor of a few for S(1)--S(6) and by a larger factor for higher-$J$ lines S(7)--S(9) (\tref{tab:fluxes_comparison}). 
The stronger suppression of higher-$J$ lines reflects their origin in the hot inner disk atmosphere.

Even reducing the dust-to-gas mass ratio to 0.0015 only modestly increases the flux: a factor of two for S(9) and less than $50\%$ for the other lines. 
Compared to the observed fluxes, 
our model significantly underpredicts the higher-$J$ line fluxes (S(7)--S(9)) in the dust-obscured case, though lower-$J$ lines remain in relatively good agreement (\fref{fig:sed} and \tref{tab:fluxes}). 
This suggests that our model may underestimate the higher-$J$ line emission from the extended wind component. We revisit this possibility in \secref{sec:discussions:pumping}. 

\begin{figure*}
    \centering
    \includegraphics[width=\linewidth, clip]{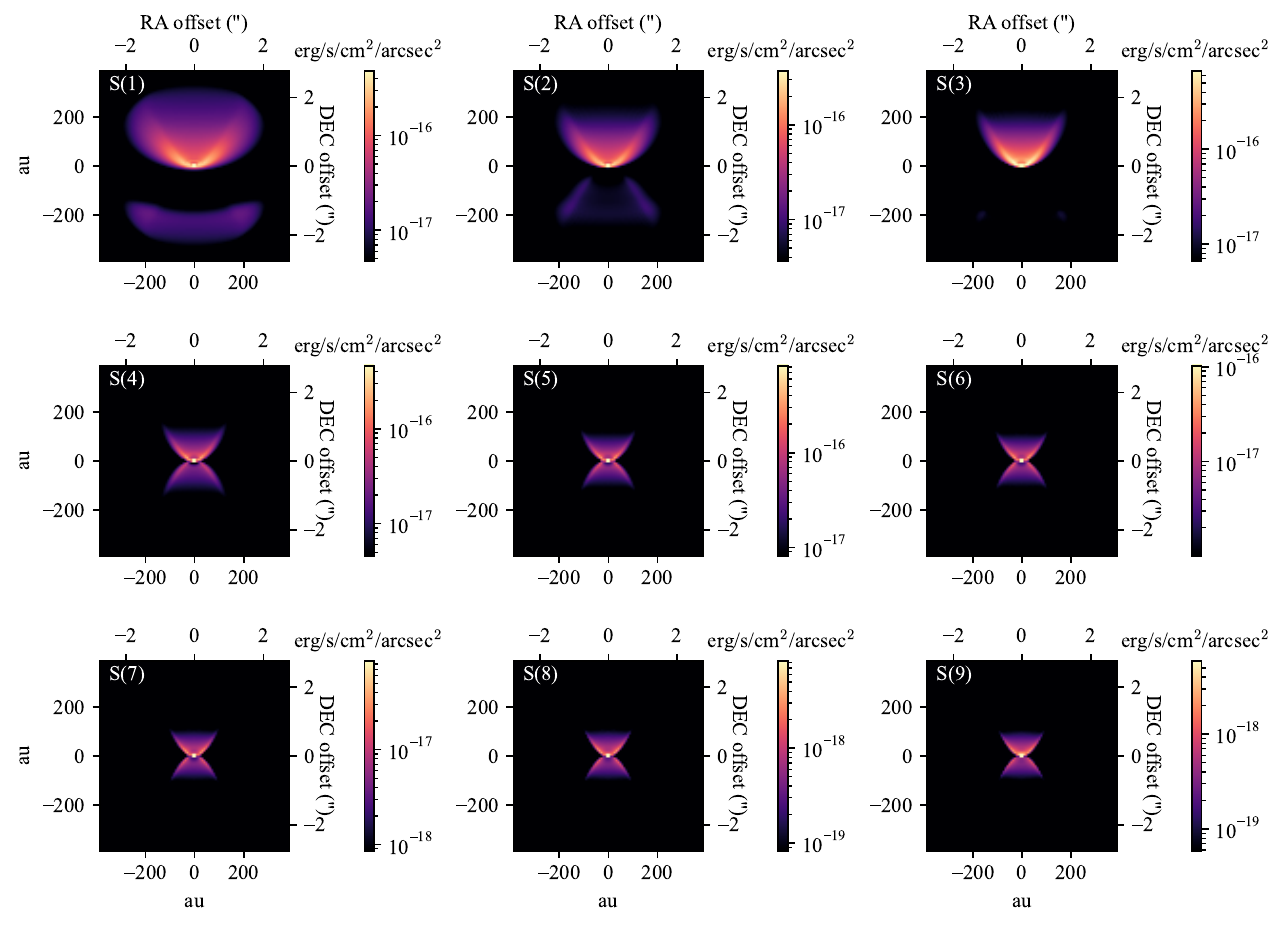}
    \caption{Same as Figures~\ref{fig:gallery} and \ref{fig:gallery_i90_wd} but for an inclination of $i= 60^\circ$ including dust extinction within the disk. 
    }
    \label{fig:gallery_i60_wd}
\end{figure*}

In less inclined disks (e.g., $i = 60^\circ$; \fref{fig:gallery_i60_wd}), attenuation primarily affects the far side of the disk. 
This is most noticeable in the S(3) image, where the far-side emission is mostly obscured, due to the silicate opacity enhancement near $\approx 10\um$. 
For lower-$J$ lines, the morphology shifts from X-shaped to more bowl-shaped with decreasing $i$.

Flux reductions with respect to the dust-free case vary across transitions, but remain relatively modest---less than a factor of two for all lines (\tref{tab:fluxes_comparison}). 
Even the higher-$J$ lines are only moderately impacted, since the inner emitting regions are not completely obscured.

To summarize, while the morphology transitions from X-shaped (edge-on) to bowl-shaped (face-on), inclination has minimal impacts on the fluxes unless disk obscuration is significant. 
Dust obscuration becomes important in edge-on systems, leading to: (1) a dark lane in images, (2) corresponding flux reduction for high-$J$ lines like S(7)--S(9) by obscuring emission from the hot inner disk atmosphere. 
For face-on to moderately inclined disks, the effects are minimal; flux reduction is less than a factor of two. 

Given 
that we may underestimate extended high-$J$ emission (\secref{sec:discussions:pumping}), the actual impact of inclination and dust obscuration could be even smaller in real systems. 
We emphasize that these conclusions apply beyond $\gtrsim 2\au$; emission from the inner regions inside our computational boundary may exhibit a stronger dependence on inclination and disk obscuration. 
The results also depend on the adopted dust opacity and the degree of settling, and therefore the trends should be interpreted qualitatively.

While edge-on sources help probe wind geometry (e.g., its semi-opening angle), disks at all inclinations may offer comparable insights for the physical conditions of photoevaporative \ce{H2} winds. Moderately inclined disks would be particularly advantageous, as they enable complementary comparisons with stellar X-ray and UV luminosities, key diagnostics for photoevaporation (\secref{sec:discussions:diagnostics}).

\revision{
\subsection{Velocity Profiles}  \label{sec:results:velocity_profiles}
We generate spectrally resolved, continuum-subtracted line spectra, using the images of the fiducial model with inclinations of $i = 0, 30, 45, 60, 90\deg$. 
The spectral window spans $\pm 30\kms$ around the line centers and is sampled with 201 wavelength points (corresponding to a spectral resolution of $R \approx 10^6$). 
In producing the spectra, we manually exclude emission from the far side of the disk for inclined cases ($i < 90\deg$), to isolate the contribution from the near-side surface. 
The resulting velocity profiles for selected lines are shown in \fref{fig:velocity_profiles}.

\begin{figure*}
    \centering
    \includegraphics[width=\linewidth]{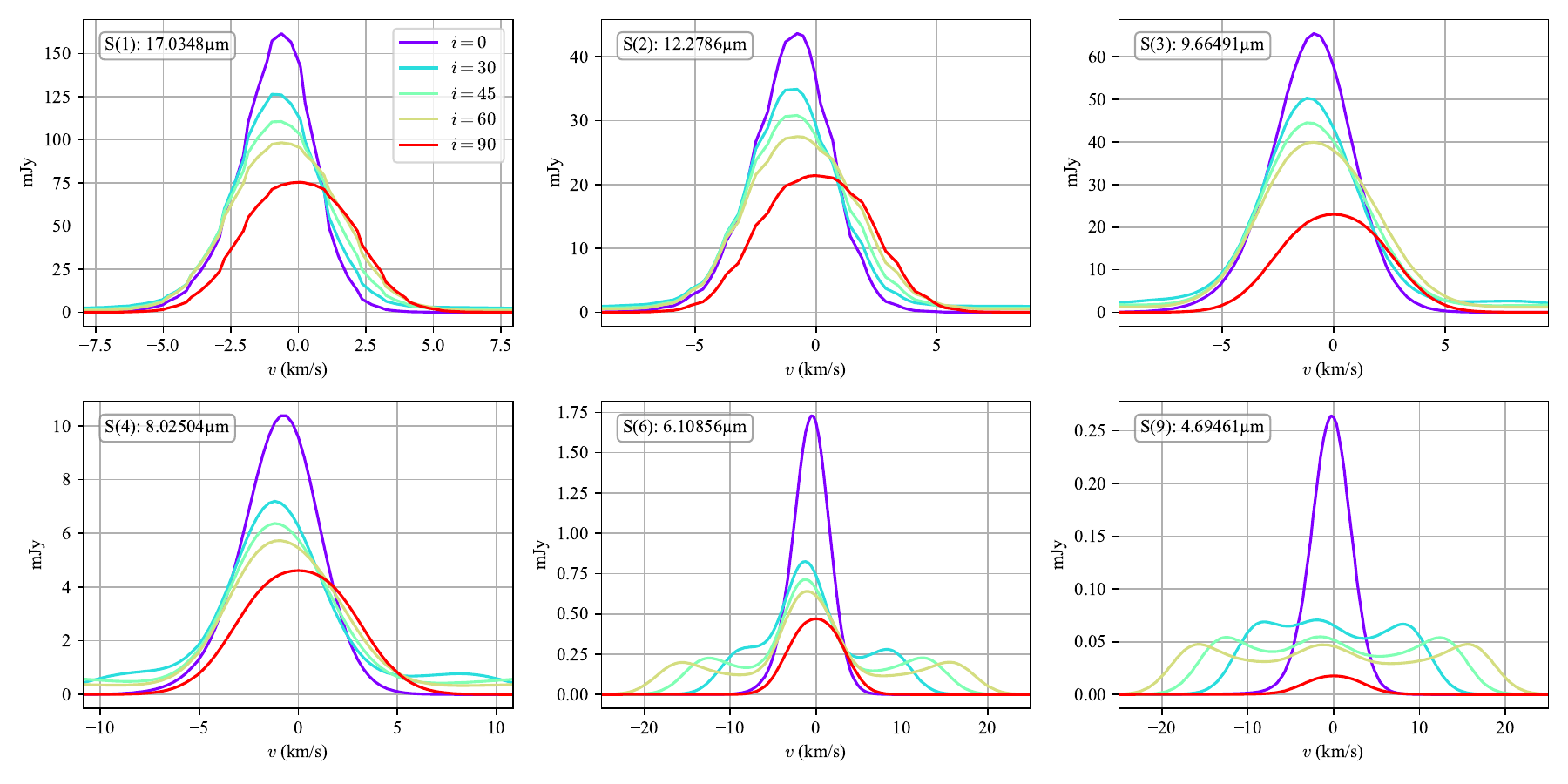}
    \caption{\revision{Velocity profiles of selected lines viewed at inclinations of $i = 0^\circ$ (purple), $30^\circ$ (cyan), $45^\circ$ (light green), $60^\circ$ (yellow), and $90\deg$ (red). For $i < 90\deg$, the profiles are computed excluding emission from the far side of the disk, isolating the contribution from the near-side surface.}  }
    \label{fig:velocity_profiles}
\end{figure*}
Overall, low-$J$ lines (S(1)--S(4)) exhibit blueshifted, single-peaked profiles, reflecting strong contributions from the extended wind component. 
For higher-$J$ lines (S(5)--S(9)), emission from the hot, dense atmosphere of the inner disk ($R \lesssim 3\au$) becomes relatively significant (see Sections~\ref{sec:results:images} and \ref{sec:results:prop_traced_by_line}). 
At $i < 90\deg$, this leads to high-velocity wings ($|v| \gtrsim 10\kms$) produced by Keplerian motion in the inner disk atmosphere. 
At intermediate inclinations ($i=30$--$60\deg$), the high-$J$ line profiles appear flatter and multi-peaked. 
However, this may partly reflect underestimated high-$J$ emission from the extended wind (see also \secref{sec:discussions:pumping}); 
including it could yield single-peaked, brighter profiles, suppressing the flatter structures. 
The profiles in \fref{fig:velocity_profiles} also exclude the contribution from the innermost region ($\lesssim 2\au$), outside our computational domain, and should therefore be interpreted as those arising from the outer disk.  

The velocity centroids $v_{\rm cent}$ are generally small ($|v_{\rm cent}|\lesssim 1\kms$) across all lines. 
The FWHMs of single-peaked lines increase with higher-$J$, ranging from $3\kms$ to $8\kms$,  
and broaden slightly with inclination up to $i \approx 60\deg$ (by only a factor of $\approx 1.5$), remaining nearly constant at higher inclinations $i \gtrsim 60\deg$. 
These small $|v_{\rm cent}|$ and narrow widths highlight the slow nature of photoevaporative winds launched from outer radii (\secref{sec:results:wind_properties}) and imply that resolving their kinematics is inherently challenging, even at the highest spectral resolutions ($R\approx 100,000$; $\Delta v \sim 3\kms$) achievable with current and future instruments such as VLT/CRIRES+ and ELT/METIS. 

While $v_{\rm cent}$ below the instrumental resolution can in principle be measured with sufficient signal-to-noise ratio (S/N),  
detecting a $\sim 1\kms$ shift at $R\sim 100,000$ requires ${\rm S/N} \gtrsim 10$.
CRIRES+ covers the S(8) and S(9) lines and can reach ${\rm S/N }\approx 10$--20 at the line centers with a few-hour integration time, assuming a typical FWHM of $ \sim 3\kms$ and line fluxes of $\approx 10^{-15}\ergs\cm^{-2}$ for the wind component (cf. SY~Cha fluxes). 
However, these high-$J$ lines may also have substantial contributions from the inner hot disk \citep{2025_SchwarzErratum}, requiring high spatial resolution as well to isolate the wind component. 
Future $>30\,{\rm m}$-class telescopes, like ELT/METIS, will be particularly valuable for disentangling the kinematic signatures of winds to probe their origins.

}

\section{Discussion}    \label{sec:discussions}

In \secref{sec:discussions:diagnostics}, we recap our main findings and discuss observational diagnostics of photoevaporative winds. 
\secref{sec:discussions:pumping} examines the potential impacts of pumping processes on our results.
We then explore how the findings may vary with different stellar properties, such as mass and UV/X-ray luminosities, in \secref{sec:discussions:luminosity_dependence}. 
In \secref{sec:jwst_obs}, we compare our models with recent JWST observations and assess whether the observed \ce{H2} winds can be explained by photoevaporation, in light of preceding results.
Finally, we highlight key caveats in \secref{sec:discussions:caveat}. 

\subsection{How Can We Tell Whether Observed \ce{H2} Winds Are Photoevaporative or Not?}  \label{sec:discussions:diagnostics}

Photoevaporation drives \ce{H2} winds with a curved dissociation front. The winds typically have \ce{H2} densities of $10^2$--$10^{5} \cm^{-3}$ and temperatures of $100$--$ 1000\Kelvin$ with the fiducial parameters. 
This structure results in an X-shaped morphology in the S(1)--S(9) emission when viewed edge-on (\fref{fig:gallery}). 
The spatial extent of the emission decreases for higher $J$, from $\sim 300\au$ to $\sim 50\au$, though the latter may be underestimated (see \secref{sec:discussions:pumping}). 
The semi-opening angle also decreases for higher $J$, from $\sim 50^\circ$ to $\sim 37^\circ$, comparable to values observed in \NameOfTauxxxx{} and SY~Cha. This suggests a photoevaporative origin remains plausible for these sources, contrasting to the current interpretation favoring MHD winds. Our model demonstrates that geometric indicators such as opening angles and spatial extent are not definitive diagnostics unless the wind is extremely collimated. 
Given the PSF-induced biases (\secref{sec:results:images}), caution is essential when using these indicators to estimate wind-launching points or infer wind-driving mechanisms.

Energetics offer a more robust constraint. As discussed in \secref{sec:results:energetics}, only $\lesssim 10\%$ of the stellar UV/X-ray luminosity is converted into heat, and a small fraction of that ($\approx 9\%$ in our model) is radiated away as \ce{H2} emission. This sets a necessary condition for a photoevaporative origin: the observed \ce{H2} line luminosity needs to satisfy $L_{\rm H2} \lesssim 0.01 L_{\rm UVX}$ (in our simulation, $L_{\rm H2}\approx 3\e{-3}L_{\rm UVX}$). Exceeding this implies an additional energy source, such as accretion powering MHD winds. 
Note that this condition applies beyond $\gtrsim 2\au$, as our model excludes energy absorbed closer in; observed efficiencies should be evaluated consistently.
Also, this criterion considers only \ce{H2} emission from thermal excitation. In practice, higher-$J$ levels can be significantly influenced by nonthermal processes (\secref{sec:discussions:pumping}). Therefore, $L_{\rm H2}$ should be evaluated using lower-$J$ lines (e.g., S(1)--S(3)), which have low critical densities. 
Even when all lines are included in estimating $L_{\rm H2}$, the total \ce{H2} luminosity would still remain below $\lesssim 0.1 L_{\rm UVX}$, \revision{a hard limit} for photoevaporative winds, given that only $\mathcal{O}(10\%)$ of the UV and X-ray luminosity is typically absorbed by the wind \revision{(noting again that not all of the absorbed energy is converted into heat, as discussed in \secref{sec:results:energetics})}.

Wind velocity provides another criterion. Photoevaporative \ce{H2} winds typically have poloidal velocities of $\sim 1$--5$\kms$. 
If the \ce{H2} emission arises from \ion{H}{I}-dominated regions, velocities may reach $\approx 10\kms$, but not significantly beyond, as acceleration is limited past the isothermal sonic point. 
Thus, a velocity threshold of $\approx 10\kms$ serves as a practical upper bound.
\revision{
Accordingly, velocity centroids are expected to remain below this limit, typically on the order of $1\kms$. 
In contrast, MHD winds can exceed this range depending on the launching radius $R_0$ and the magnetic lever arm parameter $\lambda$ \citep{1982_BlandfordPayne}. A cold MHD wind model yields an asymptotic velocity of $v_\infty \approx v_{\rm K} (R_0) \sqrt{2\lambda-3} $, with $v_{\rm K}$ being Keplerian velocity. For typical values of $\lambda = 2 $--10, this gives $v_\infty \sim 10$--$40\kms$  at $R_0 \sim 10\au$ for $M_* = 1\Msun$.
}

In summary, key {\it necessary} conditions for photoevaporative winds include: moderately wide opening angles (cf. $\openingangle \approx30^\circ$--$50^\circ$ in the current model), an energy budget satisfying $L_{\rm H2} \lesssim 0.01 L_{\rm UVX}$, and wind velocities $\lesssim 10\kms$. 
While such low velocities are very challenging to resolve with JWST, morphology and energetics provide a strong diagnostic framework. 
These criteria would not rule out MHD winds, but their interpretation remains uncertain without detailed predictions from MHD models. 

A systematic survey correlating \ce{H2} flux with both UV and X-ray luminosities would be valuable. 
Moderately inclined disks are optimal targets, enabling both morphological analysis and comparisons with stellar UV/X-ray luminosities.
\revision{As we will discuss later in \secref{sec:discussions:luminosity_dependence}}, for low accretors like the one assumed in the fiducial model,
X-rays can dominate the energy input and scale with $L_{\rm H2}$ (Sections~\ref{sec:results:wind_properties} and \ref{sec:results:energetics}).
For FUV-dominated sources, trends remain unclear: lower-$J$ fluxes (e.g., S(1)--S(4)) may weakly anti-correlate with $L_{\rm FUV}$ due to enhanced \ce{H2} photodissociation, while for higher-$J$ lines, \ce{H2} photodissociation and excitation by nonthermal processes (e.g., UV pumping) may compete, making the correlation nontrivial. Further investigation is warranted from theoretical perspectives. 

\subsection{Missing Extended High-$J$ Line Emission?} \label{sec:discussions:pumping}
As discussed in \secref{sec:results:inclinations}, 
our model has likely underpredicted higher-$J$ line fluxes, especially for S(7)--S(9). 
Morphologically, our model also predicts smaller radial extent than observed for the higher-$J$ lines \citep{2025_Schwarz,2025_Pascucci}. 
In this section, we examine whether these discrepancies could rule out a photoevaporative origin, or if they instead point to missing physical processes in the current model that might enhance extended high-$J$ emission. 

These discrepancies can arise from simplifications in our treatment of the level populations. Specifically, we omit UV and X-ray pumping, as well as \ce{H2} excitation upon formation (hereafter formation pumping or chemical pumping), 
\footnote{When \ce{H2} forms via chemical reactions, a portion of the binding energy, $\approx 4.48\eV$, can go into excitation of the molecule \citep[e.g.,][]{1979_HollenbachMcKee}. }
both of which can populate excited rovibrational levels. 
Another possible factor is our neglect of emission from the inner, hotter disk outside our computational domain that is scattered by dust at larger radii.
We assess whether these processes could reconcile the discrepancies between our model and observations. 

Regarding UV pumping, absorption in the Lyman and Werner bands ($\gtrsim 11.2\eV$) excites \ce{H2} molecules to electronic excited states, followed by radiative decay into excited rovibrational levels of the electronic ground state.

To assess whether UV pumping significantly enhances pure rotational level populations, we have performed one-zone level population calculations using the Paris-Durham code \citep{2003_Flower, 2019_Godard}, 
which can compute \ce{H2} level populations for given density, temperature, and UV field strength. We adopted physical conditions representative of the \ce{H2} wind near the dissociation front at $r\sim 200\au$, where our model appears to underpredict high-$J$ emission. 
Relevant parameters center around $\nh \sim 10^3\cm^{-3}$ ($n_{\rm H2} \sim 10^2$--$10^3\cm^{-3}$) and $T \sim 600$--$900\Kelvin$. 

We find that the impact of UV pumping on pure rotational excitation sensitively depends on the treatment of formation pumping, across a broad parameter space, $\nh = 1$--$10^4\cm^{-3}$, $T = 500$--$2000\Kelvin$, and $G_0 = 1$--$10^3$.
In the standard setup of the shock code, formation pumping follows a Boltzmann distribution at $T_{\rm ex} = 17249\Kelvin$. 
Under this assumption, UV pumping has minimal effect on pure rotational levels, consistent with previous studies of irradiated interstellar gas \citep[][]{1987_BlackVanDishoeck, 2019_Godard} and protoplanetary disks \citep{2005_NomuraMillar, 2007_Nomura_II}.
Even for high-$J$ levels ($J = 7$--11), the enhancement from UV pumping is typically within a factor of two. 

In contrast, if instead \ce{H2} is assumed to form only in the lowest levels ($(v, J)=(0,0), (0,1)$), corresponding to level population calculations in this study, UV pumping becomes the sole channel for populating higher levels. This leads to order-of-magnitude increases in the populations of high-$J$ pure rotational levels, particularly for $J \gtrsim 7$--8, when $G_0 \gtrsim 10$. 
Even at $G_0 = 1$, significant increases are seen in low-temperature, low-density gas, where collisional excitation is inefficient.

These results suggest that formation pumping, alongside UV pumping, plays a critical role in setting high-$J$ populations, particularly for $J \gtrsim 7$--8. 
Incorporating these processes could improve agreement with the observed high-$J$ fluxes and morphologies, 
indicating that the discrepancies do not rule out a photoevaporative origin.
A detailed treatment, however, lies beyond the scope of this work and will be addressed in future modeling efforts.

The strong enhancement of populations due to formation pumping contrasts with the case of vibrationally excited states, such as the lowest-$J$ levels at $v = 1$, where UV pumping is typically more effective, even though their excitation energies are comparable to those of pure rotational levels at $J\approx8$. 
This highlights the importance of including UV pumping when modeling vibrationally excited lines, such as the $v=1$--0~S(1) line at 2.12$\um$, which is commonly used to trace winds \citep[e.g.,][]{2020_Gangi, 2025_Pascucci}. 
This line has also been found to be spatially extended, similar to the pure rotational S(9) emission \citep{2025_Pascucci}. 
More recently, \citet{2025_Kalscheur} have proposed FUV fluorescent lines as additional wind tracers. 
Together, these findings reinforce the need to incorporate UV pumping for testing models.

Similar to UV pumping, X-ray absorption can also excite \ce{H2} to rovibrational levels. 
Energetic photoelectrons produced by X-ray ionization can collisionally excite ambient \ce{H2} into electronically excited states, which then decay into various rovibrational levels of the ground electronic state \citep{1995_GredelDalgarno}. 
These photoelectrons can also directly excite the molecules vibrationally, as they lose energy through collisions. 
\citet{2007_Nomura_II} modeled these effects, along with UV and chemical pumping, in a hydrostatic disk and found little enhancement in pure rotational level populations, 
possibly because chemical pumping dominates rovibrational excitation. 
Thus, the impact of X-ray pumping is expected to be similar to that of UV pumping.

%

As for the potential contribution of \revision{infrared line} scattering \revision{on small dust} in the wind, its effect on high-$J$ emission appears minor, although some uncertainty remains.
\citet{2025_Schwarz} showed that the extended \ce{H2} emission in SY~Cha lies well above the flat scattered-light surface imaged by SPHERE at $2.2\um$, suggesting that scattering is unlikely to be a major factor in this system.
In \NameOfTauxxxx{}, S(5)--S(8) emission appears spatially coincident with $5\um$ dust emission \citep{2024_Arulanantham}, whereas S(9) emission seems spatially offset from it \citep{2025_Pascucci}. 
Overall, 
we expect the contribution of scattering to high-$J$ emission to be small; this would be the case especially for photoevaporative winds given the generally dust-optically-thin nature.


\subsection{Stellar Parameter Dependence}   \label{sec:discussions:luminosity_dependence}
The properties of photoevaporative winds are generally sensitive to stellar UV and X-ray luminosities, which can strongly influence \ce{H2} emission. 
Here, we explore how varying these luminosities affects the morphologies and fluxes of the \ce{H2} lines, relative to our fiducial model. 
We also briefly consider the role of stellar mass. 
These tests provide context for our comparison with the observations in \secref{sec:jwst_obs}. 
They also highlight the potential of using correlations between \ce{H2} emission and stellar properties as a diagnostic of photoevaporation.


As $\LFUV$ increases, line emission is expected to strengthen due to enhanced heating. 
However, \ce{H2} also becomes more susceptible to FUV photodissociation, particularly in low-density, high-temperature regions ($T \gtrsim 500\Kelvin$), making the net effect on the fluxes non-trivial.
Additionally, higher $\LFUV$ broadens the \ce{H2} dissociation front, increasing the semi-opening angle.

To assess this, we have restarted the fiducial simulation with higher FUV luminosities, $\LFUV = 4.0\e{30}\ergs$ and $3.5\e{31}\ergs$ (4$\times$ and $40\times$ the fiducial value), corresponding to $\dotMacc = 10^{-9}\Msun\yr^{-1}$ and $10^{-8}\Msun\yr^{-1}$, respectively. All other parameters remained the same as in \tref{tab:fiducialmodel}.  
In these runs, FUV photodissociation contributes to \ce{H2} destruction at levels exceeding or comparable to those from X-ray photoionization in the fiducial model.

\begin{figure*}

    \centering
    \includegraphics[width=0.48\linewidth, clip]{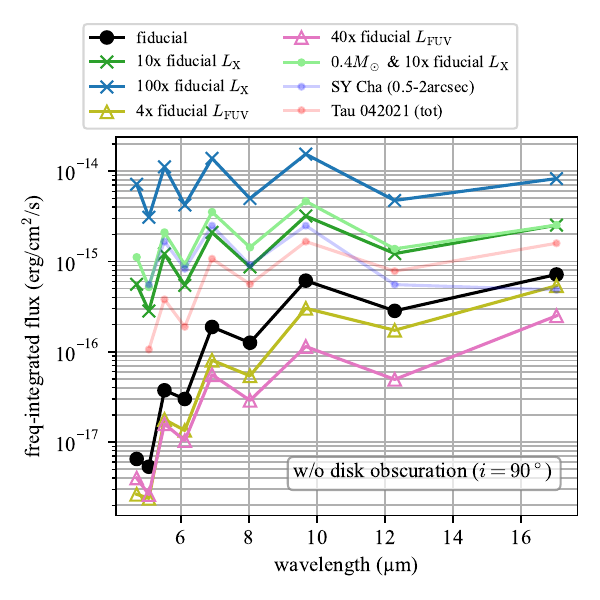}
    \includegraphics[width=0.48\linewidth, clip]{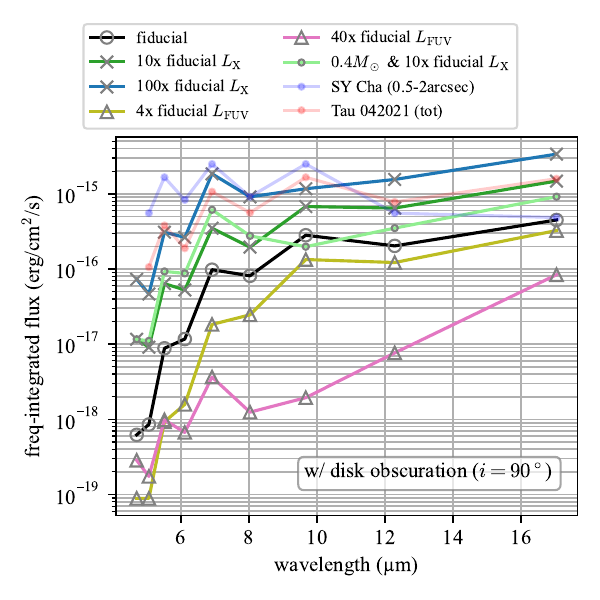}
    \\
    \includegraphics[width=0.48\linewidth, clip]{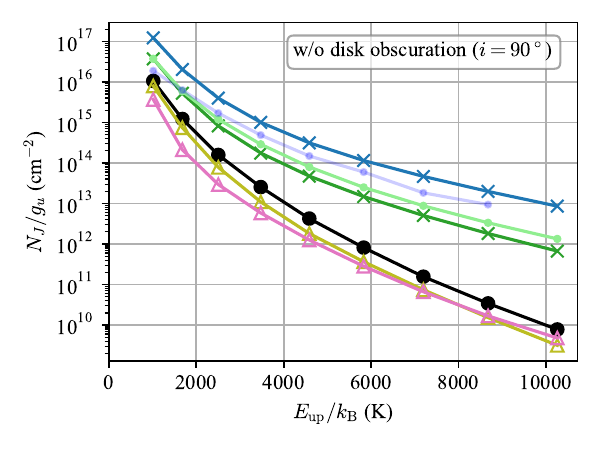}
    \includegraphics[width=0.48\linewidth, clip]{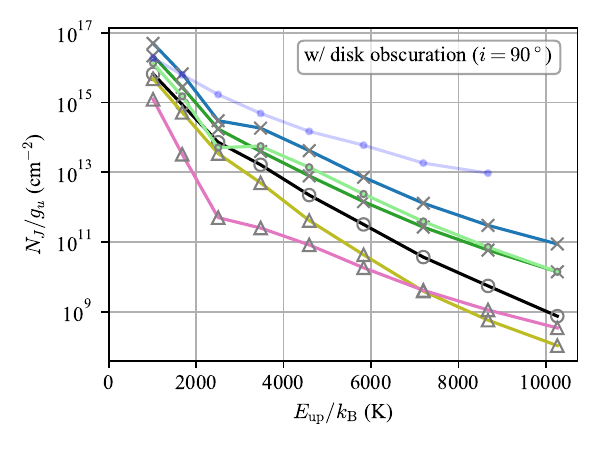}
    \caption{
    (top left): Same as \fref{fig:sed}, but showing line fluxes for models with enhanced FUV and X-ray luminosities: 10$\times$ (green) and $100\times$ (blue) the fiducial $L_{\rm X}$, and 4$\times$ (yellow) and $40\times$ (pink) the fiducial $L_{\rm FUV}$. 
    The lower-mass run ($M_* = 0.4 \Msun$ with $10\times$ the fiducial $L_{\rm X}$) is also included (light green).
    All fluxes are computed with $i = 90^\circ$ and without disk obscuration.
    \revision{(top right): Same as the top left panel, but with reduced fluxes due to disk obscuration. }
    (bottom left): Rotation diagram, same as \fref{fig:rot-diagram_wobs}, but using the fluxes from the top left panel.
    \revision{(bottom right): Rotation diagram using the fluxes with disk obscuration from the top right panel. }
    }
    \label{fig:sed_comparison}
\end{figure*}

In the $4\times$ FUV case, the overall disk structure is similar to the fiducial model (\fref{fig:hydro}), but enhanced FUV photodissociation destroys hot \ce{H2} near the dissociation front, reducing the vertical extent of high-$J$ emission, and slightly broadening the semi-opening angles to $\aveopeningangle\approx40^\circ$--$46^\circ$. Line fluxes decrease by a factor of a few (\revision{compare yellow and black lines in the top panels of \fref{fig:sed_comparison}}).
The energy efficiency is also correspondingly reduced to $L_{\rm H2}\approx0.93\e{-3}L_{\rm UVX}$. 

These trends are more pronounced in the $40\times$ FUV run. 
X-ray heating remains dominant throughout most of the \ce{H2} region, except near the wind base where FUV photoelectric heating becomes comparable. The main effect of stronger FUV is enhanced destruction of warm \ce{H2} ($T\gtrsim 300\Kelvin$). 
The overall temperature structure changes little but is slightly reduced due to increased cooling by \ion{O}{I}, whose abundance rises following stronger CO photodissociation.
These effects collectively lead to weaker \ce{H2} line fluxes \revision{(pink lines in the top panels of \fref{fig:sed_comparison}). 
When disk obscuration is included for an edge-on geometry (pink line in the top right panel), the depletion of warm \ce{H2} further reduces fluxes from the extended \ce{H2} winds.} 
The resulting energy efficiency is even lower at $L_{\rm H2} \approx 1.3\e{-4} L_{\rm UVX}$.

We also tested enhanced X-ray luminosities of $L_{\rm X} = 2.82\e{31}\ergs$ and $L_{\rm X} = 2.82\e{32}\ergs$ (10$\times$ and 100$\times$ the fiducial value; the latter for experimental purposes). 
At 10$\times$, line fluxes increase across all transitions (\revision{thick green lines in \fref{fig:sed_comparison}}), and the semi-opening angles widen to $\aveopeningangle \approx 45^\circ$--$55^\circ$. 
High-$J$ emission remains compact, leaving unresolved the issue discussed in \secref{sec:discussions:pumping}. 
At 100$\times$, fluxes rise further (thick blue lines in \fref{fig:sed_comparison}) 
and opening angles reach $\openingangle \approx 50^\circ$--$60^\circ$. 
These trends reflect faster \ce{H2} destruction and more efficient heating of denser \ce{H2} layers at higher $L_{\rm X}$.
The energy efficiency is comparable to the fiducial model: $L_{\rm H2} = 5.2\e{-3}L_{\rm UVX}$ and $4.8\e{-3} L_{\rm UVX}$ for the 10$\times$ and 100$\times$ runs, respectively. 
The relatively higher efficiencies, compared to the high-$L_{\rm FUV}$ models, indicate greater energy deposition per \ce{H2} molecule destruction. 


In summary, both \ce{H2} morphology and fluxes are sensitive to UV and X-ray luminosities. 
Our preliminary results reveal overall trends: line fluxes increase with higher $L_{\rm X}$ and lower $L_{\rm FUV}$, while semi-opening angles grow with both. 
However, the decreasing trend of high-$J$ line fluxes (e.g., S(5)--S(9)) with increasing $L_{\rm FUV}$ is uncertain---if UV and chemical pumping were included, higher $L_{\rm FUV}$ could instead enhance high-$J$ emission.
Moreover, the results from the high-FUV runs should not be interpreted as implying that \ce{H2} winds cannot be reproduced in systems with strong FUV radiation, as variations in PAH and elemental abundances may also play a role (\secref{sec:discussions:caveat}). 
These results simply illustrate relative trends when either the FUV or X-ray luminosity is held fixed.
A systematic parameter study 
is therefore needed to verify these trends and fully map the range of fluxes and morphologies that photoevaporative winds can produce.
Our current findings suggest that photoevaporative \ce{H2} winds are more readily detectable in evolved disks around low-mass, active stars with low accretion rates (particularly in low-$J$ lines).

Stellar mass can also influence wind geometry. 
For instance, stars with $M_*< 1\Msun$ may produce narrower semi-opening angles than our fiducial model, as \ce{H2} can be launched from smaller radii and the disk scale height increases due to weaker stellar gravity. 
This trend contrasts with the broader semi-opening angles seen under stronger UV and X-ray irradiation, suggesting that different combinations of $M_*$, $L_{\rm X}$, and $\LFUV$ could result in similar wind morphologies.

To test this, we run a model with $M_* = 0.4\Msun$ and $10\times$ the fiducial X-ray luminosity (with $L_* = 0.93\Lsun$ and $R_* = 2.12\Rsun$). We chose this setup, as it is expected to most closely reproduce the morphology in the fiducial model, based on the preceding exploration.
Indeed, this model yields a similar morphology to the fiducial run (see also the discussion in \secref{sec:jwst:Tau042021}). The \ce{H2} line fluxes are comparable to those in the $10\times$ X-ray run (see the light green lines in the top panels of \fref{fig:sed_comparison}).

These preliminary findings reinforce our argument in \secref{sec:discussions:diagnostics} that, without well-constrained stellar properties, geometric structure alone is insufficient to distinguish between photoevaporation and MHD winds. 
A systematic exploration of both stellar mass and luminosities is therefore essential for robust interpretation.
On the theoretical side, these variations of morphology and fluxes indicates the need of source-specific modeling in conclusively assessing the origin of observed \ce{H2} winds.

\subsection{Comparisons with JWST Observations}    \label{sec:jwst_obs}
The characteristic X-shaped morphology of the \ce{H2} emission in our model images (\fref{fig:gallery}) resembles recent JWST observations of \NameOfTauxxxx{} \citep{2024_Arulanantham} and SY~Cha \citep{2025_Schwarz}, prompting the question of how well our photoevaporation model can quantitatively reproduce these features. 
In this section, we address that question. 
Comparisons with \NameOfTauxxxx{} \citep{2024_Arulanantham} and SY~Cha \citep{2025_Schwarz} are presented in Sections~\ref{sec:jwst:Tau042021} and \ref{sec:jwst:SYcha}, respectively.

\subsubsection{\NameOfTauxxxx{}} \label{sec:jwst:Tau042021}

\NameOfTauxxxx{} (2MASS~J04202144+2813491) features an edge-on disk around an M-type star in Taurus \citep{2009_Luhman, 2014_Stapelfeldt, 2019_Simon}. 
The disk inclination is $i > 85^\circ$ \citep[][]{2020_Marion}, and the stellar mass is $M_* = 0.4\Msun$ \citep{2024_Duchene}. 
\footnote{\citet{2019_Simon} report $M_* = 0.27\Msun$. See Section~3.4 of \citet{2024_Duchene} for a discussion on this discrepancy.}
The accretion rate is estimated at $\dot{M}_{\rm acc} = 2\e{-11}\Msun$ through hydrogen recombination lines \citep{2024_Arulanantham}, though this likely underestimates the true value due to the nearly edge-on orientation. 
In fact, \citet{2025_Naman} report a much higher accretion rate $\dotMacc \sim 10^{-8}\Msun\yr^{-1}$ based on jet mass-loss rates.

Recent ALMA observations in Bands~4 and 7 have resolved the disk's vertical structure, revealing strong settling of large grains \citep{2020_Marion}. 
Meanwhile, JWST/MIRI imaging at $7.7\mum$ and $12.8\mum$ has revealed a prominent X-shaped structure extending above the midplane \citep{2024_Duchene}. 
This X-shaped emission lies above the warm molecular \ce{^12CO} layer traced by ALMA, which has a semi-opening angle of $\aveopeningangle \sim 55^\circ$ and vertical extent reaching up to $\sim 225\au$. 


More recently, \citet{2024_Arulanantham} detected the \ce{H2} S(1)--S(8) lines with JWST/MIRI. 
The S(2) image ($12.3\um$) exhibits a clear X-shaped morphology, consistent with earlier $7.7\um$ and $12.8\um$ data of \citet{2024_Duchene}, and may in part be traced by vertically extended \ce{^12CO} \citep{2025_Foucher}. 
The S(1)--S(4) lines are similarly vertically extended beyond the scattered-light continuum, whereas S(5)--S(8) appear more compact and cospatial with dust emission at $5\um$.

The semi-opening angle from the S(2) line is $\aveopeningangle = 35^\circ\pm 5^\circ$, which \citet{2024_Arulanantham} interpret as features of MHD winds.
They cite its consistency with the inclination $i\sim 35^\circ$, the angle at which the largest blueshifts of [\ion{O}{I}]~BCs and NCs are observed \citep[][]{2019_Banzatti,2022_Pascucci}
The BCs, in particular, are considered to originate from MHD winds based on their inferred emitting radii. 
This inclination is also consistent with the picture of the ``bead on a rigid wire'' MHD wind model, which requires $\theta > 30^\circ$ to launch outflows \citep[e.g.,][]{1982_BlandfordPayne}. 
However, this interpretation remains tentative, as neither the launching radius nor velocity structure can be resolved with MIRI's channel~3 pixel scale ($0.\arcsec 245\sim 34\au$) or spectral resolution ($\Delta v \sim 120\kms$ at $12\um$). 

\citet{2025_Pascucci} also favors the MHD wind scenario based on the observed wind morphologies and nested structures. The view also aligns with the presence of the narrow, collimated jet traced by [\ion{Fe}{II}] and the recently reported high accretion $\dotMacc \sim 10^{-8}\Msun\yr^{-1}$ \citep{2025_Naman}, which may lead to significant shielding of stellar UV and X-ray by inner winds, potentially making photoevaporation less efficient \citep{2020_Pascucci, 2022_Takasao}.

Nevertheless, the observed \ce{H2} line centroids lie between $-20$--$2\kms$ \citep{2024_Arulanantham}, a possible range of photoevaporative winds---though velocities along the rotational axis is not measurable in edge-on systems.
In addition, the total luminosity of the pure rotational lines (\tref{tab:fluxes}) is $L_{\rm H2} \approx 1.6\e{28}\ergs$, consistent with expectations for photoevaporative winds driven with UV and X-ray luminosities of $\sim 10^{30}$--$10^{31}\ergs$ (\secref{sec:results:energetics} and Eq.\eqref{eq:energy_efficiency}).
From these perspectives, a photoevaporative origin can remain a viable possibility, and merits comparison with our model predictions to assess its plausibility. 


\begin{figure*}
    \centering
    \includegraphics[width = \linewidth, clip]{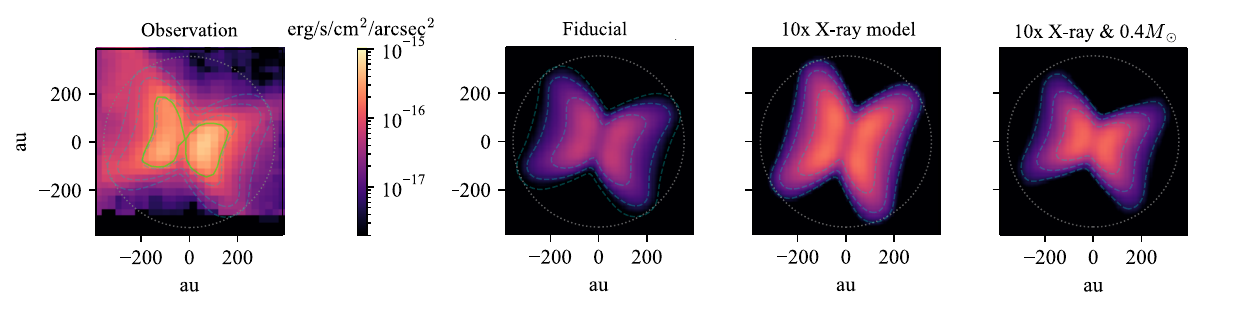}
    \caption{
    Morphology comparison between the models and observations of \NameOfTauxxxx{} for the S(2) line. 
    The left panel shows the frequency-integrated, continuum-subtracted JWST image from \citet{2024_Arulanantham}.
    The three panels on the right show the model S(2) images from the fiducial model (cf. \fref{fig:gallery_i90_wd}),  the 10$\times$ X-ray model ($L_{\rm X} \approx 3\e{31}\ergs$; see \secref{sec:discussions:luminosity_dependence}), and the test model with $M_* = 0.4M_\odot$ and $10\times $ X-ray luminosity. 
    All model images are convolved with a 2D Gaussian (FWHM $\sim 72\au$) and rotated to match the observed position angle. 
    Gray dotted circles mark the extent of our computational domain.
    Cyan dashed contours indicate 35\%, 10\%, and 3\% of the peak intensity in each convolved model image;
    those of the fiducial model are also overplotted on the observed image, which well align with the observed morphology. 
    The green contour in the observed image shows 35\% of the peak intensity, approximating the spatial integration domain used to measure the S(2) flux in \citet{2024_Arulanantham}. 
    \revision{Note that the flux integration regions differ between our models and the observations (see \fref{fig:sed} and \secref{sec:jwst:Tau042021} for details).}
    The color scale is shared for all the panels.
    }
    \label{fig:comparison_with_tau}
\end{figure*}

\fref{fig:comparison_with_tau} compares our convolved S(2) maps with the observed image from \citet{2024_Arulanantham}. The fiducial model reproduces the observed morphology well, with a semi-opening angle of $\aveopeningangle = 36.1^\circ$ (\tref{tab:geometry}), consistent with the reported $ 35^\circ \pm 5^\circ$.
A similar agreement is seen for the S(9) line ($\aveopeningangle = 37^\circ$), consistent with $\sim 35^\circ$--$40^\circ$ measured from PSF-deconvolved JWST/NIRSpec images \citep{2025_Pascucci}. 
The geometric radius $\Rgeo$ is somewhat larger in our model ($\approx 19$--$26\au$ vs. $6\pm 7 \au$), but still broadly consistent. 
We note that our fiducial model assumes $M_* = 1\Msun$ whereas \NameOfTauxxxx{} has $M_* = 0.4\Msun$; the corresponding critical radius \citep{2003_Liffman} for $\approx 10^3\Kelvin$ molecular gas is $\sim 14\au$, close to the observed $\Rgeo$. 


Despite the morphological agreement, the fiducial model is fainter than the observations. This may appear inconsistent with \fref{fig:sed} and \tref{tab:fluxes}, which show broadly consistent low-$J$ line fluxes, 
but this discrepancy arises mainly because the observed fluxes are measured over smaller emitting areas (see the green contour in the top-right panel of \fref{fig:comparison_with_tau}). 
The observed flux ($\approx7.8\e{-16}\ergs\cm^{-2}$) is measured within this contour, whereas the model flux ($\approx2.0\e{-16}\ergs\cm^{-2}$) is integrated over the full domain. 
\revision{When limited to the region where the surface brightness exceeds 35\% of the peak (the first cyan dashed contour), the model flux becomes $1.3\e{-16}\ergs\cm^{-2}$, about six times less than the observed value.}

\revision{
\fref{fig:comparison_with_tau} also compares the convolved S(2) images of the 10$\times$ X-ray luminosity model from \secref{sec:discussions:luminosity_dependence} with the observations. 
This model yield fluxes of $4.6\e{-16}\ergs\cm^{-2}$ 
within the region where the surface brightness exceeds 35\% of the peak (the first cyan dashed contour), only a factor of $\sim 1.7$ lower than the observed value. 
This suggests that higher X-ray luminosities can reproduce the observed flux levels; indeed, our 100$\times$ X-ray luminosity model yields $1.1\e{-15}\ergs\cm^{-2}$, a factor of $\sim 1.4$ higher than the observed flux. 
}

We also compare the observations with our test model using $M_* = 0.4 \Msun$, more consistent with \NameOfTauxxxx{} (see \secref{sec:discussions:luminosity_dependence}).
\footnote{\citet{2025_Naman} suggest the jet wiggling points to a close-in binary with a separation of $\sim 1.35\au$ and component masses of $\sim 0.33\Msun$ and $\sim 0.07\Msun$}
This model adopts 10$\times$ the fiducial $L_{\rm X}$ ($\sim 3\e{31}\ergs$). 
Compared to the fiducial case, it yields fluxes closer to the observed value while retaining similar morphology (bottom panels of \fref{fig:comparison_with_tau}). 
\revision{The flux within the first cyan dashed contour (where surface brightness exceeding 35\% of the peak) is $2.2\e{-16}\ergs\cm^{-2}$, about 3.5~times lower than the observed value.

These comparisons indicate that there likely exists a region of parameter space at higher X-ray luminosities where both the morphology and surface brightness could be reproduced. Although none of the (preliminary) models from \secref{sec:discussions:luminosity_dependence} achieve this simultaneously, our parameter space exploration is still limited and insufficient to draw meaningful conclusions on the plausibility of a photoevaporative origin. A detailed fitting analysis---beyond the scope of this work---would be valuable to further examine this.
}

One possible concern is that the required high $L_{\rm X}$ lies near the upper end of observed values for a $0.4\Msun$ star and is likely rare \citep{2007_Gudel, 2016_Gregory}, so caution is warranted.
\revision{Nevertheless, uncertainties remain in our treatment of X-ray heating efficiency and in both the elemental and chemical abundances of key coolants (see also \secref{sec:discussions:caveat}). For instance, if \ion{O}{I} or CO were somewhat depleted, the reduced cooling would lead to higher temperatures and densities for a given X-ray luminosity. Consequently, surface brightness levels comparable to the observations could still be reproduced with more moderate X-ray luminosities. }

Another caveat is that the recently estimated accretion rate of $\dotMacc \sim 10^{-8}\Msun\yr^{-1}$ \citep{2025_Naman} poses a potential challenge for the photoevaporation scenario, as strong accretion-generated UV could efficiently dissociate \ce{H2}. 
This is consistent with the steep short-wavelength SED slope reported by \citet{2024_Duchene}.
Our simplified high-$L_{\rm FUV}$ model with $\dotMacc = 10^{-8}\Msun\yr^{-1}$ in \secref{sec:discussions:luminosity_dependence} shows a $\sim 6\times$ drop in the S(2) line flux and a modest widening of the semi-opening angle to $\sim 47^\circ$, suggesting that reproducing the observed features through photoevaporation could become relatively challenging under intense UV conditions.
Nevertheless, given uncertainties in possible additional \ce{H2} production processes (\secref{sec:discussions:caveat}) and the uncertain factors mentioned above, the apparent tension between the model and a potentially strong UV field does not definitively rule out a photoevaporative origin. Importantly, the morphological agreement remains striking. 
Whether MHD winds can sustain \ce{H2} under such a high-UV regime remains an open and important question.

We therefore conclude that photoevaporation is viable to reproduce the key features of \ce{H2} winds in \NameOfTauxxxx{}, keeping open the possibility of a photoevaporative origin, in contrast to the current interpretation generally favoring MHD winds. 
However, this agreement alone does not constitute conclusive evidence. 
Notably, morphology and fluxes are sensitive to stellar parameters (\secref{sec:discussions:luminosity_dependence}), and therefore conclusively identifying a photoevaporative origin requires source-specific modeling---none of our models were tuned to fit \NameOfTauxxxx{}. 
Even if the observed \ce{H2} wind is photoevaporative, explaining the observed narrow [\ion{Fe}{II}] jet likely requires an additional MHD component, suggesting photoevaporation and MHD winds may coexist or operate in different parts of the disk. 




\subsubsection{SY~Cha} \label{sec:jwst:SYcha}
SY~Cha hosts a moderately inclined transition disk \citep[$i = 51^\circ$;][]{2023_Orihara} around a K-type star with $M_*=0.7\Msun$ \citep{2016_Manara, 2016_Feiden, 2021_Gaia}. 
The accretion rate has been estimated to be relatively low $\dotMacc\approx 6.6\e{-10}\Msun\yr^{-1}$ but has been recently measured to $\sim 10^{-8}\Msun\yr^{-1}$, typical for accreting stars \citep{2025_Pittman}.
Recent JWST MIRI-MRS observations revealed extended \ce{H2} emission in the S(3)--S(7) lines, with a relatively wide semi-opening angle of $\aveopeningangle \approx 50$--$60^\circ$ \citep{2025_Schwarz}.

Our model (cf. \fref{fig:gallery_i60_wd}) broadly reproduces the morphology and surface brightness of S(3) and S(4), but the observed S(5)--S(7) emission appears more spatially extended.
Conversely, the S(1) and S(2) lines show only compact emission in the observations (at least within the adopted color scales), whereas our model predicts extended conical structures when displayed on logarithmic scales. 

\revision{
Despite these morphological differences, the observed S(1) and S(2) fluxes measured with the radius aperture of $0.5$--$2\arcsec$ \citep{2025_SchwarzErratum} are consistent with those of our fiducial model (\fref{fig:sed} and \tref{tab:fluxes_comparison}) and the $10\times$ X-ray luminosity model (\fref{fig:sed_comparison}), noting that our models use $d = 140\pc $ instead of $d = 180.7\pc$. 
A similar level of agreement is also found in the rotation diagram (bottom left panel of \fref{fig:sed_comparison}). 

The observationally inferred \ce{H2} mass-loss rate ($3.77\pm 0.63 \times 10^{-10}\Msun\yr^{-1}$) agrees well with that of the fiducial model ($\approx 4\e{-10}\Msun\yr^{-1}$) but is lower than that of the $10\times$ X-ray case ($\approx 2\e{-9}\Msun\yr^{-1}$). 
The difference likely stems from the observational estimation method, which assumes a wind velocity of $10\kms$ and derives the \ce{H2} column density from hot gas tracers (S(3)--S(7)). Since the hot \ce{H2} component represent only the surface layer of the \ce{H2} dominated region, the inferred mass-loss rate is expected to be smaller than in our model, where the most of the mass loss arises from warm gas primarily traced by S(1)--S(3) with velocities of $\lesssim 5\kms$. In addition, if part of the observed high-$J$ emission originates from molecules excited by UV or chemical pumping, $\Trot$ can be the kinetic temperature, leading to an underestimated column density and thereby mass-loss rate. 
Therefore, it is unclear whether the discrepancy in the mass-loss rates is physical or merely apparent, and whether the observed low rate itself indicates a photoevaporative origin cannot be unambiguously determined within this work. 

Overall, the morphological agreement is qualitative and somewhat less compelling than for \NameOfTauxxxx{} (\secref{sec:jwst:Tau042021}), though the flux agreement is more notable.
These results suggest that a photoevaporative origin also remains plausible for the \ce{H2} winds in SY~Cha.
However, reproducing the observed [\ion{Ne}{II}] would likely require an additional MHD wind component, as in \NameOfTauxxxx{}. 
}
Furthermore, SY~Cha appears to experience markedly different irradiation conditions:
its X-ray luminosity \citep[$L_{\rm X} \approx 6.9\e{29}\ergs$;][]{2010_Gudel} is about four times lower, while its FUV luminosity between $6$--$11\eV$ from ULLYSES \citep[$\sim 9\e{30}\ergs$;][]{2025_Ullyses} is roughly nine times higher than in our fiducial setup. 
Whether this combination of conditions can fully account for the observations remains uncertain.
To properly assess the photoevaporative scenario, dedicated source-specific modeling that includes UV and chemical pumping is needed.

\subsection{Caveats and Future Refinements}    \label{sec:discussions:caveat}
In addition to the omission of pumping in our level population calculations (\secref{sec:discussions:pumping}), there are several other model limitations that should be considered in future  work and when applying our results to broader systems.  
We outline them below.

\paragraph{Chemical network} 
To keep our simulations computationally feasible, we have adopted a reduced chemical network, which may introduce some uncertainties and limit our ability to unambiguously determine the wind origin of observed \ce{H2} winds, as discussed in \secref{sec:jwst_obs}. While the network has been validated through extensive testing (Appendix~\ref{sec:detailed_description_chemistry}), some effects remain unaccounted for.

For example, \ce{H2} formation via PAH catalysis may enhance the abundance of hot \ce{H2} in the \ion{H}{I} region \citep{2012_Bruderer, 2013_Bruderer}. 
Although this depends on the poorly constrained PAH abundance, incorporating PAH-related reactions could improve model completeness, especially since PAHs have been observed in systems like \NameOfTauxxxx{} \citep[][]{2025_Dartois}. 

\revision{Similarly, an additional uncertainty lies in the \ce{H2} formation rate on warm grains \citep{2004_Habart, 2004_CazauxTielens, 2010_CazauxTielensErratum}. While this process is likely ineffective within the winds in our model, where density is relatively low, it could still affect the \ce{H2} density near the base, particularly at inner radii ($\lesssim 10\au$). 

Grain-catalyzed \ce{H2} formation is further complicated by the uncertain abundance of small grains. While we have adopted a small-dust-to-gas mass ratio of 0.0015 (assuming 85\% of the dust mass has settled), recent JWST mid-infrared imaging of edge-on disks suggests that small grains may be well-mixed and more abundant in disk surfaces than previously assumed \citep{2024_Duchene, 2025_Tazaki}. 
A higher small-grain abundance would also enhance dust-gas collisional cooling. 
Both effects are most pronounced in the high-density region and thus primarily affect the \ce{H2} density near the inner wind base. 
Future studies exploring how the small-grain abundance in disk surfaces on \ce{H2} emission may be valuable for simultaneously interpreting observations of both gas and dust.
}

Additionally, the current network includes X-ray photoionization only for H and \ce{H2} (although absorption by other species such as C and O is accounted for in the calculations of the heating rates). It may be worth exploring how including other species influences the chemical structures, especially in the context of comparing with multi-species data.

\paragraph{Heating and cooling processes}
Not all of X-ray energy goes into heating; some is used for ionization and excitation. This is typically encapsulated by the heating efficiency $f_{\rm h}$, which we set to $f_{\rm h} = 0.4$ for molecular gas and $f_{\rm h} = 0.1$ for atomic gas, following \citet{1991_XuMcCray,1996_Maloney}. 
These values represent typical efficiencies for $\sim 10^3\eV$ primary photoelectrons in neutral gas. In reality, $f_{\rm h}$ depends on both photoelectron energy and the ionization degree of the ambient gas. 

It can rise to $f_{\rm h}\sim 0.3$--$0.5$ in regions with elevated electron fractions \citep{1985_ShullSteenberg}, such as the \ion{H}{I} layer near the \ce{H2} dissociation front, where $y_{\rm e} \sim 10^{-4}$--$10^{-2}$ in our model. 
This could lead to higher temperatures and thus stronger high-$J$ emission.
A more refined treatment of X-ray heating, accounting for its energy dependence and local ionization states, would improve model accuracy and strengthen comparisons with observations. 

\revision{
Regarding cooling, uncertainties in the elemental abundances of key coolants may also influence the resulting temperature and density structures of the \ce{H2} winds, as well as the net UV and X-ray luminosities, as discussed in \secref{sec:jwst_obs}. Accounting for such variations would be particularly important in source-specific modeling and fitting analyses.
}

\paragraph{Advection effects on level populations}
The current model assumes steady-state level populations, meaning it neglects the transport of excited \ce{H2} before radiative decay. 
This approximation can break down in fast-moving gas, particularly under non-LTE conditions, where excited molecules may travel a characteristic distance of 
\[
\begin{split}
    \Delta L & \sim v_{\rm p} A_{ul}^{-1} ~\mathrm{min}\braket{1, n_{\rm cr}/n} 
    \\
    & \approx   70\au \braket{\frac{v_{\rm p}}{5\kms}}
    \braket{\frac{A_{ul}}{4.8\e{-10}\unit{s^{-1}}}} ^{-1}
        \mathrm{min}\braket{1, n_{\rm cr}/n}. 
\end{split}
\]
The nominal value of $A_{ul}$ corresponds to the S(1) transition (note that, approximately, $A_{ul}\propto\nu_{ul}^5$ for quadrupole lines). 

In our photoevaporation model, this effect is likely minor due to the slow flow speed and the near-LTE conditions of the lowest-$J$ lines, at least within the computational domain. 
It is also likely unimportant for high-$J$ lines such as S(6)--S(9), given $A_{ul} > 10^{-7}\unit{s}^{-1}$.
However, in general, this advection effect should be considered in models or observations involving emission from faster outflows.



\section{Summary}   \label{sec:summary}
While recent JWST observations have revealed the spatially-resolved properties of \ce{H2} winds from PPDs \citep[e.g.,][]{2024_Duchene, 2024_Arulanantham, 2025_Schwarz, 2025_Pascucci},
there are still no theoretical models to date directly comparable with these observations, leaving the origins of the winds uncertain.

To address this gap, we have constructed the first model of photoevaporative \ce{H2} winds directly comparable with JWST data.
The core questions of this study have been the morphologies and fluxes of \ce{H2} pure rotational lines, S(1)--S(9), from photoevaporative winds.
Using radiation hydrodynamics simulations coupled with chemistry, we derived steady-state structures of photoevaporative \ce{H2} winds (\fref{fig:hydro}). We then post-processed the simulations to calculate level populations and line transfer, considering collisional (de)excitation and spontaneous decay. 

In our fiducial model with $M_* = 1\Msun$ and UV/X-ray luminosities of $\sim 10^{30}\ergs$ (\tref{tab:fiducialmodel}; \fref{fig:stellar_spectrum}), the synthetic images exhibit the X-shaped morphology with radial extents of $\gtrsim 50$--$300\au$ and semi-opening angles of $\aveopeningangle \approx 37^\circ$--$50^\circ$ (\fref{fig:gallery} and \tref{tab:geometry}). 
The predicted line fluxes are $\sim 10^{-16}$--$10^{-15}\ergs\cm^{-2}$ for the S(1)--S(5) lines, and $\sim 10^{-17}\ergs\cm^{-2}$ for the S(6)--S(9) lines, broadly consistent with JWST measurements (\tref{tab:fluxes}; \fref{fig:sed}).
However, we have likely underestimated higher-$J$ emission from extended \ce{H2} due to our neglect of UV, X-ray, or chemical pumping (\secref{sec:discussions:pumping}). 

We have also discussed how morphology and fluxes can vary with stellar parameters. Line fluxes increase with higher X-ray luminosity and lower FUV luminosity, while $\aveopeningangle$ widen with increasing UV and X-ray luminosities (\secref{sec:discussions:luminosity_dependence}; \fref{fig:sed_comparison}). A lower stellar mass produces more collimated \ce{H2} winds by allowing launching from inner radii. 
These trends are promising but still preliminary, and more extensive modeling is needed to achieve robust conclusions.

We have compared our predictions with JWST observations of \NameOfTauxxxx{} and SY~Cha \citep{2024_Arulanantham, 2025_Pascucci, 2025_Schwarz, 2025_SchwarzErratum} and have found remarkable agreement in morphology, along with broad consistency in fluxes, especially for lower-$J$ lines (Figures~\ref{fig:sed} and \ref{fig:comparison_with_tau}; \tref{tab:fluxes}). 
For high-$J$ lines (S(5)--S(9)), again, the fluxes and spatial extents are likely underestimated (\secref{sec:discussions:pumping}).
In contrast to the current interpretation, these comparisons support the plausibility of a photoevaporative origin for the observed \ce{H2} winds, with some remaining challenges (\secref{sec:jwst:Tau042021}).

Although commonly used, semi-opening angles by themselves are insufficient to distinguish between MHD and photoevaporative winds, unless winds are extremely collimated.
Notably, PSF effects can introduce substantial biases in geometric indicators (\secref{sec:results:images} and \tref{tab:geometry}).
To better diagnose wind origins, complementary data on stellar UV/X-ray luminosities and accretion rates are essential. 
They would enable assessing whether the observed \ce{H2} line luminosities are energetically feasible for photoevaporation (Sections~\ref{sec:results:energetics}, \ref{sec:discussions:diagnostics}, and \ref{sec:jwst_obs}). 
In this context, moderately inclined disks are ideal targets to explore the possibility of photoevaporative winds, as they enable both morphological and energetics analysis (\secref{sec:discussions:diagnostics} for details on diagnostics of photoevaporative winds).

Overall, photoevaporation can reproduce the key features of \ce{H2} winds.
However, conclusively identifying the wind origin for individual sources requires source-specific modeling, as morphology and fluxes are sensitive to stellar parameters.

Future improvements should include UV, X-ray, and chemical pumping in level population calculations, and a broader exploration of stellar parameters.
Detailed predictions from MHD models are also warranted.
Together, these advances will construct a more complete framework for interpreting current and upcoming observations of disk winds.

\begin{acknowledgements}
We thank Ewine van Dishoeck, Sylvie Cabrit, and Marion Villenave for insightful discussions; Simon Bruderer for providing access to DALI; Nicole Arulanantham for sharing the JWST data of \NameOfTauxxxx{} and for discussions about the source; and Kamber Schwartz for helpful conversations on SY~Cha; Rolf Kuiper and Alexandre Faure for their technical support with the simulations and post-processing. 
We are also grateful to the anonymous referee for providing the comments that have improved the quality of this paper. 
This research has made use of spectroscopic and collisional data from the EMAA database (\url{https://emaa.osug.fr} and \url{https://dx.doi.org/10.17178/EMAA}). 
EMAA is supported by the Observatoire des Sciences de l’Univers de Grenoble (OSUG). 
This work has made use of the Paris-Durham public shock code V1.1, distributed by the CNRS-INSU National Service ``ISM Platform'' at the Paris Observatory Data Center (\url{http://ism.obspm.fr}). 
Based on observations obtained with the NASA/ESA Hubble Space Telescope, retrieved from the Mikulski Archive for Space Telescopes (MAST) at the Space Telescope Science Institute (STScI). STScI is operated by the Association of Universities for Research in Astronomy, Inc. under NASA contract NAS 5-26555.
Numerical computations were carried out on the Cray XC50 at the Center for Computational Astrophysics, National Astronomical Observatory of Japan, and on the Gattaca-2 platform at Jet Propulsion Laboratory.
R.N. and G.R. acknowledge support from the European Union (ERC Starting Grant DiscEvol, project number 101039651) and from Fondazione Cariplo, grant No. 2022-1217. 
A.D.S. acknowledges support from the ERC grant 101019751 MOLDISK. 
Views and opinions expressed are, however, those of the author(s) only and do not necessarily reflect those of the European Union or the European Research Council. Neither the European Union nor the granting authority can be held responsible for them.
\end{acknowledgements}

%
%
\bibliographystyle{aa}
\bibliography{bibsamples}

\begin{thebibliography}{147}
\expandafter\ifx\csname natexlab\endcsname\relax\def\natexlab#1{#1}\fi

\bibitem[{{Agra-Amboage} {et~al.}(2014){Agra-Amboage}, {Cabrit}, {Dougados},
  {Kristensen}, {Ibgui}, \& {Reunanen}}]{2014_Agra-Amboage}
{Agra-Amboage}, V., {Cabrit}, S., {Dougados}, C., {et~al.} 2014, \aap, 564, A11

\bibitem[{{Alexander} {et~al.}(2014){Alexander}, {Pascucci}, {Andrews},
  {Armitage}, \& {Cieza}}]{2014_Alexander}
{Alexander}, R., {Pascucci}, I., {Andrews}, S., {Armitage}, P., \& {Cieza}, L.
  2014, in Protostars and Planets VI, ed. H.~{Beuther}, R.~S. {Klessen}, C.~P.
  {Dullemond}, \& T.~{Henning}, 475

\bibitem[{{Alexander}(2008)}]{2008_Alexander}
{Alexander}, R.~D. 2008, \mnras, 391, L64

\bibitem[{{Anderson} {et~al.}(2024){Anderson}, {Williams}, {Blake},
  {Pontoppidan}, {Salyk}, {Boogert}, {Ross}, \& {Cleeves}}]{2024_Anderson}
{Anderson}, A.~R., {Williams}, J.~P., {Blake}, G.~A., {et~al.} 2024, \apj, 977,
  213

\bibitem[{{Anninos} {et~al.}(1997){Anninos}, {Zhang}, {Abel}, \&
  {Norman}}]{1997_Anninos}
{Anninos}, P., {Zhang}, Y., {Abel}, T., \& {Norman}, M.~L. 1997, \nat, 2, 209

\bibitem[{{Appenzeller} {et~al.}(1984){Appenzeller}, {Oestreicher}, \&
  {Jankovics}}]{1984_Appenzeller}
{Appenzeller}, I., {Oestreicher}, R., \& {Jankovics}, I. 1984, \aap, 141, 108

\bibitem[{{Arulanantham} {et~al.}(2024){Arulanantham}, {McClure},
  {Pontoppidan}, {Beck}, {Sturm}, {Harsono}, {Boogert}, {Cordiner}, {Dartois},
  {Drozdovskaya}, {Espaillat}, {Melnick}, {Noble}, {Palumbo}, {Pendleton},
  {Terada}, \& {van Dishoeck}}]{2024_Arulanantham}
{Arulanantham}, N., {McClure}, M.~K., {Pontoppidan}, K., {et~al.} 2024, \apjl,
  965, L13

\bibitem[{{Bai}(2016)}]{2016_Bai}
{Bai}, X.-N. 2016, \apj, 821, 80

\bibitem[{{Bajaj} {et~al.}(2025){Bajaj}, {Pascucci}, {Beck}, {Edwards},
  {Cabrit}, {Najita}, {Schwarz}, {Semenov}, {Salyk}, {Gorti}, {Brittain},
  {Krijt}, {Ruaud}, \& {Page}}]{2025_Naman}
{Bajaj}, N.~S., {Pascucci}, I., {Beck}, T.~L., {et~al.} 2025, \aj, 169, 296

\bibitem[{{Bajaj} {et~al.}(2024){Bajaj}, {Pascucci}, {Gorti}, {Alexander},
  {Sellek}, {Morrison}, {Gaspar}, {Clarke}, {Xie}, {Ballabio}, \&
  {Deng}}]{2024_Bajaj}
{Bajaj}, N.~S., {Pascucci}, I., {Gorti}, U., {et~al.} 2024, \aj, 167, 127

\bibitem[{{Bakes} \& {Tielens}(1994)}]{1994_BakesTielens}
{Bakes}, E.~L.~O. \& {Tielens}, A.~G.~G.~M. 1994, \apj, 427, 822

\bibitem[{{Ballabio} {et~al.}(2020){Ballabio}, {Alexander}, \&
  {Clarke}}]{2020_Ballabio}
{Ballabio}, G., {Alexander}, R.~D., \& {Clarke}, C.~J. 2020, \mnras, 496, 2932

\bibitem[{{Banzatti} {et~al.}(2022){Banzatti}, {Abernathy}, {Brittain},
  {Bosman}, {Pontoppidan}, {Boogert}, {Jensen}, {Carr}, {Najita}, {Grant},
  {Sigler}, {Sanchez}, {Kern}, \& {Rayner}}]{2022_Banzatti}
{Banzatti}, A., {Abernathy}, K.~M., {Brittain}, S., {et~al.} 2022, \aj, 163,
  174

\bibitem[{{Banzatti} {et~al.}(2019){Banzatti}, {Pascucci}, {Edwards}, {Fang},
  {Gorti}, \& {Flock}}]{2019_Banzatti}
{Banzatti}, A., {Pascucci}, I., {Edwards}, S., {et~al.} 2019, \apj, 870, 76

\bibitem[{{Beck} \& {Bary}(2019)}]{2019_BeckBary}
{Beck}, T.~L. \& {Bary}, J.~S. 2019, \apj, 884, 159

\bibitem[{{Beck} {et~al.}(2008){Beck}, {McGregor}, {Takami}, \&
  {Pyo}}]{2008_Beck}
{Beck}, T.~L., {McGregor}, P.~J., {Takami}, M., \& {Pyo}, T.-S. 2008, \apj,
  676, 472

\bibitem[{{Black}(1987)}]{1987_Black}
{Black}, J.~H. 1987, {Heating and Cooling of the Interstellar Gas}, ed. D.~J.
  {Hollenbach} \& J.~{Thronson}, Harley~A., Vol. 134, 731

\bibitem[{{Black} \& {van Dishoeck}(1987)}]{1987_BlackVanDishoeck}
{Black}, J.~H. \& {van Dishoeck}, E.~F. 1987, \apj, 322, 412

\bibitem[{{Blandford} \& {Payne}(1982)}]{1982_BlandfordPayne}
{Blandford}, R.~D. \& {Payne}, D.~G. 1982, \mnras, 199, 883

\bibitem[{{Brown} {et~al.}(2013){Brown}, {Pontoppidan}, {van Dishoeck},
  {Herczeg}, {Blake}, \& {Smette}}]{2013_Brown}
{Brown}, J.~M., {Pontoppidan}, K.~M., {van Dishoeck}, E.~F., {et~al.} 2013,
  \apj, 770, 94

\bibitem[{{Bruderer}(2013)}]{2013_Bruderer}
{Bruderer}, S. 2013, \aap, 559, A46

\bibitem[{{Bruderer} {et~al.}(2009){Bruderer}, {Doty}, \&
  {Benz}}]{2009_Bruderer}
{Bruderer}, S., {Doty}, S.~D., \& {Benz}, A.~O. 2009, \apjs, 183, 179

\bibitem[{{Bruderer} {et~al.}(2012){Bruderer}, {van Dishoeck}, {Doty}, \&
  {Herczeg}}]{2012_Bruderer}
{Bruderer}, S., {van Dishoeck}, E.~F., {Doty}, S.~D., \& {Herczeg}, G.~J. 2012,
  \aap, 541, A91

\bibitem[{{Cabrit} {et~al.}(1999){Cabrit}, {Ferreira}, \& {Raga}}]{1999_Cabrit}
{Cabrit}, S., {Ferreira}, J., \& {Raga}, A.~C. 1999, \aap, 343, L61

\bibitem[{{Cazaux} \& {Tielens}(2002)}]{2002_CazauxTielens}
{Cazaux}, S. \& {Tielens}, A.~G.~G.~M. 2002, \apjl, 575, L29

\bibitem[{{Cazaux} \& {Tielens}(2004)}]{2004_CazauxTielens}
{Cazaux}, S. \& {Tielens}, A.~G.~G.~M. 2004, \apj, 604, 222

\bibitem[{{Cazaux} \& {Tielens}(2010)}]{2010_CazauxTielensErratum}
{Cazaux}, S. \& {Tielens}, A.~G.~G.~M. 2010, \apj, 715, 698

\bibitem[{{Clarke} {et~al.}(2001){Clarke}, {Gendrin}, \&
  {Sotomayor}}]{2001_Clarke}
{Clarke}, C.~J., {Gendrin}, A., \& {Sotomayor}, M. 2001, \mnras, 328, 485

\bibitem[{{Dalgarno} {et~al.}(1999){Dalgarno}, {Yan}, \& {Liu}}]{1999_Dalgarno}
{Dalgarno}, A., {Yan}, M., \& {Liu}, W. 1999, \apjs, 125, 237

\bibitem[{{Dartois} {et~al.}(2025){Dartois}, {Noble}, {McClure}, {Sturm},
  {Beck}, {Arulanantham}, {Drozdovskaya}, {Espaillat}, {Harsono}, {Palumbo},
  {Pendleton}, \& {Pontoppidan}}]{2025_Dartois}
{Dartois}, E., {Noble}, J.~A., {McClure}, M.~K., {et~al.} 2025, arXiv e-prints,
  arXiv:2503.24309

\bibitem[{{de Valon} {et~al.}(2020){de Valon}, {Dougados}, {Cabrit}, {Louvet},
  {Zapata}, \& {Mardones}}]{2020_deValon}
{de Valon}, A., {Dougados}, C., {Cabrit}, S., {et~al.} 2020, \aap, 634, L12

\bibitem[{{Draine} \& {Bertoldi}(1996)}]{1996_DraineBertoldi}
{Draine}, B.~T. \& {Bertoldi}, F. 1996, \apj, 468, 269

\bibitem[{{Duch{\^e}ne} {et~al.}(2024){Duch{\^e}ne}, {M{\'e}nard},
  {Stapelfeldt}, {Villenave}, {Wolff}, {Perrin}, {Pinte}, {Tazaki}, \&
  {Padgett}}]{2024_Duchene}
{Duch{\^e}ne}, G., {M{\'e}nard}, F., {Stapelfeldt}, K.~R., {et~al.} 2024, \aj,
  167, 77

\bibitem[{{Dullemond} {et~al.}(2012){Dullemond}, {Juhasz}, {Pohl}, {Sereshti},
  {Shetty}, {Peters}, {Commercon}, \& {Flock}}]{2012_Dullemond}
{Dullemond}, C.~P., {Juhasz}, A., {Pohl}, A., {et~al.} 2012, {RADMC-3D: A
  multi-purpose radiative transfer tool}, Astrophysics Source Code Library,
  record ascl:1202.015

\bibitem[{{Edwards} {et~al.}(1987){Edwards}, {Cabrit}, {Strom}, {Heyer},
  {Strom}, \& {Anderson}}]{1987_Edwards}
{Edwards}, S., {Cabrit}, S., {Strom}, S.~E., {et~al.} 1987, \apj, 321, 473

\bibitem[{{Fang} {et~al.}(2018){Fang}, {Pascucci}, {Edwards}, {Gorti},
  {Banzatti}, {Flock}, {Hartigan}, {Herczeg}, \& {Dupree}}]{2018_Fang}
{Fang}, M., {Pascucci}, I., {Edwards}, S., {et~al.} 2018, \apj, 868, 28

\bibitem[{{Fang} {et~al.}(2023){Fang}, {Wang}, {Herczeg}, {Hashimoto}, {Xu},
  {Nemer}, {Pascucci}, {Haffert}, \& {Aoyama}}]{2023_FangNature}
{Fang}, M., {Wang}, L., {Herczeg}, G.~J., {et~al.} 2023, Nature Astronomy, 7,
  905

\bibitem[{{Feiden}(2016)}]{2016_Feiden}
{Feiden}, G.~A. 2016, \aap, 593, A99

\bibitem[{{Flores-Rivera} {et~al.}(2023){Flores-Rivera}, {Flock}, {Kurtovic},
  {Husemann}, {Banzatti}, {Ringqvist}, {Kamann}, {M{\"u}ller}, {Fendt},
  {Garc{\'\i}a Lopez}, {Marleau}, {Henning}, {Carrasco-Gonz{\'a}lez}, {van
  Boekel}, {Keppler}, {Launhardt}, \& {Aoyama}}]{2023_FloresRivera}
{Flores-Rivera}, L., {Flock}, M., {Kurtovic}, N.~T., {et~al.} 2023, \aap, 670,
  A126

\bibitem[{{Flower} {et~al.}(2000){Flower}, {Le Bourlot}, {Pineau des
  For{\^e}ts}, \& {Roueff}}]{2000_Flower}
{Flower}, D.~R., {Le Bourlot}, J., {Pineau des For{\^e}ts}, G., \& {Roueff}, E.
  2000, \mnras, 314, 753

\bibitem[{{Flower} \& {Pineau des For{\^e}ts}(2003)}]{2003_Flower}
{Flower}, D.~R. \& {Pineau des For{\^e}ts}, G. 2003, \mnras, 343, 390

\bibitem[{{Flower} {et~al.}(2021){Flower}, {Pineau des For{\^e}ts},
  {Hily-Blant}, {Faure}, {Lique}, \& {Gonz{\'a}lez-Lezana}}]{2021_Flower}
{Flower}, D.~R., {Pineau des For{\^e}ts}, G., {Hily-Blant}, P., {et~al.} 2021,
  \mnras, 507, 3564

\bibitem[{{Foucher} {et~al.}(2025){Foucher}, {Dutrey}, {Pi{\'e}tu},
  {Guilloteau}, {Chapillon}, {Denis-Alpizar}, {Dartois}, {Di Folco}, {Gavino},
  {Gorti}, {Henning}, {K{\'o}sp{\'a}l}, {Le Petit}, {Majumdar}, {Meshaka},
  {Phuong}, {Ruaud}, {Semenov}, {Tang}, \& {Wolf}}]{2025_Foucher}
{Foucher}, C., {Dutrey}, A., {Pi{\'e}tu}, V., {et~al.} 2025, arXiv e-prints,
  arXiv:2510.04677

\bibitem[{{Gaia Collaboration} {et~al.}(2021){Gaia Collaboration}, {Smart},
  {Sarro}, {Rybizki}, {Reyl{\'e}}, {Robin}, {Hambly}, {Abbas}, {Barstow}, {de
  Bruijne}, {Bucciarelli}, {Carrasco}, {Cooper}, {Hodgkin}, {Masana},
  {Michalik}, {Sahlmann}, {Sozzetti}, {Brown}, {Vallenari}, {Prusti},
  {Babusiaux}, {Biermann}, {Creevey}, {Evans}, {Eyer}, {Hutton}, {Jansen},
  {Jordi}, {Klioner}, {Lammers}, {Lindegren}, {Luri}, {Mignard}, {Panem},
  {Pourbaix}, {Randich}, {Sartoretti}, {Soubiran}, {Walton}, {Arenou},
  {Bailer-Jones}, {Bastian}, {Cropper}, {Drimmel}, {Katz}, {Lattanzi}, {van
  Leeuwen}, {Bakker}, {Casta{\~n}eda}, {De Angeli}, {Ducourant}, {Fabricius},
  {Fouesneau}, {Fr{\'e}mat}, {Guerra}, {Guerrier}, {Guiraud}, {Jean-Antoine
  Piccolo}, {Messineo}, {Mowlavi}, {Nicolas}, {Nienartowicz}, {Pailler},
  {Panuzzo}, {Riclet}, {Roux}, {Seabroke}, {Sordo}, {Tanga}, {Th{\'e}venin},
  {Gracia-Abril}, {Portell}, {Teyssier}, {Altmann}, {Andrae}, {Bellas-Velidis},
  {Benson}, {Berthier}, {Blomme}, {Brugaletta}, {Burgess}, {Busso}, {Carry},
  {Cellino}, {Cheek}, {Clementini}, {Damerdji}, {Davidson}, {Delchambre},
  {Dell'Oro}, {Fern{\'a}ndez-Hern{\'a}ndez}, {Galluccio}, {Garc{\'\i}a-Lario},
  {Garcia-Reinaldos}, {Gonz{\'a}lez-N{\'u}{\~n}ez}, {Gosset}, {Haigron},
  {Halbwachs}, {Harrison}, {Hatzidimitriou}, {Heiter}, {Hern{\'a}ndez},
  {Hestroffer}, {Holl}, {Jan{\ss}en}, {Jevardat de Fombelle}, {Jordan},
  {Krone-Martins}, {Lanzafame}, {L{\"o}ffler}, {Lorca}, {Manteiga}, {Marchal},
  {Marrese}, {Moitinho}, {Mora}, {Muinonen}, {Osborne}, {Pancino}, {Pauwels},
  {Recio-Blanco}, {Richards}, {Riello}, {Rimoldini}, {Roegiers}, {Siopis},
  {Smith}, {Ulla}, {Utrilla}, {van Leeuwen}, {van Reeven}, {Abreu Aramburu},
  {Accart}, {Aerts}, {Aguado}, {Ajaj}, {Altavilla}, {{\'A}lvarez}, {{\'A}lvarez
  Cid-Fuentes}, {Alves}, {Anderson}, {Anglada Varela}, {Antoja}, {Audard},
  {Baines}, {Baker}, {Balaguer-N{\'u}{\~n}ez}, {Balbinot}, {Balog}, {Barache},
  {Barbato}, {Barros}, {Bartolom{\'e}}, {Bassilana}, {Bauchet},
  {Baudesson-Stella}, {Becciani}, {Bellazzini}, {Bernet}, {Bertone}, {Bianchi},
  {Blanco-Cuaresma}, {Boch}, {Bombrun}, {Bossini}, {Bouquillon}, {Bragaglia},
  {Bramante}, {Breedt}, {Bressan}, {Brouillet}, {Burlacu}, {Busonero},
  {Butkevich}, {Buzzi}, {Caffau}, {Cancelliere}, {C{\'a}novas},
  {Cantat-Gaudin}, {Carballo}, {Carlucci}, {Carnerero}, {Casamiquela},
  {Castellani}, {Castro-Ginard}, {Castro Sampol}, {Chaoul}, {Charlot},
  {Chemin}, {Chiavassa}, {Cioni}, {Comoretto}, {Cornez}, {Cowell}, {Crifo},
  {Crosta}, {Crowley}, {Dafonte}, \& {Dapergolas}}]{2021_Gaia}
{Gaia Collaboration}, {Smart}, R.~L., {Sarro}, L.~M., {et~al.} 2021, \aap, 649,
  A6

\bibitem[{{Galli} \& {Palla}(1998)}]{1998_GalliPalla}
{Galli}, D. \& {Palla}, F. 1998, \aap, 335, 403

\bibitem[{{Gangi} {et~al.}(2020){Gangi}, {Nisini}, {Antoniucci}, {Giannini},
  {Biazzo}, {Alcal{\'a}}, {Frasca}, {Munari}, {Arkharov}, {Harutyunyan},
  {Manara}, {Rigliaco}, \& {Vitali}}]{2020_Gangi}
{Gangi}, M., {Nisini}, B., {Antoniucci}, S., {et~al.} 2020, \aap, 643, A32

\bibitem[{{Geers} {et~al.}(2007){Geers}, {van Dishoeck}, {Visser},
  {Pontoppidan}, {Augereau}, {Habart}, \& {Lagrange}}]{2007_Geers}
{Geers}, V.~C., {van Dishoeck}, E.~F., {Visser}, R., {et~al.} 2007, \aap, 476,
  279

\bibitem[{{Godard} {et~al.}(2019){Godard}, {Pineau des For{\^e}ts}, {Lesaffre},
  {Lehmann}, {Gusdorf}, \& {Falgarone}}]{2019_Godard}
{Godard}, B., {Pineau des For{\^e}ts}, G., {Lesaffre}, P., {et~al.} 2019, \aap,
  622, A100

\bibitem[{Gonz^^c3^^a1lez-Lezana {et~al.}(2021)Gonz^^c3^^a1lez-Lezana,
  Hily-Blant, \& Faure}]{2021_Gonzalez-lezana}
Gonz^^c3^^a1lez-Lezana, T., Hily-Blant, P., \& Faure, A. 2021, The Journal of
  Chemical Physics, 154, 054310

\bibitem[{{Gorti} {et~al.}(2009){Gorti}, {Dullemond}, \&
  {Hollenbach}}]{2009_Gorti}
{Gorti}, U., {Dullemond}, C.~P., \& {Hollenbach}, D. 2009, \apj, 705, 1237

\bibitem[{{Gorti} \& {Hollenbach}(2004)}]{2004_Gorti}
{Gorti}, U. \& {Hollenbach}, D. 2004, \apj, 613, 424

\bibitem[{{Gredel} \& {Dalgarno}(1995)}]{1995_GredelDalgarno}
{Gredel}, R. \& {Dalgarno}, A. 1995, \apj, 446, 852

\bibitem[{{Gregory} {et~al.}(2016){Gregory}, {Adams}, \&
  {Davies}}]{2016_Gregory}
{Gregory}, S.~G., {Adams}, F.~C., \& {Davies}, C.~L. 2016, \mnras, 457, 3836

\bibitem[{{Gressel} {et~al.}(2020){Gressel}, {Ramsey}, {Brinch}, {Nelson},
  {Turner}, \& {Bruderer}}]{2020_Gressel}
{Gressel}, O., {Ramsey}, J.~P., {Brinch}, C., {et~al.} 2020, arXiv e-prints,
  arXiv:2005.03431

\bibitem[{{Gressel} {et~al.}(2015){Gressel}, {Turner}, {Nelson}, \&
  {McNally}}]{2015_Gressel}
{Gressel}, O., {Turner}, N.~J., {Nelson}, R.~P., \& {McNally}, C.~P. 2015,
  \apj, 801, 84

\bibitem[{{G{\"u}del} {et~al.}(2007){G{\"u}del}, {Briggs}, {Arzner}, {Audard},
  {Bouvier}, {Feigelson}, {Franciosini}, {Glauser}, {Grosso}, {Micela},
  {Monin}, {Montmerle}, {Padgett}, {Palla}, {Pillitteri}, {Rebull}, {Scelsi},
  {Silva}, {Skinner}, {Stelzer}, \& {Telleschi}}]{2007_Gudel}
{G{\"u}del}, M., {Briggs}, K.~R., {Arzner}, K., {et~al.} 2007, \aap, 468, 353

\bibitem[{{G{\"u}del} {et~al.}(2018){G{\"u}del}, {Eibensteiner}, {Dionatos},
  {Audard}, {Forbrich}, {Kraus}, {Rab}, {Schneider}, {Skinner}, \&
  {Vorobyov}}]{2018_Gudel}
{G{\"u}del}, M., {Eibensteiner}, C., {Dionatos}, O., {et~al.} 2018, \aap, 620,
  L1

\bibitem[{{G{\"u}del} {et~al.}(2010){G{\"u}del}, {Lahuis}, {Briggs}, {Carr},
  {Glassgold}, {Henning}, {Najita}, {van Boekel}, \& {van
  Dishoeck}}]{2010_Gudel}
{G{\"u}del}, M., {Lahuis}, F., {Briggs}, K.~R., {et~al.} 2010, \aap, 519, A113

\bibitem[{{Habart} {et~al.}(2004){Habart}, {Boulanger}, {Verstraete},
  {Walmsley}, \& {Pineau des For{\^e}ts}}]{2004_Habart}
{Habart}, E., {Boulanger}, F., {Verstraete}, L., {Walmsley}, C.~M., \& {Pineau
  des For{\^e}ts}, G. 2004, \aap, 414, 531

\bibitem[{{Hartigan} {et~al.}(1995){Hartigan}, {Edwards}, \&
  {Ghandour}}]{1995_Hartigan}
{Hartigan}, P., {Edwards}, S., \& {Ghandour}, L. 1995, \apj, 452, 736

\bibitem[{{Heays} {et~al.}(2017){Heays}, {Bosman}, \& {van
  Dishoeck}}]{2017_Heays}
{Heays}, A.~N., {Bosman}, A.~D., \& {van Dishoeck}, E.~F. 2017, \aap, 602, A105

\bibitem[{{Hollenbach} {et~al.}(1994){Hollenbach}, {Johnstone}, {Lizano}, \&
  {Shu}}]{1994_Hollenbach}
{Hollenbach}, D., {Johnstone}, D., {Lizano}, S., \& {Shu}, F. 1994, \apj, 428,
  654

\bibitem[{{Hollenbach} \& {McKee}(1979)}]{1979_HollenbachMcKee}
{Hollenbach}, D. \& {McKee}, C.~F. 1979, \apjs, 41, 555

\bibitem[{{Hollenbach} \& {McKee}(1989)}]{1989_HollenbachMcKee}
{Hollenbach}, D. \& {McKee}, C.~F. 1989, \apj, 342, 306

\bibitem[{{Hu} {et~al.}(2025){Hu}, {Bae}, {Zhu}, \& {Wang}}]{2025_Hu}
{Hu}, X., {Bae}, J., {Zhu}, Z., \& {Wang}, L. 2025, \apj, 986, 161

\bibitem[{{Jankovics} {et~al.}(1983){Jankovics}, {Appenzeller}, \&
  {Krautter}}]{1983_Jankovics}
{Jankovics}, I., {Appenzeller}, I., \& {Krautter}, J. 1983, \pasp, 95, 883

\bibitem[{{Jonkheid} {et~al.}(2004){Jonkheid}, {Faas}, {van Zadelhoff}, \& {van
  Dishoeck}}]{2004_Jonkheid}
{Jonkheid}, B., {Faas}, F.~G.~A., {van Zadelhoff}, G.~J., \& {van Dishoeck},
  E.~F. 2004, \aap, 428, 511

\bibitem[{{Jonkheid} {et~al.}(2006){Jonkheid}, {Kamp}, {Augereau}, \& {van
  Dishoeck}}]{2006_Jonkheid}
{Jonkheid}, B., {Kamp}, I., {Augereau}, J.~C., \& {van Dishoeck}, E.~F. 2006,
  \aap, 453, 163

\bibitem[{{Kalscheur} {et~al.}(2025){Kalscheur}, {France}, {Nisini},
  {Schneider}, {Alexander}, {Eisl{\"o}ffel}, {Campbell-White}, {Shang},
  {Gangi}, {Guo}, \& {Chang}}]{2025_Kalscheur}
{Kalscheur}, M., {France}, K., {Nisini}, B., {et~al.} 2025, \aj, 169, 240

\bibitem[{{Komaki} {et~al.}(2021){Komaki}, {Nakatani}, \&
  {Yoshida}}]{2021_Komaki}
{Komaki}, A., {Nakatani}, R., \& {Yoshida}, N. 2021, \apj, 910, 51

\bibitem[{{Kuiper} {et~al.}(2010){Kuiper}, {Klahr}, {Dullemond}, {Kley}, \&
  {Henning}}]{2010_Kuiper}
{Kuiper}, R., {Klahr}, H., {Dullemond}, C., {Kley}, W., \& {Henning}, T. 2010,
  \aap, 511, A81

\bibitem[{{Kuiper} \& {Klessen}(2013)}]{2013_Kuiper}
{Kuiper}, R. \& {Klessen}, R.~S. 2013, \aap, 555, A7

\bibitem[{{Kuiper} {et~al.}(2020){Kuiper}, {Yorke}, \& {Mignone}}]{2020_Kuiper}
{Kuiper}, R., {Yorke}, H.~W., \& {Mignone}, A. 2020, \apjs, 250, 13

\bibitem[{{Kunitomo} {et~al.}(2020){Kunitomo}, {Suzuki}, \&
  {Inutsuka}}]{2020_Kunitomo}
{Kunitomo}, M., {Suzuki}, T.~K., \& {Inutsuka}, S.-i. 2020, \mnras, 492, 3849

\bibitem[{{Kwan} \& {Tademaru}(1995)}]{1995_KwanTademasu}
{Kwan}, J. \& {Tademaru}, E. 1995, \apj, 454, 382

\bibitem[{{Launhardt} {et~al.}(2023){Launhardt}, {Pavlyuchenkov}, {Akimkin},
  {Dutrey}, {Gueth}, {Guilloteau}, {Henning}, {Pi{\'e}tu}, {Schreyer},
  {Semenov}, {Stecklum}, \& {Bourke}}]{2023_Launhardt}
{Launhardt}, R., {Pavlyuchenkov}, Y.~N., {Akimkin}, V.~V., {et~al.} 2023, \aap,
  678, A135

\bibitem[{{Law} {et~al.}(2023){Law}, {E. Morrison}, {Argyriou}, {Patapis},
  {{\'A}lvarez-M{\'a}rquez}, {Labiano}, \& {Vandenbussche}}]{2023_Law}
{Law}, D.~R., {E. Morrison}, J., {Argyriou}, I., {et~al.} 2023, \aj, 166, 45

\bibitem[{{Lee} {et~al.}(1996){Lee}, {Bettens}, \& {Herbst}}]{1996_Lee}
{Lee}, H.-H., {Bettens}, R.~P.~A., \& {Herbst}, E. 1996, \aaps, 119, 111

\bibitem[{{Lesur} {et~al.}(2023){Lesur}, {Flock}, {Ercolano}, {Lin}, {Yang},
  {Barranco}, {Benitez-Llambay}, {Goodman}, {Johansen}, {Klahr}, {Laibe},
  {Lyra}, {Marcus}, {Nelson}, {Squire}, {Simon}, {Turner}, {Umurhan}, \&
  {Youdin}}]{2023_Lesur}
{Lesur}, G., {Flock}, M., {Ercolano}, B., {et~al.} 2023, in Astronomical
  Society of the Pacific Conference Series, Vol. 534, Protostars and Planets
  VII, ed. S.~{Inutsuka}, Y.~{Aikawa}, T.~{Muto}, K.~{Tomida}, \& M.~{Tamura},
  465

\bibitem[{{Liffman}(2003)}]{2003_Liffman}
{Liffman}, K. 2003, PASA, 20, 337

\bibitem[{{Lique}(2015)}]{2015_Lique}
{Lique}, F. 2015, \mnras, 453, 810

\bibitem[{{Louvet} {et~al.}(2018){Louvet}, {Dougados}, {Cabrit}, {Mardones},
  {M{\'e}nard}, {Tabone}, {Pinte}, \& {Dent}}]{2018_Louvet}
{Louvet}, F., {Dougados}, C., {Cabrit}, S., {et~al.} 2018, \aap, 618, A120

\bibitem[{{Luhman} {et~al.}(2009){Luhman}, {Mamajek}, {Allen}, \&
  {Cruz}}]{2009_Luhman}
{Luhman}, K.~L., {Mamajek}, E.~E., {Allen}, P.~R., \& {Cruz}, K.~L. 2009, \apj,
  703, 399

\bibitem[{{Maloney} {et~al.}(1996){Maloney}, {Hollenbach}, \&
  {Tielens}}]{1996_Maloney}
{Maloney}, P.~R., {Hollenbach}, D.~J., \& {Tielens}, A.~G.~G.~M. 1996, \apj,
  466, 561

\bibitem[{{Manara} {et~al.}(2023){Manara}, {Ansdell}, {Rosotti}, {Hughes},
  {Armitage}, {Lodato}, \& {Williams}}]{2023_Manara}
{Manara}, C.~F., {Ansdell}, M., {Rosotti}, G.~P., {et~al.} 2023, in
  Astronomical Society of the Pacific Conference Series, Vol. 534, Protostars
  and Planets VII, ed. S.~{Inutsuka}, Y.~{Aikawa}, T.~{Muto}, K.~{Tomida}, \&
  M.~{Tamura}, 539

\bibitem[{{Manara} {et~al.}(2016){Manara}, {Fedele}, {Herczeg}, \&
  {Teixeira}}]{2016_Manara}
{Manara}, C.~F., {Fedele}, D., {Herczeg}, G.~J., \& {Teixeira}, P.~S. 2016,
  \aap, 585, A136

\bibitem[{{McElroy} {et~al.}(2013){McElroy}, {Walsh}, {Markwick}, {Cordiner},
  {Smith}, \& {Millar}}]{2012_UMIST}
{McElroy}, D., {Walsh}, C., {Markwick}, A.~J., {et~al.} 2013, \aap, 550, A36

\bibitem[{{McGinnis} {et~al.}(2018){McGinnis}, {Dougados}, {Alencar},
  {Bouvier}, \& {Cabrit}}]{2018_Mcginnis}
{McGinnis}, P., {Dougados}, C., {Alencar}, S.~H.~P., {Bouvier}, J., \&
  {Cabrit}, S. 2018, \aap, 620, A87

\bibitem[{{Melnikov} {et~al.}(2023){Melnikov}, {Boley}, {Nikonova}, {Caratti o
  Garatti}, {Garcia Lopez}, {Stecklum}, {Eisl{\"o}ffel}, \&
  {Weigelt}}]{2023_Melnikov}
{Melnikov}, S., {Boley}, P.~A., {Nikonova}, N.~S., {et~al.} 2023, \aap, 673,
  A156

\bibitem[{{Mignone} {et~al.}(2007){Mignone}, {Bodo}, {Massaglia}, {Matsakos},
  {Tesileanu}, {Zanni}, \& {Ferrari}}]{2007_Mignone}
{Mignone}, A., {Bodo}, G., {Massaglia}, S., {et~al.} 2007, \apjs, 170, 228

\bibitem[{{Nakatani} {et~al.}(2018{\natexlab{a}}){Nakatani}, {Hosokawa},
  {Yoshida}, {Nomura}, \& {Kuiper}}]{2018_Nakatani}
{Nakatani}, R., {Hosokawa}, T., {Yoshida}, N., {Nomura}, H., \& {Kuiper}, R.
  2018{\natexlab{a}}, \apj, 857, 57

\bibitem[{{Nakatani} {et~al.}(2018{\natexlab{b}}){Nakatani}, {Hosokawa},
  {Yoshida}, {Nomura}, \& {Kuiper}}]{2018_Nakatanib}
{Nakatani}, R., {Hosokawa}, T., {Yoshida}, N., {Nomura}, H., \& {Kuiper}, R.
  2018{\natexlab{b}}, \apj, 865, 75

\bibitem[{{Nakatani} {et~al.}(2021){Nakatani}, {Kobayashi}, {Kuiper}, {Nomura},
  \& {Aikawa}}]{2021_Nakatani}
{Nakatani}, R., {Kobayashi}, H., {Kuiper}, R., {Nomura}, H., \& {Aikawa}, Y.
  2021, \apj, 915, 90

\bibitem[{{Nakatani} {et~al.}(2024){Nakatani}, {Turner}, \&
  {Takasao}}]{2024_Nakatani}
{Nakatani}, R., {Turner}, N.~J., \& {Takasao}, S. 2024, \apj, 974, 281

\bibitem[{{Nomura} {et~al.}(2007){Nomura}, {Aikawa}, {Tsujimoto}, {Nakagawa},
  \& {Millar}}]{2007_Nomura_II}
{Nomura}, H., {Aikawa}, Y., {Tsujimoto}, M., {Nakagawa}, Y., \& {Millar}, T.~J.
  2007, \apj, 661, 334

\bibitem[{{Nomura} \& {Millar}(2005)}]{2005_NomuraMillar}
{Nomura}, H. \& {Millar}, T.~J. 2005, \aap, 438, 923

\bibitem[{{Oliveira} {et~al.}(2010){Oliveira}, {Pontoppidan}, {Mer{\'{\i}}n},
  {van Dishoeck}, {Lahuis}, {Geers}, {J{\o}rgensen}, {Olofsson}, {Augereau}, \&
  {Brown}}]{2010_Oliveira}
{Oliveira}, I., {Pontoppidan}, K.~M., {Mer{\'{\i}}n}, B., {et~al.} 2010, \apj,
  714, 778

\bibitem[{{Omukai}(2000)}]{2000_Omukai}
{Omukai}, K. 2000, \apj, 534, 809

\bibitem[{{Omukai} {et~al.}(2010){Omukai}, {Hosokawa}, \&
  {Yoshida}}]{2010_Omukai}
{Omukai}, K., {Hosokawa}, T., \& {Yoshida}, N. 2010, \apj, 722, 1793

\bibitem[{{Orihara} {et~al.}(2023){Orihara}, {Momose}, {Muto}, {Hashimoto},
  {Liu}, {Tsukagoshi}, {Kudo}, {Takahashi}, {Yang}, {Hasegawa}, {Dong},
  {Konishi}, \& {Akiyama}}]{2023_Orihara}
{Orihara}, R., {Momose}, M., {Muto}, T., {et~al.} 2023, \pasj, 75, 424

\bibitem[{{Osterbrock}(1989)}]{1989_Osterbrockbook}
{Osterbrock}, D.~E. 1989, {Astrophysics of gaseous nebulae and active galactic
  nuclei} (Mill Valley, CA: Univ. Science Books)

\bibitem[{{Panoglou} {et~al.}(2012){Panoglou}, {Cabrit}, {Pineau Des
  For{\^e}ts}, {Garcia}, {Ferreira}, \& {Casse}}]{2012_Panoglou}
{Panoglou}, D., {Cabrit}, S., {Pineau Des For{\^e}ts}, G., {et~al.} 2012, \aap,
  538, A2

\bibitem[{{Pascucci} {et~al.}(2020){Pascucci}, {Banzatti}, {Gorti}, {Fang},
  {Pontoppidan}, {Alexander}, {Ballabio}, {Edwards}, {Salyk}, {Sacco},
  {Flaccomio}, {Blake}, {Carmona}, {Hall}, {Kamp}, {K{\"a}ufl}, {Meeus},
  {Meyer}, {Pauly}, {Steendam}, \& {Sterzik}}]{2020_Pascucci}
{Pascucci}, I., {Banzatti}, A., {Gorti}, U., {et~al.} 2020, \apj, 903, 78

\bibitem[{{Pascucci} {et~al.}(2025){Pascucci}, {Beck}, {Cabrit}, {Bajaj},
  {Edwards}, {Louvet}, {Najita}, {Skinner}, {Gorti}, {Salyk}, {Brittain},
  {Krijt}, {Muzerolle Page}, {Ruaud}, {Schwarz}, {Semenov}, {Duch{\^e}ne}, \&
  {Villenave}}]{2025_Pascucci}
{Pascucci}, I., {Beck}, T.~L., {Cabrit}, S., {et~al.} 2025, Nature Astronomy,
  9, 81

\bibitem[{{Pascucci} {et~al.}(2023){Pascucci}, {Cabrit}, {Edwards}, {Gorti},
  {Gressel}, \& {Suzuki}}]{2022_Pascucci}
{Pascucci}, I., {Cabrit}, S., {Edwards}, S., {et~al.} 2023, in Astronomical
  Society of the Pacific Conference Series, Vol. 534, Protostars and Planets
  VII, ed. S.~{Inutsuka}, Y.~{Aikawa}, T.~{Muto}, K.~{Tomida}, \& M.~{Tamura},
  567

\bibitem[{{Pascucci} \& {Sterzik}(2009)}]{2009_PascucciSterzik}
{Pascucci}, I. \& {Sterzik}, M. 2009, \apj, 702, 724

\bibitem[{{Pascucci} {et~al.}(2011){Pascucci}, {Sterzik}, {Alexander},
  {Alencar}, {Gorti}, {Hollenbach}, {Owen}, {Ercolano}, \&
  {Edwards}}]{2011_Pascucci}
{Pascucci}, I., {Sterzik}, M., {Alexander}, R.~D., {et~al.} 2011, \apj, 736, 13

\bibitem[{{Pelletier} \& {Pudritz}(1992)}]{1992_PelletierPudritz}
{Pelletier}, G. \& {Pudritz}, R.~E. 1992, \apj, 394, 117

\bibitem[{{Perrin} {et~al.}(2014){Perrin}, {Sivaramakrishnan}, {Lajoie},
  {Elliott}, {Pueyo}, {Ravindranath}, \& {Albert}}]{2014_Perrin}
{Perrin}, M.~D., {Sivaramakrishnan}, A., {Lajoie}, C.-P., {et~al.} 2014, in
  Society of Photo-Optical Instrumentation Engineers (SPIE) Conference Series,
  Vol. 9143, Space Telescopes and Instrumentation 2014: Optical, Infrared, and
  Millimeter Wave, ed. J.~M. {Oschmann}, Jr., M.~{Clampin}, G.~G. {Fazio}, \&
  H.~A. {MacEwen}, 91433X

\bibitem[{{Pittman} {et~al.}(2025){Pittman}, {Espaillat}, {Robinson},
  {Thanathibodee}, {Lopez}, {Calvet}, {Zhu}, {Walter}, {Wendeborn}, {Manara},
  {Campbell-White}, {Claes}, {Fang}, {Frasca}, {Gameiro}, {Gangi},
  {Hern{\'a}ndez}, {K{\'o}sp{\'a}l}, {Mauc{\'o}}, {Muzerolle}, {Siwak},
  {Tychoniec}, \& {Venuti}}]{2025_Pittman}
{Pittman}, C.~V., {Espaillat}, C.~C., {Robinson}, C.~E., {et~al.} 2025, arXiv
  e-prints, arXiv:2507.01162

\bibitem[{{Pontoppidan} {et~al.}(2011){Pontoppidan}, {Blake}, \&
  {Smette}}]{2011_Pontoppidan}
{Pontoppidan}, K.~M., {Blake}, G.~A., \& {Smette}, A. 2011, \apj, 733, 84

\bibitem[{{Rab} {et~al.}(2022){Rab}, {Weber}, {Grassi}, {Ercolano}, {Picogna},
  {Caselli}, {Thi}, {Kamp}, \& {Woitke}}]{2022_Rab}
{Rab}, C., {Weber}, M., {Grassi}, T., {et~al.} 2022, \aap, 668, A154

\bibitem[{{Rigliaco} {et~al.}(2013){Rigliaco}, {Pascucci}, {Gorti}, {Edwards},
  \& {Hollenbach}}]{2013_Rigliaco}
{Rigliaco}, E., {Pascucci}, I., {Gorti}, U., {Edwards}, S., \& {Hollenbach}, D.
  2013, \apj, 772, 60

\bibitem[{{R{\"o}llig} {et~al.}(2007){R{\"o}llig}, {Abel}, {Bell}, {Bensch},
  {Black}, {Ferland}, {Jonkheid}, {Kamp}, {Kaufman}, {Le Bourlot}, {Le Petit},
  {Meijerink}, {Morata}, {Ossenkopf}, {Roueff}, {Shaw}, {Spaans}, {Sternberg},
  {Stutzki}, {Thi}, {van Dishoeck}, {van Hoof}, {Viti}, \&
  {Wolfire}}]{2007_Rollig}
{R{\"o}llig}, M., {Abel}, N.~P., {Bell}, T., {et~al.} 2007, \aap, 467, 187

\bibitem[{{Roman-Duval} {et~al.}(2025){Roman-Duval}, {Fischer}, {Fullerton},
  {Taylor}, {Plesha}, {Proffitt}, {Monroe}, {Fischer}, {Aloisi}, {Bouret},
  {Britt}, {Calvet}, {Carlberg}, {Crowther}, {De Rosa}, {Dixon}, {Espaillat},
  {Evans}, {Fox}, {France}, {Garcia}, {Fleming}, {Frazer}, {G{\'o}mez de
  Castro}, {Herczeg}, {Hernandez}, {Hirschauer}, {James}, {Johns-Krull},
  {Leitherer}, {Lockwood}, {Najita}, {Oey}, {Oliveira}, {Pauly}, {Reid},
  {Riedel}, {Rodriguez}, {Sahnow}, {Sankrit}, {Sembach}, {Shaw}, {Smith},
  {Sohn}, {Som}, {{\'U}beda}, \& {Welty}}]{2025_Ullyses}
{Roman-Duval}, J., {Fischer}, W.~J., {Fullerton}, A.~W., {et~al.} 2025, \apj,
  985, 109

\bibitem[{{Roueff} {et~al.}(2019){Roueff}, {Abgrall}, {Czachorowski},
  {Pachucki}, {Puchalski}, \& {Komasa}}]{2019_Roueff}
{Roueff}, E., {Abgrall}, H., {Czachorowski}, P., {et~al.} 2019, \aap, 630, A58

\bibitem[{{Santoro} \& {Shull}(2006)}]{2006_SantoroShull}
{Santoro}, F. \& {Shull}, J.~M. 2006, \apj, 643, 26

\bibitem[{{Schneider} {et~al.}(2013){Schneider}, {Eisl{\"o}ffel}, {G{\"u}del},
  {G{\"u}nther}, {Herczeg}, {Robrade}, \& {Schmitt}}]{2013_Schneider}
{Schneider}, P.~C., {Eisl{\"o}ffel}, J., {G{\"u}del}, M., {et~al.} 2013, \aap,
  557, A110

\bibitem[{{Schwarz} {et~al.}(2025{\natexlab{a}}){Schwarz}, {Samland},
  {Olofsson}, {Henning}, {Sellek}, {G{\"u}del}, {Tabone}, {Kamp}, {Lagage},
  {van Dishoeck}, {Caratti o Garatti}, {Glauser}, {Ray}, {Arabhavi},
  {Christiaens}, {Franceschi}, {Gasman}, {Grant}, {Kanwar}, {Kaeufer},
  {Kurtovic}, {Perotti}, {Temmink}, \& {Vlasblom}}]{2025_SchwarzErratum}
{Schwarz}, K.~R., {Samland}, M., {Olofsson}, G., {et~al.} 2025{\natexlab{a}},
  \apj, 991, 232

\bibitem[{{Schwarz} {et~al.}(2025{\natexlab{b}}){Schwarz}, {Samland},
  {Olofsson}, {Henning}, {Sellek}, {G{\"u}del}, {Tabone}, {Kamp}, {Lagage},
  {van Dishoeck}, {Caratti o Garatti}, {Glauser}, {Ray}, {Arabhavi},
  {Christiaens}, {Franceschi}, {Gasman}, {Grant}, {Kanwar}, {Kaeufer},
  {Kurtovic}, {Perotti}, {Temmink}, \& {Vlasblom}}]{2025_Schwarz}
{Schwarz}, K.~R., {Samland}, M., {Olofsson}, G., {et~al.} 2025{\natexlab{b}},
  \apj, 980, 148

\bibitem[{{Sellek} {et~al.}(2024{\natexlab{a}}){Sellek}, {Bajaj}, {Pascucci},
  {Clarke}, {Alexander}, {Xie}, {Ballabio}, {Deng}, {Gorti}, {Gaspar}, \&
  {Morrison}}]{2024_SellekBajaj}
{Sellek}, A.~D., {Bajaj}, N.~S., {Pascucci}, I., {et~al.} 2024{\natexlab{a}},
  \aj, 167, 223

\bibitem[{{Sellek} {et~al.}(2024{\natexlab{b}}){Sellek}, {Grassi}, {Picogna},
  {Rab}, {Clarke}, \& {Ercolano}}]{2024_Sellek}
{Sellek}, A.~D., {Grassi}, T., {Picogna}, G., {et~al.} 2024{\natexlab{b}},
  \aap, 690, A296

\bibitem[{{Shoda} {et~al.}(2025){Shoda}, {Nakatani}, \& {Takasao}}]{2025_Shoda}
{Shoda}, M., {Nakatani}, R., \& {Takasao}, S. 2025, \aap, 696, L4

\bibitem[{{Shu} {et~al.}(1993){Shu}, {Johnstone}, \& {Hollenbach}}]{1993_Shu_b}
{Shu}, F.~H., {Johnstone}, D., \& {Hollenbach}, D. 1993, \icarus, 106, 92

\bibitem[{{Shull} \& {van Steenberg}(1985)}]{1985_ShullSteenberg}
{Shull}, J.~M. \& {van Steenberg}, M.~E. 1985, \apj, 298, 268

\bibitem[{{Siess} {et~al.}(2000){Siess}, {Dufour}, \& {Forestini}}]{2000_Siess}
{Siess}, L., {Dufour}, E., \& {Forestini}, M. 2000, \aap, 358, 593

\bibitem[{{Simon} {et~al.}(2019){Simon}, {Guilloteau}, {Beck}, {Chapillon}, {Di
  Folco}, {Dutrey}, {Feiden}, {Grosso}, {Pi{\'e}tu}, {Prato}, \&
  {Schaefer}}]{2019_Simon}
{Simon}, M., {Guilloteau}, S., {Beck}, T.~L., {et~al.} 2019, \apj, 884, 42

\bibitem[{{Simon} {et~al.}(2016){Simon}, {Pascucci}, {Edwards}, {Feng},
  {Gorti}, {Hollenbach}, {Rigliaco}, \& {Keane}}]{2016_Simon}
{Simon}, M.~N., {Pascucci}, I., {Edwards}, S., {et~al.} 2016, \apj, 831, 169

\bibitem[{{Spitzer}(1978)}]{1978_Spitzer}
{Spitzer}, L. 1978, {Physical processes in the interstellar medium} (New York:
  Wiley-Interscience)

\bibitem[{{Stapelfeldt} {et~al.}(2014){Stapelfeldt}, {Duch{\^e}ne}, {Perrin},
  {Wolff}, {Krist}, {Padgett}, {M{\'e}nard}, \& {Pinte}}]{2014_Stapelfeldt}
{Stapelfeldt}, K.~R., {Duch{\^e}ne}, G., {Perrin}, M., {et~al.} 2014, in IAU
  Symposium, Vol. 299, Exploring the Formation and Evolution of Planetary
  Systems, ed. M.~{Booth}, B.~C. {Matthews}, \& J.~R. {Graham}, 99--103

\bibitem[{{Suzuki} \& {Inutsuka}(2009)}]{2009_SuzukiInutsuka}
{Suzuki}, T.~K. \& {Inutsuka}, S.-i. 2009, \apjl, 691, L49

\bibitem[{{Suzuki} {et~al.}(2016){Suzuki}, {Ogihara}, {Morbidelli}, {Crida}, \&
  {Guillot}}]{2016_Suzuki}
{Suzuki}, T.~K., {Ogihara}, M., {Morbidelli}, A., {Crida}, A., \& {Guillot}, T.
  2016, \aap, 596, A74

\bibitem[{{Tabone} {et~al.}(2022){Tabone}, {Rosotti}, {Lodato}, {Armitage},
  {Cridland}, \& {van Dishoeck}}]{2022_Tabone}
{Tabone}, B., {Rosotti}, G.~P., {Lodato}, G., {et~al.} 2022, \mnras, 512, L74

\bibitem[{{Takasao} {et~al.}(2022){Takasao}, {Tomida}, {Iwasaki}, \&
  {Suzuki}}]{2022_Takasao}
{Takasao}, S., {Tomida}, K., {Iwasaki}, K., \& {Suzuki}, T.~K. 2022, \apj, 941,
  73

\bibitem[{{Tazaki} {et~al.}(2025){Tazaki}, {M{\'e}nard}, {Duch{\^e}ne},
  {Villenave}, {Ribas}, {Stapelfeldt}, {Perrin}, {Pinte}, {Wolff}, {Padgett},
  {Ma}, {Martinien}, \& {Roumesy}}]{2025_Tazaki}
{Tazaki}, R., {M{\'e}nard}, F., {Duch{\^e}ne}, G., {et~al.} 2025, \apj, 980, 49

\bibitem[{{Tielens} \& {Hollenbach}(1985)}]{1985_TielensHollenbach}
{Tielens}, A.~G.~G.~M. \& {Hollenbach}, D. 1985, \apj, 291, 722

\bibitem[{{Vicente} {et~al.}(2013){Vicente}, {Bern{\'e}}, {Tielens},
  {Hu{\'e}lamo}, {Pantin}, {Kamp}, \& {Carmona}}]{2013_Vicente}
{Vicente}, S., {Bern{\'e}}, O., {Tielens}, A.~G.~G.~M., {et~al.} 2013, \apjl,
  765, L38

\bibitem[{{Villenave} {et~al.}(2020){Villenave}, {M{\'e}nard}, {Dent},
  {Duch{\^e}ne}, {Stapelfeldt}, {Benisty}, {Boehler}, {van der Plas}, {Pinte},
  {Telkamp}, {Wolff}, {Flores}, {Lesur}, {Louvet}, {Riols}, {Dougados},
  {Williams}, \& {Padgett}}]{2020_Marion}
{Villenave}, M., {M{\'e}nard}, F., {Dent}, W.~R.~F., {et~al.} 2020, \aap, 642,
  A164

\bibitem[{{Wang} {et~al.}(2019){Wang}, {Bai}, \& {Goodman}}]{2019_Wang}
{Wang}, L., {Bai}, X.-N., \& {Goodman}, J. 2019, \apj, 874, 90

\bibitem[{{Wang} \& {Goodman}(2017)}]{2017_Wang}
{Wang}, L. \& {Goodman}, J. 2017, \apj, 847, 11

\bibitem[{{Weber} {et~al.}(2020){Weber}, {Ercolano}, {Picogna}, {Hartmann}, \&
  {Rodenkirch}}]{2020_Weber}
{Weber}, M.~L., {Ercolano}, B., {Picogna}, G., {Hartmann}, L., \& {Rodenkirch},
  P.~J. 2020, \mnras, 496, 223

\bibitem[{{Weder} {et~al.}(2023){Weder}, {Mordasini}, \&
  {Emsenhuber}}]{2023_Weder}
{Weder}, J., {Mordasini}, C., \& {Emsenhuber}, A. 2023, arXiv e-prints,
  arXiv:2304.12380

\bibitem[{{Whelan} {et~al.}(2021){Whelan}, {Pascucci}, {Gorti}, {Edwards},
  {Alexander}, {Sterzik}, \& {Melo}}]{2021_Whelan}
{Whelan}, E.~T., {Pascucci}, I., {Gorti}, U., {et~al.} 2021, \apj, 913, 43

\bibitem[{{Wilms} {et~al.}(2000){Wilms}, {Allen}, \& {McCray}}]{2000_Wilms}
{Wilms}, J., {Allen}, A., \& {McCray}, R. 2000, \apj, 542, 914

\bibitem[{{Woitke} {et~al.}(2019){Woitke}, {Kamp}, {Antonellini}, {Anthonioz},
  {Baldovin-Saveedra}, {Carmona}, {Dionatos}, {Dominik}, {Greaves},
  {G{\"u}del}, {Ilee}, {Liebhardt}, {Menard}, {Min}, {Pinte}, {Rab}, {Rigon},
  {Thi}, {Thureau}, \& {Waters}}]{2019_Woitke}
{Woitke}, P., {Kamp}, I., {Antonellini}, S., {et~al.} 2019, \pasp, 131, 064301

\bibitem[{{Woitke} {et~al.}(2009){Woitke}, {Kamp}, \& {Thi}}]{2009_Woitke}
{Woitke}, P., {Kamp}, I., \& {Thi}, W.~F. 2009, \aap, 501, 383

\bibitem[{{Xu} \& {McCray}(1991)}]{1991_XuMcCray}
{Xu}, Y. \& {McCray}, R. 1991, \apj, 375, 190

\end{thebibliography}

\begin{appendix}
\section{Details of Simulation setup} \label{sec:detailed_sim_setup}

\subsection{Initial Disk Configuration} \label{sec:hydro:initial_condition}
The system is assumed to be axisymmetric and midplane symmetric, surrounding a central star with a mass of $M_*$. 
The disk is initially in hydrostatic equilibrium with a vertically isothermal temperature profile, and thus the density distribution is 
\begin{equation}
    \nh = n_{\rm m}(R) 
    \exp \left[ -\frac{z^2}{2H^2} \right] , \label{eq:inidenstr}
\end{equation}
where $(R,z)$ are the radial distance and the vertical height in cylindrical coordinates, respectively;
$\nh$ is the number density of hydrogen nuclei;
$n_{\rm m} (R)$ is that at the midplane ($z= 0$);
and $H$ is the pressure scale height defined as the ratio of the local isothermal sound speed $\cs$ to the orbital frequency $\Omega \equiv \sqrt{GM_*/R^3}$. 
Integrating \eqnref{eq:inidenstr} w.r.t. $z$ gives the corresponding surface density profile,
\begin{equation}
        		\Sigma (R) =  \sqrt{2\pi} H \rho_{\rm m} (R) , 
\end{equation}
	where $\rho_{\rm m}$ is mass density on the midplane 
	and relates to $n_{\rm m}$ in terms of mean gas mass per hydrogen nucleus, $m$, 
	as $\rho_{\rm m} = m  n_{\rm m}$. 
    The initial radial profiles of $\Sigma$ is
	\begin{equation}
		\Sigma = \Sigma_0 \braket{\frac{R}{R_0}}^{-1} 
		\label{eq:sigma}	
	\end{equation}
	with $R_0$ and $\Sigma_0$ being a scale radius and reference surface density at $R_0$.
    The product of $\Sigma_0$ and $R_0$ is set by the total disk mass
	\begin{equation}
		M_{\rm disk} = \int_{R_{\rm min} }^{R_{\rm max}} 2\pi R \Sigma \, \dd R
		= 2\pi  (R_{\rm max} - R_{\rm min}) R_0\Sigma_0. 
		\label{eq:diskmass}
	\end{equation}
	with an inner and outer truncated radii, $R_{\rm min}$ and $R_{\rm max}$, respectively.
    We use $R_{\rm min} \approx 1\au$ and $R_{\rm max} = 40 R_{\rm g} \approx 350\au$ in our fiducial model. 

    The initial temperature is assumed to be $T = 100\Kelvin (R/1\au)^{-0.5}$ but is immediately updated from the initial value according to the local thermochemical processes after the run starts. 
    The density structure is also adjusted accordingly and reaches a steady state on a timescale of several thousand years. 
    The results are thus insensitive to the initial condition.


\subsection{Hydrodynamics}     \label{sec:hydro:equations}

    The dynamical evolution is followed by the publicly available hydrodynamics simulation code, PLUTO \citep{2007_Mignone}. 
    The code is suitably modified by implementing a variety of physics such as UV and X-ray photoheating, photochemical reactions, and multispecies chemical network \citep{2018_Nakatani, 2018_Nakatanib, 2021_Nakatani}. 
    
    The simulation is performed in 2D spherical polar coordinates $(r, \theta)$.
    We follow the time evolution of gas density $\rho$,
	velocities $\vec{v} = (v_r,~ v_\theta, ~ v_\phi)$,
	total gas energy density,
	and chemical abundances $\{y_i | i = \ce{H, H^+, H2, ...}\}$.
    The basic equations are
    \begin{gather}
                \frac{\partial \rho}{\partial t} + \nabla \cdot \rho \vec{v}                                             =  0 ,                  \\
                \frac{\partial \rho v_r}{\partial t} + \nabla \cdot \left( \rho v_r \vec{v} \right)              =  -\frac{\partial p}{\partial r}
                                -\rho \frac{GM_*}{r^2} + \rho \frac{v_\theta^2 + v_\phi^2}{r}                   ,                       \\
                \frac{\partial \rho v_\theta}{\partial t} + \nabla \cdot \left( \rho v_\theta \vec{v} \right)    = - \frac{1}{r}\frac{\partial p}{\partial \theta }
                                - \rho \frac{v_\theta v_r}{r} + \frac{\rho v_\phi^2}{r} \cot \theta                     ,                       \\
                \frac{\partial \rho v_\phi}{\partial t} + \nabla ^l \cdot \left( \rho v_\phi \vec{v} \right)     = 0     ,       \label{eq:euler_phi}            \\              
                \begin{split}
                &\frac{\partial }{\partial t}\left(\frac{1}{2} \rho v^2 + \frac{p}{\gamma-1}\right) + \nabla \cdot \left[\left(\frac{1}{2} \rho v^2 + \frac{\gamma p}{\gamma-1}  \right)\vec{v}\right]                                     
                \\ 
                &\quad\quad\quad= - \rho v_r \frac{ GM_* }{r^2} +\rho \left( \Gamma -\Lambda    \right),   
                \end{split} \label{eq:energy_eq}
                 \\
                \text{and} \quad
                \frac{\partial \nh y_i }{\partial t} + \nabla \cdot \left( \nh y_i \vec{v} \right)               = \nh R_i       .                       \label{eq:chemevoeq}
    \end{gather}
    Here $p$ denotes the gas pressure; $\gamma$ is specific heat ratio;
    $\Gamma$ is the total heating rate per unit mass (specific heating rate),
    and $\Lambda$ is the total cooling rate per unit mass (specific cooling rate);
    and $R_i$ is the total reaction rate for the corresponding chemical species $i$.
    \eqnref{eq:euler_phi} is written in the angular momentum conserving form.  
    For more detail regarding hydrodynamics setup, we refer the readers to \cite{2018_Nakatani, 2018_Nakatanib}.

\section{Chemistry}  \label{sec:detailed_description_chemistry}
The chemical reactions included in our numerical simulations are listed in \tref{tab:chem_reac_list}. 
They are selected to reproduce the PDR benchmark results of \citet{2007_Rollig} (Appendix~\ref{sec:benchmark:roellig}) while keeping the hydrodynamics simulations computationally feasible, and have also been validated through comparisons with DALI (Appendix~\ref{sec:benchmark:dali}).

Reaction rates for FUV-driven photochemistry follow \citet{2012_UMIST} and are expressed as $k = k_0 \chi \exp[-C A_{\rm V}]$,
where $\chi$ is the local FUV flux normalized to the Draine field ($F_{\rm D} \approx 2.8\e{-3}\ergs\cm^{-2}$),
$k_0$ is the unattenuated reaction rate coefficient at $\chi = 1$, $C$ is a numerical constant, and $A_{\rm V}$ is the visual extinction.
Self-shielding is included for \ce{H2} and CO photodissociation and \ion{C}{I} photoionization \citep{1996_DraineBertoldi, 1996_Lee, 1985_TielensHollenbach}.
We compute $A_{\rm V}$ using
$A_{\rm V} = 5.32\e{-22} \, \mathrm{mag\,cm^2} \times N_{\rm H} \, (\dgratio / 0.01)$,
where $N_{\rm H}$ is the hydrogen nucleus column density and $\dgratio$ is the small-dust-to-gas mass ratio (see \secref{sec:methods:hydro}).

The unattenuated rate coefficient $k_0$ is calculated by integrating the product of cross-section $\sigma_\nu$ and the stellar flux $F_{*,\nu}$:
\[
k_0 =   \braket{\frac{F_{*, \rm FUV}}{F_{\rm D}}}^{-1} \int {\rm d}\nu \, \sigma_\nu \frac{F_{*, \nu} }{h\nu}
\]
where $F_{*, \rm FUV}$ is the total stellar flux in the FUV band. 
The cross-section data are taken from the Leiden database \citep{2017_Heays}, and the integration is carried out over all frequencies up to the Lyman limit ($h\nu \approx 13.6\eV$).

\subsection{Benchmark: Plane-Parallel PDR}  \label{sec:benchmark:roellig}
We benchmark our chemical network against the results from various PDR codes presented in \citet{2007_Rollig}.
We perform 1D plane-parallel PDR calculations using the parameters of the V1--V4 models of \citet{2007_Rollig} and compare the resulting temperature and chemical abundance profiles with the published benchmarks. 
Here, we adopt a grain-catalyzed \ce{H2} formation rate coefficient of $3\e{-18}\sqrt{T/1\Kelvin}\cm^3\unit{s}^{-1}$, for consistency with the benchmarks.

\begin{figure*}
    \centering
    \includegraphics[width=\linewidth]{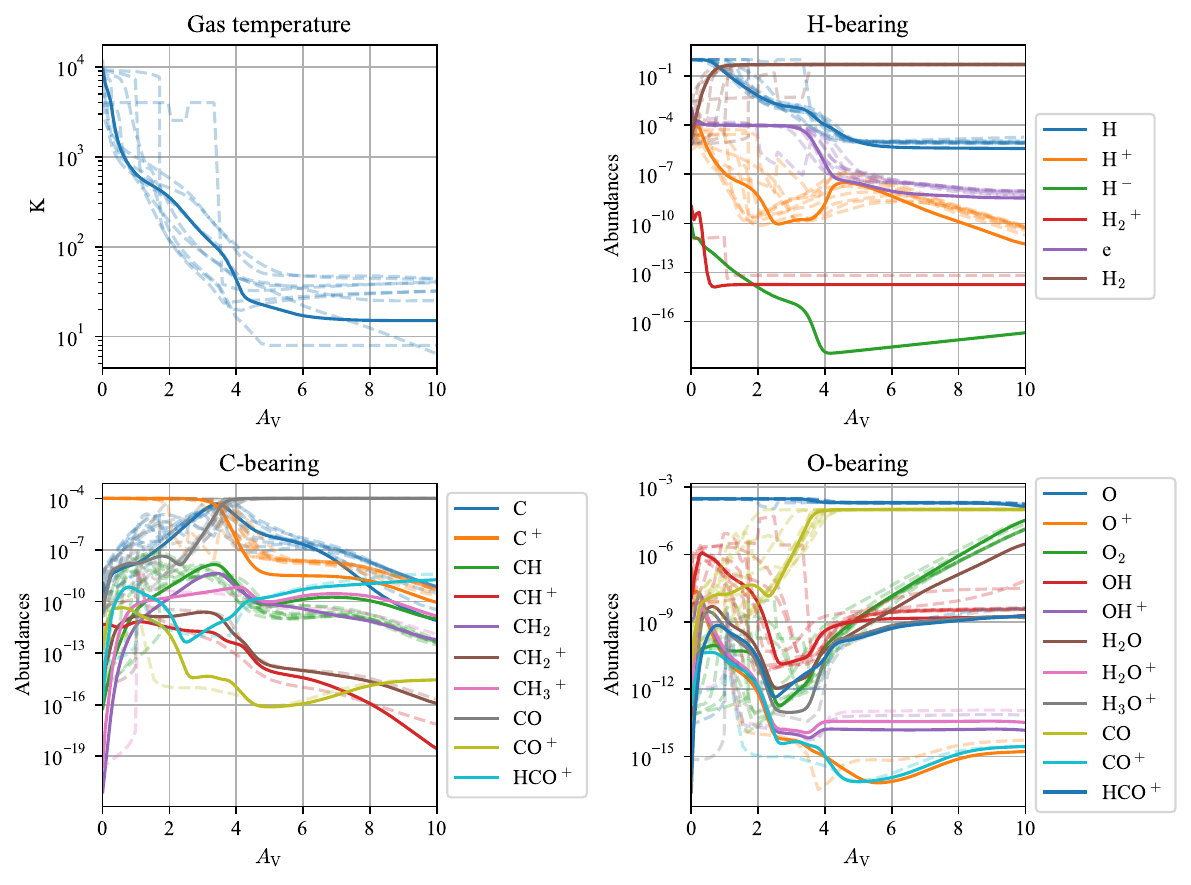}
    \caption{
    Comparison of temperature and chemical abundances between our thermochemical model (thick solid lines) and those from various PDR codes (thin dashed lines) from \citet{2007_Rollig}.
    }
    \label{fig:benchmark:v4}
\end{figure*}
\fref{fig:benchmark:v4} shows the comparison for the V4 model (high density and strong UV field).
Our reduced chemical network reproduces the benchmark results with reasonable accuracy. 
We have confirmed a similar level of qualitative agreements for the V1--V3 models, particularly for $A_{\rm V} \lesssim 1$, the regime most relevant to our study. 
In the low-density (V1 and V2) models, however, our network overestimates CO and underestimates \ion{C}{I} near the CO photodissociation front, an area for potential improvement in future updates.

\subsection{Benchmark: Protoplanetary Disk} \label{sec:benchmark:dali}
We also benchmark our chemical network by comparing 2D thermochemical structures with those generated by the DALI code \citep{2009_Bruderer, 2012_Bruderer, 2013_Bruderer}.
For consistency, we adopt a PAH abundance set to 10\% of the ISM value throughout the comparisons.
To align with DALI’s setup, we disable EUV and adopt its default X-ray spectrum over $10^3\eV \leq E \leq 10^5\eV$: $L_{\rm X}(E)\propto \exp(-E/kT_{\rm X})$ with the X-ray plasma temperature of $T_{\rm X} = 7\e{7}\Kelvin$ and a total luminosity of $10^{30}\ergs$. 
Other stellar parameters are taken from \tref{tab:fiducialmodel}.

\begin{figure*}
    \centering
    \includegraphics[width=\linewidth]{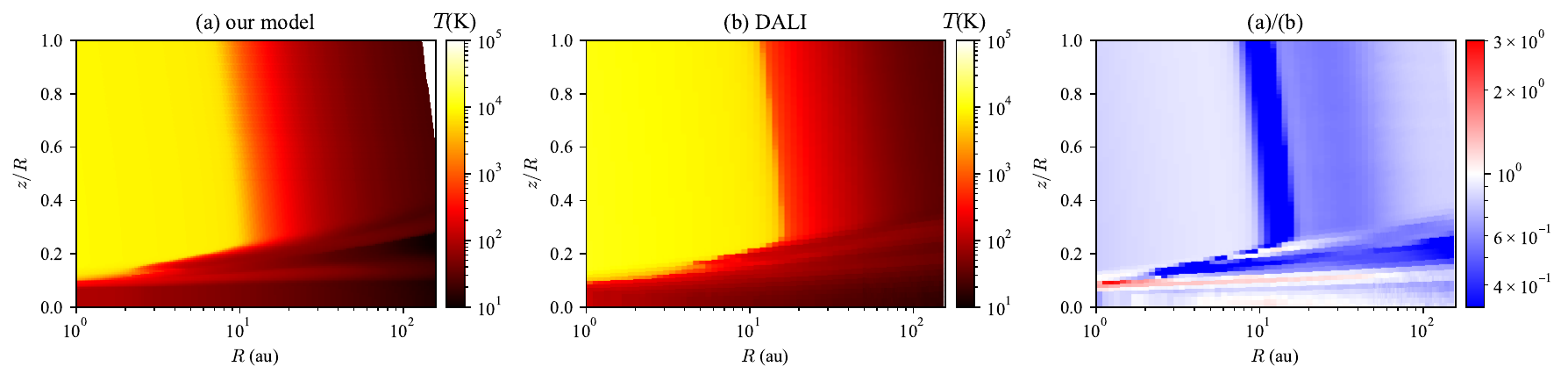}
    \includegraphics[width=\linewidth]{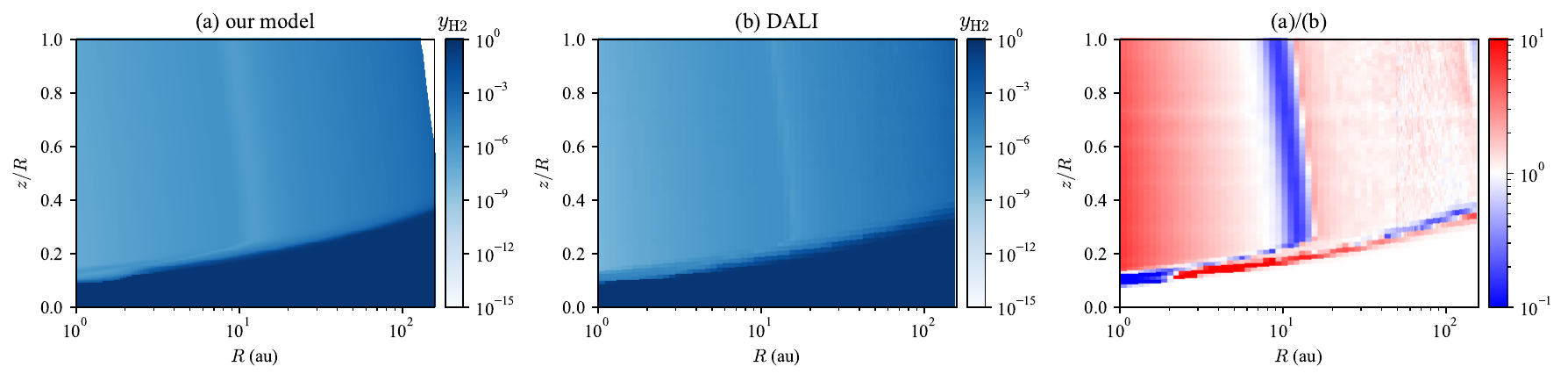}
    \caption{
    Comparison of gas temperature (top row) and \ce{H2} abundance (bottom row) between our model (left) and DALI (middle).
    Right panels show the ratio of our model to DALI.
    }
    \label{fig:benchmark:dali}
\end{figure*}

\fref{fig:benchmark:dali} shows the temperature and \ce{H2} abundance distributions for a case where the density is fixed to the initial condition of our simulation.
The right panels show our reduced network reproduces the temperature and \ce{H2} abundance structures of DALI reasonably well, despite notable differences in chemical networks and computational methods.
The high-temperature region (yellowish) is slightly more extended in DALI (by $\sim 5\au$), due to stronger X-ray heating in the low-column-density region, likely stemming from its use of a more detailed efficiency prescription from \citet{1999_Dalgarno}, compared to our simplified treatment (see \secref{sec:discussions:caveat}).
Overall, the agreement supports the validity of our thermochemical model.

We have performed the same comparisons while varying $\dotMacc$ (i.e., $L_{\rm FUV}$) and bolometric luminosity, and have found a similar level of qualitative agreement. 
However, the match in the \ce{H2} abundance deteriorates when PAH-catalyzed \ce{H2} formation is included.
This reaction enhances \ce{H2} abundances in hotter, \ion{H}{I}-dominated region at larger radii \citep{2012_Bruderer,2013_Bruderer}. 

Including this reaction potentially increases high-$J$ emission while retaining lower-$J$ fluxes, which could reinforce our interpretation that the observed \ce{H2} winds are of photoevaporative origin. 
Nevertheless, the overall impact remains uncertain, as the formation rate is sensitive to PAH abundances, which are poorly constrained in general. 
Still, given the detection of PAHs in \NameOfTauxxxx{} \citep{2025_Dartois}, it would be worthwhile to explore this effect on the flux and morphology in future work.

Additionally, DALI treats \ce{H2} pumping heating in much greater detail than the simplified approach used in the current model \citep{1979_HollenbachMcKee}, leading to differences in the temperature structure at higher accretion rates ($\dotMacc \sim 10^{-8}\Msun\yr^{-1}$). 
While this does not affect our main conclusions in the present study, it highlights an area for future improvement.

\begin{table*}
\caption{\label{tab:chem_reac_list}List of Chemical Reactions Included in Hydrodynamics Simulations}
\centering
\small
\begin{tabular}{cr | c r | cr}
\hline\hline
Reaction & Reference & Reaction & Reference & Reaction & Reference\\ 
\hline
 \ce{ H    +  e-       -> H+   +   e-  + e- }  & 1          &  \ce{ CO   +  H2+      -> CO+  +   H2 }        & 2    &  \ce{ O2   +{$\gamma$} -> O    +   O }         & 2\\     
 \ce{ H+   +  e-       -> H    +   \gamma}     & 1          &  \ce{ CO   +  H2+      -> HCO+ +   H }         & 2    &  \ce{ O+   +  H        -> O    +   H+ }        & 2\\     
 \ce{ H    +  e-       -> H-   +   \gamma}     & 1          &  \ce{ CO   +  H3+      -> HCO+ +   H2 }        & 2    &  \ce{ O+   +  H2       -> OH+  +   H }         & 2\\     
 \ce{ H-   +  H        -> H2   +   e- }        & 1          &  \ce{ CO   +  CRP      -> CO+  +   e- }        & 2    &  \ce{ O+   +  e-       -> O    +   \gamma}     & 2\\    
 \ce{ H    +  H+       -> H2+  +   \gamma}     & 1          &  \ce{ CO   +  CRPHOT   -> C    +   O }         & 2    &  \ce{ OH   +  H+       -> OH+  +   H }         & 2\\     
 \ce{ H2+  +  H        -> H2   +   H+ }        & 1          &  \ce{ CO   +{$\gamma$} -> C    +   O}          & 2, 3 &  \ce{ OH   +{$\gamma$} -> O    +   H }         & 2\\    
 \ce{ H2   +  H+       -> H    +   H2+ }       & 1          &  \ce{ CO+  +  H        -> CO   +   H+ }        & 2    &  \ce{ OH+  +  H2       -> H2O+ +   H }         & 2\\   
 \ce{ H2   +  e-       -> H    +   H + e- }    & 1          &  \ce{ CO+  +  H2       -> HCO+ +   H }         & 2    &  \ce{ OH+  +  e-       -> O    +   H}          & 2\\    
 \ce{ H2   +  H        -> H    +   H + H }     & 1          &  \ce{ CO+  +  C        -> CO   +   C+ }        & 2    &  \ce{ OH+  +{$\gamma$} -> O+   +   H }         & 2\\     
 \ce{ H-   +  e-       -> H    +   e- + e- }   & 1          &  \ce{ CO+  +  O        -> CO   +   O+ }        & 2    &  \ce{ H2O  +  H+       -> H2O+ +   H }         & 2\\     
 \ce{ H-   +  H+       -> H    +   H }         & 1          &  \ce{ CO+  +  e-       -> O    +   C }         & 2    &  \ce{ H2O  +  C+       -> HCO+ +   H }         & 2\\     
 \ce{ H-   +  H+       -> H2+  +   e- }        & 1          &  \ce{ CO+  +{$\gamma$} -> C+   +   O }         & 2    &  \ce{ H2O  +{$\gamma$} -> OH   +   H }         & 2\\    
 \ce{ H2+  +  e-       -> H    +   H }         & 1          &  \ce{ CH   +  H        -> C    +   H2 }        & 2    &  \ce{ H2O+ +  H2       -> H3O+ +   H }         & 2\\    
 \ce{ H2+  +  H-       -> H2   +   H }         & 1          &  \ce{ CH   +  H+       -> CH+  +   H }         & 2    &  \ce{ H2O+ +  e-       -> OH   +   H }         & 2\\     
 \ce{ H    +  H   +  H -> H2   +   H }         & 1          &  \ce{ CH   +  C+       -> CH+  +   C }         & 2    &  \ce{ H2O+ +  e-       -> O    +   H2 }        & 2\\   
 \ce{ H    +  H + H2   -> H2   +   H2 }        & 1          &  \ce{ CH   +  O        -> CO   +   H }         & 2    &  \ce{ H2O+ +  e-       -> O    +   H + H }     & 2\\    
 \ce{ H2   +  H2       -> H2   +   H + H }     & 1          &  \ce{ CH   +  O        -> HCO+ +   e- }        & 2    &  \ce{ H3O+ +  e-       -> O    +   H2 + H }    & 2\\    
 \ce{ H    +  H        -> H+   +   e- + H }    & 1          &  \ce{ CH   +{$\gamma$} -> CH+  +   e- }        & 2    &  \ce{ H3O+ +  e-       -> OH   +   H2 }        & 2\\   
 \ce{ H    +  H        ->[dust] H2 }           & 1          &  \ce{ CH   +{$\gamma$} -> C    +   H }         & 2    &  \ce{ H3O+ +  e-       -> OH   +   H + H }     & 2\\    
 \ce{ H    +  CRP      -> H+   +   e- }        & 2          &  \ce{ CH+  +  H        -> C+   +   H2 }        & 2    &  \ce{ H3O+ +  e-       -> H2O  +   H }         & 2\\   
 \ce{ H    +  CRPHOT   -> H+   +   e- }        & 2          &  \ce{ CH+  +  H2       -> CH2+ +   H }         & 2\\   
 \ce{ H    + {$\gamma_{\rm EUV}$}    -> H+   +   e-}   &  3 &  \ce{ CH+  +  O        -> CO+  +   H }         & 2\\ 
 \ce{ H    + {$\gamma_{\rm X}$}      -> H+   +   e-}   &  4 &  \ce{ CH+  +  e-       -> C    +   H }         & 2\\    
 \ce{ H2   +  CRP      -> H+   +   H- }        & 2          &  \ce{ CH+  +{$\gamma$} -> C    +   H+ }        & 2\\   
 \ce{ H2   +  CRP      -> H+   +   H  + e- }   & 2          &  \ce{ CH2  +  H        -> CH   +   H2 }        & 2\\   
 \ce{ H2   +  CRP      -> e-   +   H2+ }       & 2          &  \ce{ CH2  +  H+       -> CH+  +   H2 }        & 2\\   
 \ce{ H2   +  CRP      -> H    +   H }         & 2          &  \ce{ CH2  +  H+       -> CH2+ +   H }         & 2\\   
 \ce{ H2   + {$\gamma$} -> H    +   H}          & 2, 3      &  \ce{ CH2  +  H2+      -> CH2+ +   H2 }        & 2\\   
 \ce{ H2   + {$\gamma_{\rm EUV}$}    -> H2+  +   e-}   &  3 &  \ce{ CH2  +  C+       -> CH2+ +   C }         & 2\\  
 \ce{ H2   + {$\gamma_{\rm X}$}      -> H2+  +   e-}   &  4 &  \ce{ CH2  +  O        -> CO   +   H + H }     & 2\\ 
 \ce{ H-   + {$\gamma$} -> H    +   e- }       & 2          &  \ce{ CH2  +  O        -> CO   +   H2 }        & 2\\   
 \ce{ H2+  +  H2       -> H3+  +   H }         & 2          &  \ce{ CH2  + CRPHOT    -> CH   +   H }         & 2\\   
 \ce{ H3+  +  e-       -> H2   +   H }         & 2          &  \ce{ CH2  + CRPHOT    -> CH2+ +   e- }        & 2\\   
 \ce{ H3+  +  e-       -> H    +   H + H }     & 2          &  \ce{ CH2  +{$\gamma$} -> CH   +   H }         & 2\\      
 \ce{ He   +  CRP      -> He+  +   e- }        & 2          &  \ce{ CH2  +{$\gamma$} -> CH2+ +   e- }        & 2\\  
 \ce{ He   +  CRPHOT   -> He+  +   e- }        & 2          &  \ce{ CH2+ +  H2       -> CH3+ +   H }         & 2\\    
 \ce{ He+  +  H        -> He   +   H+ }        & 2          &  \ce{ CH2+ +  e-       -> CH   +   H }         & 2\\  
 \ce{ He+  +  H2       -> He   +   H2+ }       & 2          &  \ce{ CH2+ +  e-       -> C    +   H2 }        & 2\\  
 \ce{ He+  +  H2       -> He   +   H+ + H }    & 2          &  \ce{ CH2+ +  e-       -> C    +   H + H }     & 2\\  
 \ce{ He+  +  e-       -> He   +   \gamma}     & 2          &  \ce{ CH2+ +{$\gamma$} -> CH   +   H+ }        & 2\\  
 \ce{ He+  +  C        -> C+   +   He }        & 2          &  \ce{ CH2+ +{$\gamma$} -> CH+  +   H }         & 2\\   
 \ce{ He+  +  CO       -> O    +   C+ + He }   & 2          &  \ce{ CH2+ +{$\gamma$} -> C+   +   H2 }        & 2\\   
 \ce{ C    +  H        -> CH   +   \gamma }    & 2          &  \ce{ CH3+ +  e-       -> CH2  +   H }         & 2\\   
 \ce{ C    +  H2       -> CH2  +   \gamma}     & 2          &  \ce{ CH3+ +  e-       -> CH   +   H2 }        & 2\\    
 \ce{ C    +  O        -> CO   +   \gamma }    & 2          &  \ce{ CH3+ +  e-       -> CH   +   H + H }     & 2\\  
 \ce{ C    +  O+       -> CO+  +   \gamma}     & 2          &  \ce{ CH3+ +{$\gamma$} -> CH+  +   H2 }        & 2\\  
 \ce{ C    +  O2       -> CO   +   O }         & 2          &  \ce{ CH3+ +{$\gamma$} -> CH2+ +   H }         & 2\\  
 \ce{ C    +  OH       -> CO   +   H }         & 2          &  \ce{ HCO+ +  e-       -> CO   +   H }         & 2\\    
 \ce{ C    +  CRP      -> C+   +   e- }        & 2          &  \ce{ HCO+ +{$\gamma$} -> CO+  +   H }         & 2\\  
 \ce{ C    +  CRPHOT   -> C+   +   e- }        & 2          &  \ce{ O    +  H        -> OH   +   \gamma}     & 2\\   
 \ce{ C    +{$\gamma$} -> C+   +   e-}         & 2, 5       &  \ce{ O    +  H+       -> O+   +   H }         & 2\\  
 \ce{ C+   +  H        -> CH+  +   \gamma}     & 2          &  \ce{ O    +  H2       -> OH   +   H }         & 2\\   
 \ce{ C+   +  H2       -> CH2+ +   \gamma}     & 2          &  \ce{ O    +  H3+      -> OH+  +   H2 }        & 2\\  
 \ce{ C+   +  O        -> CO+  +   \gamma}     & 2          &  \ce{ O    +  H3+      -> H2O+ +   H }         & 2\\       
 \ce{ C+   +  O2       -> CO   +   O+ }        & 2          &  \ce{ O    +  O         -> O2   +   \gamma}    & 2\\   
 \ce{ C+   +  O2       -> CO+  +   O }         & 2          &  \ce{ O    +  OH       -> O2   +   H }         & 2\\   
 \ce{ C+   +  OH       -> CO   +   H+ }        & 2          &  \ce{ O    + CRP       -> O+   +   e- }        & 2\\  
 \ce{ C+   +  OH       -> CO+  +   H }         & 2          &  \ce{ O    + CRPHOT    -> O+   +   e- }        & 2\\    
 \ce{ C+   +  e-       -> C    +   \gamma }    & 2          &  \ce{ O2   + CRPHOT    -> O    +   O }         & 2\\   
\hline
\end{tabular}
\tablefoot
{References: (1) \citet{2000_Omukai} (2) UMIST database \citep{2012_UMIST} (3) \cite{2018_Nakatani} (4) \cite{2018_Nakatanib} (5) \citet{2021_Nakatani}.  \\
On the left-hand side of reactions, $\gamma$ indicates photons with energies below the Lyman limit unless specified by subscripts. For X-ray photoionization of H and \ce{H2}, secondary ionization is also incorporated, though not listed here.
}
\end{table*}

\section{Streamlines} \label{sec:streamlines}
Here, we present the details of our streamline analysis, following \secref{sec:results:wind_properties}. 

\subsection{Definition of the Base} \label{sec:streamlines:base}
A steady-state photoevaporative wind originates in a subsonic region and gradually accelerates to supersonic velocities.  
The flow is typically launched from a region with steep gradients in density, temperature, and heating rate, beginning with a very small---but nonzero---Mach number, as required by mass conservation ($\nabla\cdot\rho\vec{v} = 0$). 
Identifying the precise wind-launching point, or clearly distinguishing the wind from the underlying disk, is generally nontrivial. 
While one could define the wind base at the sonic surface or where the Bernoulli parameter ($E_{\rm tot}$ defined in Appendix~\ref{sec:streamlines:physical_properties}) becomes positive, steady-state winds often appear to originate deeper, below these surfaces. 

In this study, we define the wind base as the point where the Mach number reaches $\mathcal{M}_{\rm base} = 0.025$. 
This value is somewhat arbitrary but chosen to ensure the wind exhibits (quasi-)steady features beyond the base ($s \geq 0$, where $s$ is the coordinate along the streamline), such as monotonically increasing velocity and Mach number, nearly conserved specific angular momentum, and consistent energy balance. 
As expected, the total heating rate $\Gamma$ exceeds the total cooling rate $\Lambda$ for $s \geq 0$; otherwise, the energy input would be insufficient to sustain a wind. 

\begin{figure*}[htbp]
    \centering
    \includegraphics[width=\linewidth, clip]{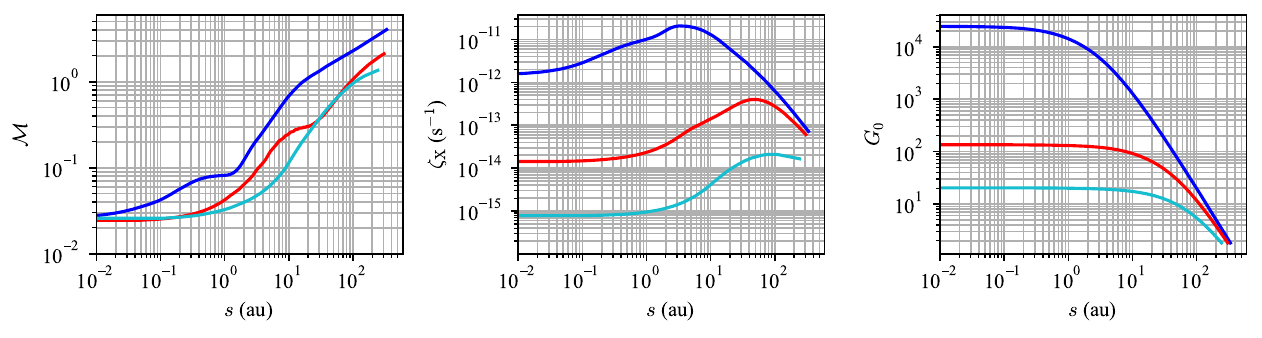}
    \caption{
    Same as \fref{fig:streamlines} but for Mach number $\mathcal{M}$, X-ray ionization rate coefficient, and $G_0$ along the three streamlines.
    }
    \label{fig:streamlines_others}
\end{figure*}

For streamlines launched at larger radii (e.g., the red and cyan streamlines in \fref{fig:streamlines}), these features are also observed for $-3\au \lesssim s \lesssim 0$ (i.e., where $\mathcal{M}<0.025$), but shifting the origin to $s = -3\au$ does not significantly affect the profiles shown in Figures~\ref{fig:streamlines}, \ref{fig:streamlines_heatcool}, \ref{fig:streamlines_others}, and \ref{fig:streamlines_heatcool_detailed}.

\subsection{Physical Properties}    \label{sec:streamlines:physical_properties}
\begin{figure*}[htbp]
    \centering
    \includegraphics[width=\linewidth, clip]{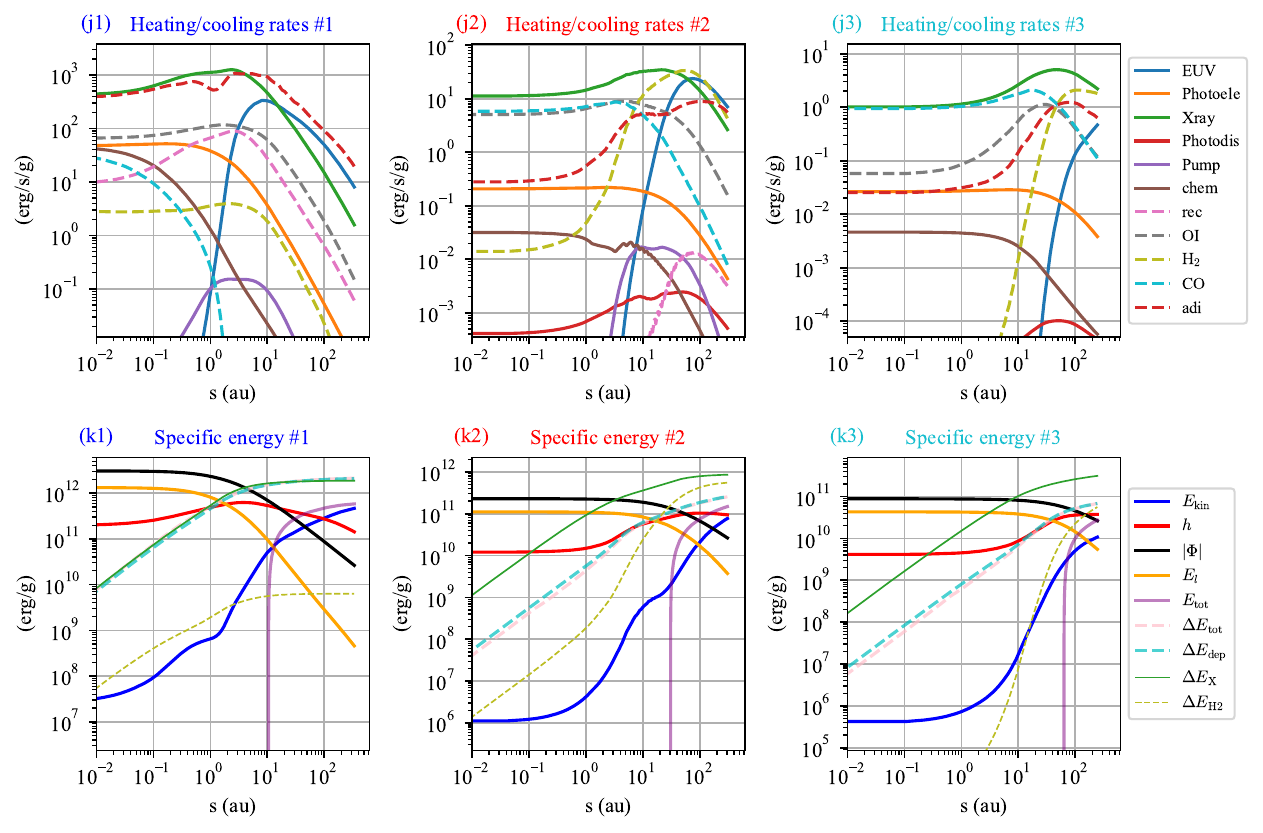}
    \caption{
    (top) Specific rates of major heating and cooling processes along the three streamlines in \fref{fig:streamlines}, shown from left to right for the blue, red, and cyan streamlines. Labels indicate: EUV photoionization heating (``EUV''); FUV grain photoelectric heating (``Photoele''); X-ray photoionization heating (``X-ray''); \ce{H2} photodissociation heating (``Photodis''); \ce{H2} pumping followed by collisional deexcitation (``Pump''); chemical heating from dust-catalyzed \ce{H2} formation (``chem''); radiative recombination cooling (``rec''); line cooling from \ion{O}{I}, \ce{H2}, and CO; and adiabatic cooling (``adi'').
    (bottom) Specific energy components along the same streamlines. Blue, red, black, and orange lines show the poloidal kinetic energy, enthalpy, gravitational potential (absolute value), and centrifugal potential, respectively. The purple line indicates total mechanical energy $E_{\rm tot}$, which varies from negative to positive. Dashed pink line shows the total energy increment, $\Delta E_{\rm tot} = E_{\rm tot}(s)-E_{\rm tot}(0)$, which closely overlaps the dashed cyan line showing the total net deposited energy from heating and cooling. Thin green and dashed yellow lines represent energy input from X-ray heating and losses via \ce{H2} cooling. On the left panel, the green, cyan, and pink lines nearly coincide.
    }
    \label{fig:streamlines_heatcool_detailed}
\end{figure*}
We provide additional details on the heating, cooling, and energetics along the representative streamlines in \fref{fig:streamlines}. 
The top panels of \fref{fig:streamlines_heatcool_detailed} reproduce \fref{fig:streamlines_heatcool} but include selected minor heating and cooling processes for reference.

The bottom panels show specific kinetic energy, enthalpy, gravitational potential, and centrifugal potential:
\[
\begin{gathered}
    E_{\rm kin} = \frac{1}{2}v_{\rm p}^2, \quad h = \frac{\gamma}{\gamma-1}c_{\rm s}^2
    \quad
    \Phi = -\frac{GM_*}{r}, \quad E_l = \frac{1}{2}v_\phi^2. 
\end{gathered}
\]
We also plot the total mechanical energy $E_{\rm tot} = E_{\rm kin} + h + \Phi + E_l$, which varies from negative to positive, along with the energy increment, $\Delta E_{\rm tot} = E_{\rm tot}(s)-E_{\rm tot}(0)$. 
This closely matches the net deposited energy
\[
    \Delta E_{\rm dep} = \int _0^s \frac{\dd s}{v_p}\braket{\Gamma - \Lambda}, 
\]
indicating the wind has reached a quasi-steady state. 
Also shown are the cumulative X-ray heating and \ce{H2} cooling: 
\[
    \Delta E_{\rm X} = \int _0^s \frac{\dd s}{v_p}\Gamma_{\rm X}, \quad 
    \Delta E_{\rm H2} = \int _0^s \frac{\dd s}{v_p}\Lambda_{\rm H2},
\]
for comparison.


\section{Mass-Loss Rates}   \label{sec:mass-loss_rates}

Here, we evaluate the cumulative mass-loss rates $\dot{M}(R)$ and surface mass-loss rates $\dot{\Sigma}$ from our fiducial simulation. 

The cumulative mass-loss rate $\dot{M}(R)$ represents the total mass lost through gas launched from $< R$. 
To compute it, we first calculate the total mass flux through the outer computational boundary as a function of $\theta$:
\[
    \dot{M}_{\rm out} (\theta) = \int_{0}^\theta \dd \theta \, 4\pi \rho v_r r^2 \sin\theta .
\]
We then trace a streamline back from the outer boundary $(r_{\rm max}, \theta)$ to identify the corresponding wind-launching radius $R$, defined following the method in Appendix~\ref{sec:streamlines:base}. 
The cumulative mass-loss rate $\dot{M}(R)$ is then given by $\dot{M}_{\rm out}(\theta)$. 
This estimate excludes contributions from gas that does not satisfy the unbound condition before reaching the outer boundary along the streamline: $E_{\rm tot} > 0$. 
The corresponding surface mass-loss rates is defined as 
\[
    \dot{\Sigma} \equiv \frac{1}{2\pi R}\frac{\dd \dot{M}}{\dd R}. 
\]
We similarly compute the cumulative and surface mass-loss rates for \ce{H2}. 

\begin{figure*}[htbp]
    \centering
    \includegraphics[width=\linewidth]{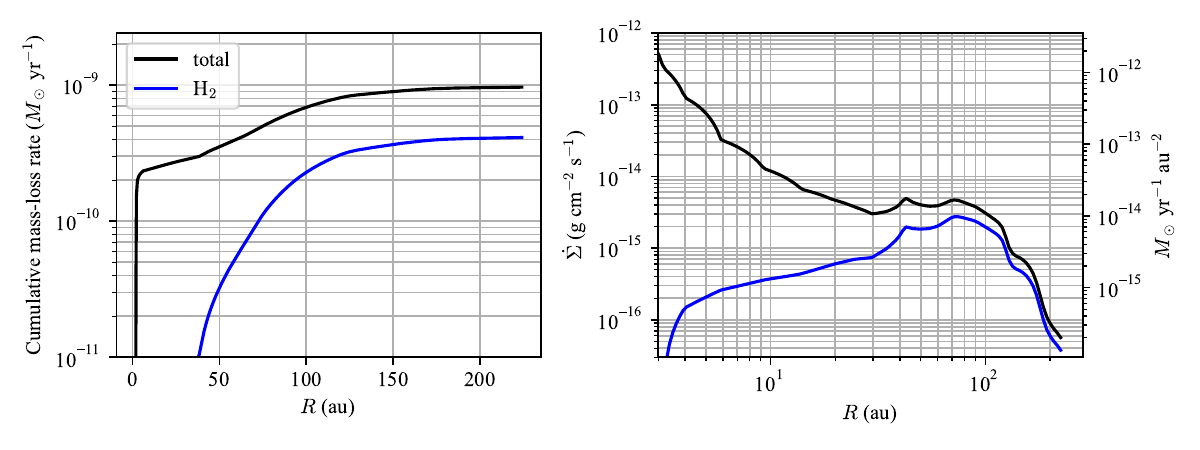}
    \caption{Cumulative (left) and surface (right) mass-loss rates, with black and blue lines indicating the total and \ce{H2} mass-loss contributions, respectively. 
    Note that the total mass-loss rates include the contributions from helium-bearing species. 
    }
    \label{fig:mass-loss_rates}
\end{figure*}

These results are shown in \fref{fig:mass-loss_rates}. 
Note that the total mass-loss rates (black lines) include helium contributions. 
Therefore, \ce{H2} is responsible for almost all the surface mass loss of hydrogen-bearing species at $\gtrsim 40\au$. 

The sharp cutoff in $\dot{M}$ and $\dot{\Sigma}$ at $R\sim 220\au$ reflects the computational boundary, beyond which launched gas do not meet the unbound condition.
This is a numerical artifact; a larger domain would likely extend the wind-launching region farther out. 

Toward the inner disk, the total $\dot{\Sigma}$ increases with decreasing radius.
This again highlights a limitation of our computational domain, which does not resolve the critical radius ($\sim 1\au$), where $\dot{\Sigma}$ is expected to plateau and sharply drop inward.


\end{appendix}

\end{document}